# A STUDY OF MULTIPLE REFRACTIVE SCATTERING OF MONOENERGETIC X-RAYS FROM ENSEMBLES OF MONODISPERSE SPHERES

Masterarbeit

zur Erlangung des akademischen Grades

Master of Science in Physics

(M.Sc.)

dem Fachbereich Physik der

Universität Siegen

vorgelegt von

*Anastasiya Khromova*

Januar 2005

Dedicated to my husband Francesco,

who was with me in sorrow and happiness all this time

TABLE OF CONTENTS

LIST OF FIGURES

x

# LIST OF TABLES

# ACKNOWLEDGMENTS

This work was supported by the UA under the contract HPRI-CT-1999-50008 and by an individual DAAD grant.

*Chapter 1*

# INTRODUCTION

Over the last years worldwide research on new methods of X-ray imaging has been carried out. The major use of X-rays has been, and still is, the visualization of the inside of systems which are not transparent to visible light. The resulting applications range from whole-body imaging to studies of atomic and molecular structures in vivo and in vitro.

The photon energy range used in practice is typically 17-150 keV. At these energies the important X-ray photon interactions are photoelectric absorption and scattering. Scattering is divided into resonant and non-resonant part, where resonant scattering goes via an intermediate state. The intermediate state relaxes by production of fluorescence radiation and Auger electrons. Non-resonant scattering can be elastic (coherent) and inelastic (incoherent, Compton scattering). All these types of interaction give rise to different imaging techniques which will be shortly discussed below. The performance of the imaging system and its components can be specified in terms of contrast, sharpness, radiation dose and noise [1]. The contrast is defined as

$$C = \frac{N_1 - N_2}{(N_1 + N_2)/2}, \quad N_1 - N_2 \ll N_1 \tag{1.1}$$

where $N_1$ and $N_2$ are the average counts per pixel measured on the background and on the detail of the image, respectively. The sources of contrast can be divided in those that change the intensity (based upon the changes in the attenuation coefficient) and the phase of the beam traversing the sample, and those that arise from shifting the scattering away from the transmitted beam. The minimum observable contrast is limited by the noise of the signal. The signal-to-noise ratio (SNR) measures the visibility of the detail, and in the case of Poisson distribution it is between

$$SNR = \frac{N_1 - N_2}{\sqrt{N_1 + N_2}} \text{ and } SNR = \frac{N_1 - N_2}{\sqrt{N_1}}, \tag{1.2}$$

depending on the number of background pixels contributing.

The SNR is improved when the incident intensity is increased, but the maximum allowed dose sets a limit. As a function of energy the SNR achieves a maximum at a certain energy depending on the object. With increasing energy the SNR decreases because the difference between the absorption coefficient becomes smaller, but at the same time the statistical contribution becomes better due to smaller absorption and therefore an increased number of photons detected. Enhancement of the contrast is essential in medical and biological imaging applications in order to limit the radiation dose.

The interactions of X-rays with matter can be expressed through the complex refractive index n:

$$n = 1 - \delta - i\beta \tag{1.3}$$

The real part δ corresponds to the phase shift due to scattering and the imaginary part β is related to the absorption of the wave. The real and imaginary parts can be written in terms of the atomic scattering factor $f = f_0(\theta) + f' + if''$:

$$\delta = \frac{\lambda^2}{2\pi} r_0 N_A \sum_i \frac{\rho_i}{A_i}\left(f_{0,i} + f_i'\right) = \frac{\lambda^2}{2\pi} r_0 \rho_e, \tag{1.4}$$

$$\beta = \frac{\lambda^2}{2\pi} r_0 N_A \sum_i \frac{\rho_i}{A_i}\left|f_i''\right| = \frac{\lambda \mu_X}{4\pi}, \tag{1.5}$$

where $N_A = 6.0221415 \times 10^{23} \, mol^{-1}$ is Avogadro's number, $r_0 = 2.82 \times 10^{-13} \, cm$ is the classical electron radius, λ is the wavelength of the incident radiation, $\rho_i$ is the density in the sample of element i with atomic weight $A_i$, $\rho_e$ is an electron density and $\mu_x$ is the linear attenuation coefficient. For forward scattering $f_0$ is equal to the atomic number Z. The last part of relation (1.4) holds above the highest K-edge of the object.

One can give an approximate form for the real part calculation where $\rho$ is the object density:

$$\frac{r_0}{2\pi} N_A = 2.72 \times 10^{10} \, [cm/mol]$$
$$\delta = 2.72 \times 10^{10} \, (\frac{Z}{A}) \rho [gcm^{-3}] \lambda^2 [cm^2] \tag{1.6}$$

Conventional (absorption) radiographs are still by far the most common X-ray images, both for medical and technical needs. The most recent techniques are the coronary angiography [2]-[10], bronchography [11], computed tomography [12]-[13] and mammography [14]-[16]. The conventional approach relies on the X-ray absorption term β as the main source of contrast. In absorption imaging, the contrast results from the variation of X-ray absorption arising from density differences and from variation in the thickness and composition of the specimen. Absorption contrast works well in distinguishing between hard and soft tissue: heavier elements, like calcium in bones and teeth, have a much higher absorption cross section than the lighter elements that constitute soft tissues. However, in many clinical situations, such as mammography, there is a need to distinguish between different kinds of soft tissues: between tumors and normal tissue for example. Because the absorption is low to begin with, and differences in density and composition are small, standard absorption X-ray imaging is not successful at this task. Moreover, the scattering components (both coherent and incoherent), which are always present in the conventional radiography, contribute to loss of contrast and spatial resolution (thus reducing the SNR). This effect of the scattered radiation can be reduced by using an anti-scatter grid. In this way only roughly 10% of the scattered radiation is reaching the detector, but at the cost of increasing the dose by about a factor of two. It is possible also to reduce the field of view to a single pencil beam or to put the detector at some distance from the body in order to allow the scattered photons to escape (to create the so called air gap) to reduce the scatter in the image. Using synchrotron radiation (SR), having a high natural collimation at least in the vertical direction, together with a monochromator, which selects the energy band, it became possible to reduce the scattering in the image without utilization of anti-scatter grid.

Apart from affecting the image quality, the scattered radiation gives essential information about the electronic structure of the sample. Different features of the scattering processes are exploited in medical imaging and therapy, and it is very important to understand their roles. For example, synchrotron X-ray fluorescence scattering has been applied to breast tissue analysis [17]. Small and wide-angle X-ray scattering (SAXS and WAXS) is a useful and complementary method for determining the size, size distribution and structure of a wide range of partially ordered (non-crystalline or semi-crystalline) materials and specimens. Examples include polymers, liquid crystals, oils, suspensions and biological samples like fibers or protein molecules in solution. The typical length scale is 0.1-200 nm. Small-angle X-ray scattering (SAXS) [18], [19] covers the range 2-200 nm and occurs at low scattering angles (1-10°) while wide-angle X-ray scattering (WAXS), responsible for the imaging of smaller structures, conventionally covers the angular range 7-60°. This technique can be used in conjunction with methods that influence and/or change the samples' structural characteristics (e.g. during temperature changes, under shear forces or upon the application of electric/magnetic field stimuli). Another methodology, called bone mineral densitometry, measures the amount of bone material (calcium hydroxyapatite) per unit volume of bone tissue. There exist several methods based on different physical processes, but one of the most successful involves measurement of the amount of radiation coherently and incoherently scattered by the bone tissue [20]. Since the amount that is coherently scattered is dependent on $Z^2$ (Z is the atomic number) while that is incoherently

scattered – on Z, the ratio coherent/incoherent scatter is very sensitive to the bone mineral density. Recently, the method has been extended to tomographical imaging of the Z-distribution using high-energy SR [21]. Several tomographic imaging applications based on scattered radiation detection have also been devised [22]-[24].

Besides absorption and scattering, an X-ray traversing an object picks up information on its refraction properties (phase shift information) that can be also utilized as a source of contrast for displaying the internal properties of the sample. These effects are important especially for small objects (30-100 μm) with either low absorption coefficient or low contrast quality: the arterial system in the heart, the bronchi, lung alveoli, clusters of micro-calcifications and low-contrast masses that are possible indicators of early breast cancer.

The phase shift φ(x, y) the lowest order, i.e. above the K-edge of the heaviest material in the object, can be expressed as:

$$\varphi(x,y) = \frac{2\pi}{\lambda}\int \delta(x,y,z)dz = -r_0 \lambda \int \rho_e(x,y,z)dz \qquad (1.7)$$

where $\rho_e(x,y,z)$ is the electron density at point $(x,y,z)$ and X-rays are propagating in the **z**-direction.

Refraction causes a deflection of the X-rays in the direction of the phase shift gradient. In one of the transverse variables, for instance x, the wave is refracted by an angle Δθ with

$$\Delta\theta \approx \frac{\lambda}{2\cdot\pi}\cdot\frac{\partial\varphi(x,y)}{\partial x}. \qquad (1.8)$$

Equations (1.7) and (1.8) can be derived from (1.4). For typical mammographic energies (15-25 keV) the real part of n-1, δ, is 1000 times larger than the absorption term β ($10^{-7}$ to $10^{-11}$) [25]. Hence, it is possible to see phase contrast when absorption contrast is undetectable. Phase contrast depends less strongly on energy than absorption contrast ($\mu \propto E^{-3}$, while $\varphi \propto E^{-1}$). Therefore, imaging can be done at higher energies while the absorbed radiation dose can be decreased, thus reducing damage to tissues.

Different techniques have been developed to exploit phase information (based on the detection of the phase variations), which can be classified into three categories: diffractometry or diffraction enhanced imaging (DEI), interferometry and in-line holography [25]. This thesis will focus on the DEI technique; for an extended discussion of the other methods we refer the reader to [26-28]. Diffraction Enhanced Imaging was first measured and reported by Förster et al. [29] for measuring with micrometric precision the wall thickness of spherical targets for laser nuclear fusion. Later several groups have implemented this concept for mammography studies [30-33].

Diffraction Enhanced Imaging (DEI) is a recent phase sensitive technique based on the use of an analyzer crystal placed between the sample and the detector. Chapter 2 is totally dedicated to the description of this technique. The original DEI algorithm proposed by Chapman et al. [30] is based on the acquisition of two images, one on each side of the analyzer crystal rocking curve (RC) at 50% of the reflectivity (on the slopes). The images are then processed to obtain the so called *apparent absorption image* and *refraction image*. The sources of contrast in the first are absorption and extinction (scatter rejection). In the refraction image the intensity in each pixel is mainly determined by the angle of refraction in the object plane. With such a setup it became possible to improve the visibility of small low absorbing and low contrast structures like lung alveoli not visible with conventional (absorption) X-ray imaging.

As it is known from anatomy, each lung contains around three hundred million alveoli with size from 30 to 100 μm (in case of exhalation and inhalation, respectively), not detectable separately. A big amount of such small structures give rise to the effect of *multiple refractive scattering*. Multiple scattering in DEI, like any scattered radiation, reduces the spatial resolution and, correspondingly, image contrast. Multiple scattering in DEI covers the angular range up to about 100 μradians for medically relevant objects. As scattered radiation of larger angles, multiple scattering in DEI could give also additional information on the internal structure of the object.

When the effect of multiple scattering is present, but substantially smaller than the RC width, the signal is proportional to the second derivative of the RC and, therefore, drastically decreases on the half slope regions used in normal DEI. How the scattering information is extracted will be discussed in Chapter 2.

In our case to study the effect of multiple scattering of X-ray photons in the experiment a phantom to model lung tissue was built by filling a flat Plexiglas box with monodisperse PMMA (poly-methyl-methacrylate $(C_5H_8O_2)_n$) microspheres. The box was milled in the form of a stair with five steps of different depth: 0.5, 1, 2, 3 and 5mm. Microspheres of three different sizes were used: 6, 30 and 100 μm. Each sample was exposed to 17, 25 and 30 keV. X-rays multiple scattering is responsible for the broadening of the rocking curve when the rays are passing through the object. Having different stair depth allows to investigate the influence of the object thickness, and correspondingly of the multiple scattering, on the image contrast and resolution as well as to find out which system parameters (as energy, exposure time, analyzer position etc.) are optimal to use with each specific kind of sample.

With the samples described above measurements of the broadening of the rocking curve (RC) (refraction angle distribution) have been done. To demonstrate that the behavior of the experimental data can be understood, a detailed Monte Carlo simulation, dealing with the propagation of X-rays through objects composed of layers of microspheres with different diameter, has been developed and compared to experimental data taken at the medical beamline at ELETTRA (Trieste, Italy).

This thesis is organized as follows:

Chapter 2 will explain the main principles of the Diffraction Enhanced Imaging algorithm and the ways of its utilization. It also addresses the problem of multiple scattering in the fine structure samples and the new methods like improvement of the existing DEI method, created to enhance the contrast in the images of such objects.

In Chapter 3 the experimental measurements made at the SYRMEP beamline at ELETTRA (Trieste, Italy) will be presented. In this chapter will be described:

1. The phantoms used in the experiment which models the highly scattering sample for the investigation of the effect of the multiple refractive scattering
2. The DEI set-up at ELETTRA.
3. Image processing using the DEI technique
4. The Data analysis

In Chapter 4 will be described the Monte Carlo photon transport program for simulation of multiple refractive scattering. In this chapter will be presented:

1. The simulated phantoms with similar properties as the experimental one
2. The Monte Carlo program algorithm

In Chapter 5 will be shown the results obtained by the Monte Carlo program simulations. A comparison between the experimental and simulated data will be discussed in Chapter 6.

At the end of the thesis in Appendix A the Monte Carlo program used for the simulation is shown.

Appendix B is dedicated to an X-ray tube spectrum simulation, an independent study and development. The theory used and the correspondent programs are presented.

*Chapter 2*

# Diffraction Enhanced Imaging (DEI) Technique

## 2.1  THE MAIN PRINCIPLES OF THE DEI TECHNIQUE – PROBLEM OF MULTIPLE REFRACTIVE SCATTERING

Diffraction Enhanced Imaging is a phase sensitive technique based on the use of an analyzer crystal placed between the sample and the detector (see Figure 2.1.1). With such a setup it became possible to improve the visibility of low absorbing and low contrast structures, not visible in the conventional X-ray images.

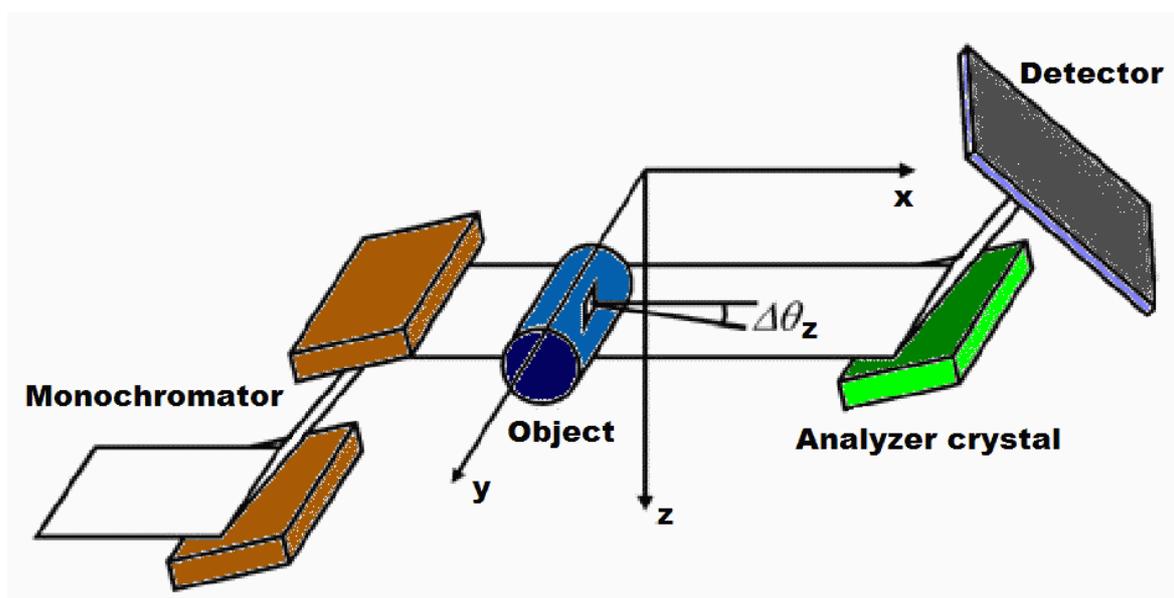

*Figure 2.1.1*: Sketch of the DEI setup

A monochromatic beam, received by a double-crystal monochromator, traverses the sample and falls onto the analyzer crystal plane. If the incident rays lie within the angular acceptance window of the analyzer crystal, they are diffracted onto the detector. In this way the analyzer crystal works like an angle and energy filter. The angular acceptance curve is called the rocking curve (RC) of the crystal (that is the angular sensitivity function defined by the reflectivity of the crystal system). It characterizes the X-ray output as a function of angular position of the analyzer when no object is present in the beam. An example of the rocking curve for Si (1, 1, 1) at 17 keV, generated by XOP 2.0 program [34], is shown in Figure 2.1.2. For the X-ray energies and crystal reflections used here, the width of this curve is a few microradians.

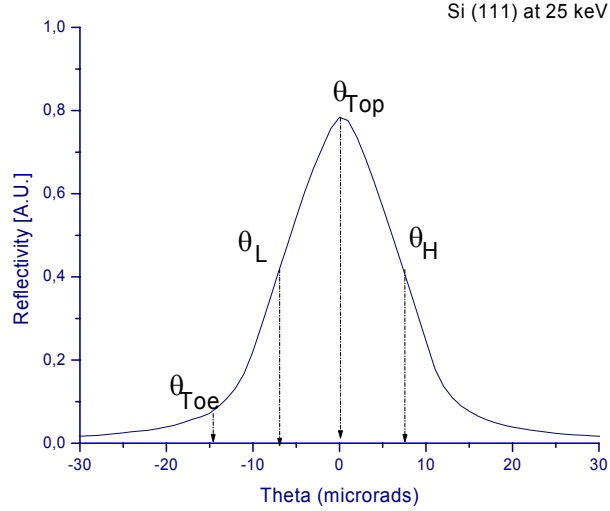

*Figure 2.1.2:* Rocking curve of the Si (1,1,1) at 25 keV with thickness 0.026cm simulated by the XOP2.0 program

In the sample the X-rays propagate by absorption and deviate slightly from the initial direction. These deviations arise from two factors: the refraction of X-rays due to refractive index gradients in the phantom and the ultra-small angle X-ray scattering (USAXS [13], multiple refractive scattering), generated by very small and unresolved structures. At the scale of the RC width this can be understood in the following way: the refraction characteristics of the object are represented by measuring the displacement, caused by the presence of the sample, of the centroid of the refraction angle distribution (RC) compared to the RC without the sample in the beam; the deviations due to the multiple scattering cause the refraction angular distribution to broaden and become less highly peaked. The scattering in the milliradian range (so called small-angle scattering), resulted from the X-ray diffraction by the organized structures within the sample, is totally removed due to the narrow acceptance width of the analyzer crystal (in the microradian range). It is this rejection of small-angle scattering that gives rise to the tremendous contrast improvement in the DEI technique.

The intensity of the beam is the number of photons per unit time and per unit area. $I_R$ will represent the intensity of those photons which leave the sample and fall within the filtering window of the analyzer crystal (in the microradian range). All other photons are rejected (see above). The analyzer crystal orientation is chosen to diffract the X-ray beam in the vertical plane at an angle $\Delta\theta_Z$ with respect to the initial direction.

Therefore, the intensity $I_B$ reaching the detector and modulated by the analyzer crystal rocking curve is equal to:

$$I_B = I_R R(\theta_B + \theta' + \Delta\theta_Z) \qquad (2.1.1)$$

where $R(\theta)$ is the analyzer reflectivity (rocking curve), $\theta_B$ is a Bragg angle and $\theta'$ is a misaligned angle with respect to the monochromator.

The Bragg angle is the angle between the incident X-ray photon direction and a set of crystal (in our case the analyzer crystal) diffraction planes. Since the analyzer crystal can be slightly misaligned with respect to the monochromator (in equation 2.1.1 this angle is called $\theta'$) the angle between the lattice planes of the analyzer crystal and the direction of the photons becomes $\theta_T = \theta_B + \theta' + \Delta\theta_Z$. Bragg's law refers to the simple equation: $n\lambda = 2d\sin\theta_B$. The variable $d$ is the distance between atomic layers in a crystal, the variable $\lambda$ is the wavelength of the incident X-ray beam, $n$ is an integer giving the reflection order and $2\theta_B$ is the scattering angle. In Figure 2.1.3 Bragg's law is depicted.

In the DEI algorithm proposed by Chapman [30] two images are acquired on the slopes (reflectivity is 50 %) of the rocking curve, at the angular positions $\theta_L = \theta_B - \dfrac{\Delta\theta_D}{2}$ (smaller angles) and $\theta_H = \theta_B + \dfrac{\Delta\theta_D}{2}$ (higher angles), where $\Delta\theta_D$ is the full width at half maximum (FWHM) of the rocking curve (see Figure 2.1.2). In this case $\pm\dfrac{\Delta\theta_D}{2}$ is the misaligned angle $\theta'$.

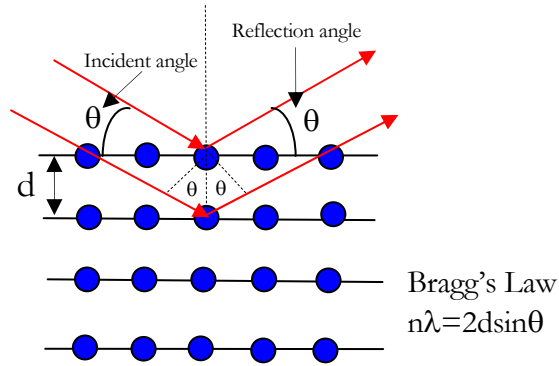

*Figure 2.1.3* Bragg's law. Here $\theta = \theta_B$ (Bragg angle), d is the distance between the atomic layers in a crystal, λ is the X-rays wavelength, n is an integer giving the reflection order

Equation (2.1.1) is solved for the two unknowns, $I_R$ and $\Delta\theta_Z$, using a first order Taylor expansion for $\theta_T = \theta_B + \theta' + \Delta\theta_Z$, leading to the two equations :

$$I_R = \frac{I_L \cdot \left(\frac{dR}{d\theta_T}\right)\bigg|_{\theta_H} - I_H \cdot \left(\frac{dR}{d\theta_T}\right)\bigg|_{\theta_L}}{R(\theta_L) \cdot \left(\frac{dR}{d\theta_T}\right)\bigg|_{\theta_H} - R(\theta_H) \cdot \left(\frac{dR}{d\theta_T}\right)\bigg|_{\theta_L}} \quad (2.1.2)$$

$$\Delta\theta_Z = \frac{I_H \cdot R(\theta_L) - I_L \cdot R(\theta_H)}{I_L \cdot \left(\frac{dR}{d\theta_T}\right)\bigg|_{\theta_H} - I_H \cdot \left(\frac{dR}{d\theta_T}\right)\bigg|_{\theta_L}} \quad (2.1.3)$$

$I_R$ is called an *apparent absorption image* and $\Delta\theta_Z$ is a *refraction image*. Sources of contrast in the *apparent absorption image* are absorption and extinction (scatter rejection). Extinction means here missing of those rays that are refracted under angles higher than the acceptance of the analyzer (see above). In the *refraction image* the intensity in each pixel is mainly determined by the angle of refraction in the object plane.

However, the method described above does not consider the effect of *multiple refractive scattering* (USAXS). In other words, this method works only in the case when each detector pixel is focused on a specific part of the sample where a well-defined refraction angle $\Delta\theta_Z$ is expected (Figure 2.1.4). But for the big amount of fine structures each pixel collects several photons, deviated at different refraction angles (Figure 2.1.5) and the signal on the slopes becomes very weak.

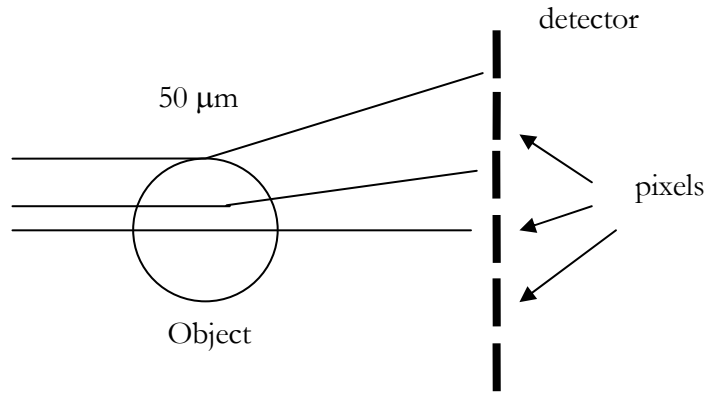

*Figure 2.1.4*: Pixel aperture resolving the structure of the object

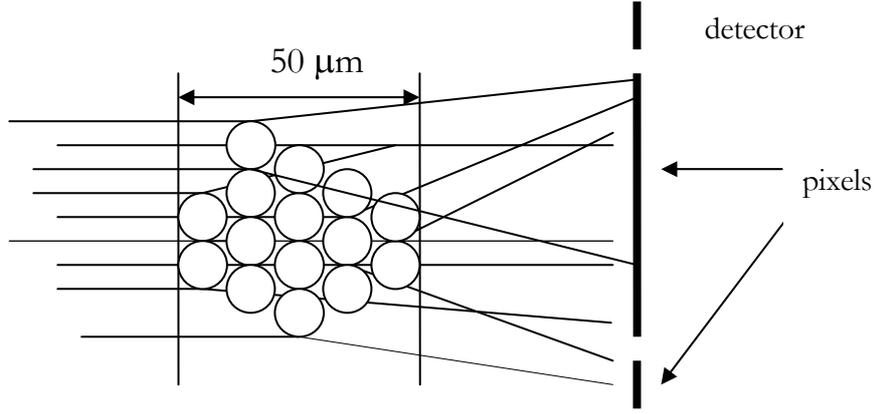

*Figure 2.1.5*: Pixel aperture sensitive to the fine structure of the object

On the other hand, a good signal for the fine structure samples is achieved on the top and toes of the RC: 100 % and 10% of the reflectivity, respectively. In this case, the signal is proportional to the second derivative of the RC. Based on this evidence a new algorithm has been proposed [35]. In comparison to the earlier method, the new algorithm measures the standard deviation $\sigma^2_{\Delta\theta_Z}$ of the refractive angle distribution of X-rays leaving the sample. In this case, the detector pixel collects a whole statistical distribution of angles instead of a single well-defined angle $\Delta\theta_Z$:

$$I_R = \frac{I_{TOE} \cdot \left.\frac{\partial^2 R}{\partial \theta_T^2}\right|_{\theta_{TOP}} - I_{TOP} \cdot \left.\frac{\partial^2 R}{\partial \theta_T^2}\right|_{\theta_{TOE}}}{R(\theta_{TOE}) \cdot \left.\frac{\partial^2 R}{\partial \theta_T^2}\right|_{\theta_{TOP}} - R(\theta_{TOP}) \cdot \left.\frac{\partial^2 R}{\partial \theta_T^2}\right|_{\theta_{TOE}}} \quad (2.1.4)$$

$$\sigma^2_{\Delta\theta_Z} = 2 \cdot \frac{I_{TOP} \cdot R(\theta_{TOE}) - I_{TOE} \cdot R(\theta_{TOP})}{I_{TOE} \cdot \left.\frac{\partial^2 R}{\partial \theta_T^2}\right|_{\theta_{TOP}} - I_{TOP} \cdot \left.\frac{\partial^2 R}{\partial \theta_T^2}\right|_{\theta_{TOE}}} \quad (2.1.5)$$

Equation (2.1.4), like equation (2.1.2), gives $I_R$ and thus can be called *apparent absorption image*. The image calculated according to equation (2.1.5) can be called *refraction image* (standard deviation, responsible for the multiple refractive scattering information).

An example of images taken at different positions on the rocking curve for a 100 μm sphere phantom at 25 keV is shown in Figure 2.1.6a-2.1.6e.

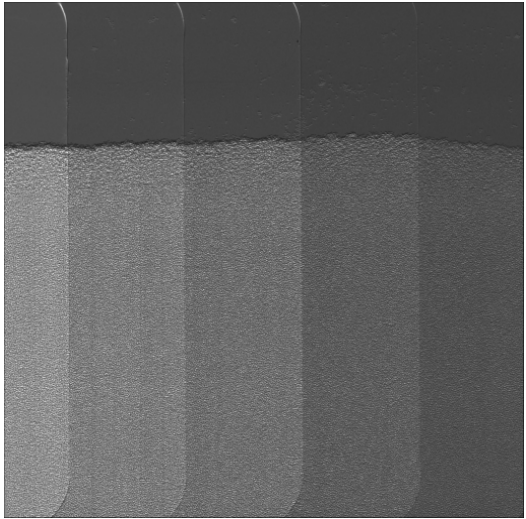

*Figure 2.1.6a:* Image at 10% of the reflectivity (left toe)

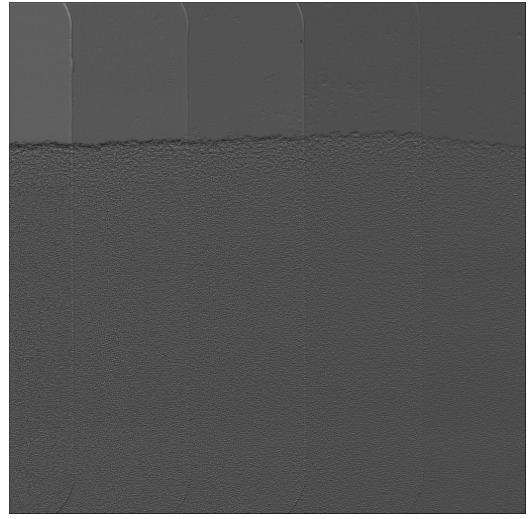

*Figure 2.1.6b:* Image at 50% of the reflectivity (left slope)

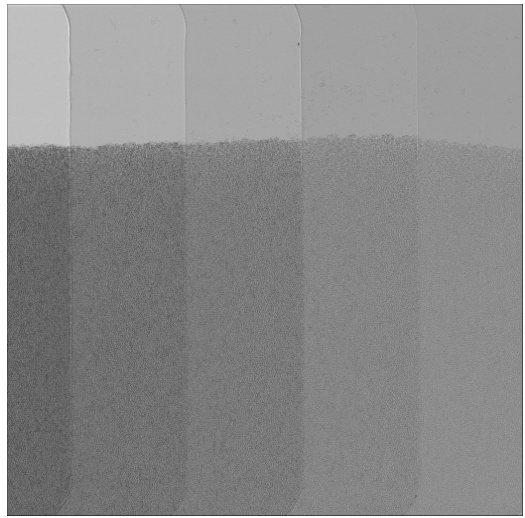

*Figure 2.1.6c:* Image at the top of the rocking curve (100% of the reflectivity)

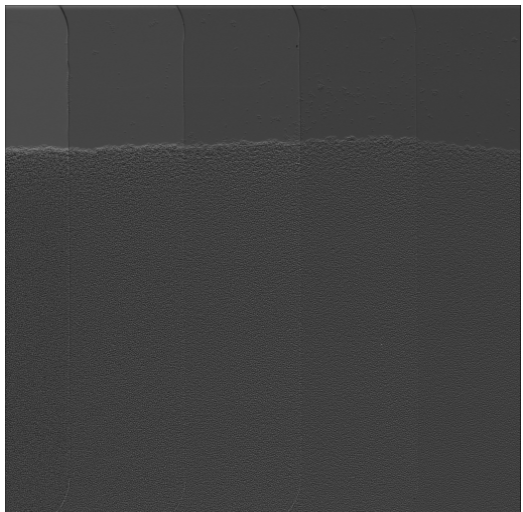

*Figure 2.1.6 d:* Image at 50% of the reflectivity (right slope)

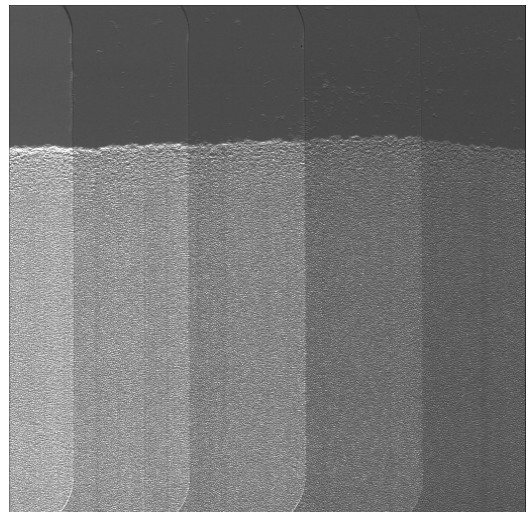

*Figure 2.1.6e:* Image at 10% of the reflectivity (right toe)

The image in Figure 2.1.6c is produced setting the analyzer crystal on the top of the rocking curve (100 % reflectivity). One can see that all the different thickness of the phantom are clearly visible, including the 0.5 mm thick layer. In this position of the analyzer crystal a very strong contrast enhancement has been observed due to the rejection of the multiple scattering, originated from the refraction on the microspheres.

Figures 2.1.6b and 2.1.6d present the phantom acquired at 50% reflectivity position on the rocking curve. One can see here that the contrast decreases in all stair depth in comparison to the image on the top. The reason for this lies in the fact that a balance between the number of refracted rays that are rejected and that are accepted is achieved. Ideally we should see on such kind of images the edge enhancement of the spheres, but due to the multiple scattering effects this enhancement is blurred. This is typical for all the samples featuring a big amount of small unresolved structures and this is the reason why the original Chapman DEI algorithm does not suit this case.

The images 2.1.6a and 2.1.6e correspond to a position about 10% of the reflectivity curve. Here the scattered photons are mainly collected, which contribute to a positive signal and a complete reverse contrast is observed.

Recently other methods have been proposed [36], [37], which are an improvement on the previous DEI algorithms. In these methods multiple images are acquired (Multiple Image Radiography (MIR)) at N different positions of the analyzer RC.

This method allows extracting scattering images separately from the absorption and refraction images and gives more complete representation of the object's effect on the X-rays photons beam. A MIR data set is a multidimensional image with two spatial coordinates (x and y) and one angular coordinate. This means the following: images taken at each angular position of the analyzer crystal are two dimensional ((x, y)); they are stacked to each other to receive a box-like structure (the z coordinate represents the taken image number, in ascending order); in this way for each pixel (spatial location) in the image the individual angular distribution is obtained. The angular distribution of the intensity reflected by the analyzer crystal is the convolution of the object angular spectrum (*sample* RC) with the analyzer RC (*reference* RC):

$$\sigma^2{}_{measured} = \sigma^2{}_{RC} + \sigma^2{}_{sample} \qquad (2.1.6)$$

The information from the MIR method can be analysed at different stages. At a basic level we can be interested in determining, for each pixel in the image, the angular distribution of the X-ray photons directly after the sample. This is a very important source of information to describe the sample's properties.

For each pixel we have the zero-, first-, second-order moments $M_i = \sum_{j=1...N}(\theta_j)^i \cdot R(\theta_j)$ (i=0, 1, 2) and the maximum of the RC (see [36], [37] for a more precise description). $R(\theta_j)$ is the intensity in a pixel for the angular setting $\theta_j$. This corresponds to the integrated intensity, the center of mass, the standard deviation and the maximum reflectivity, respectively.

- The ratio of the integrated intensities of the sample and the reference RC is the integrated absorption;
- The angular displacement of the center of mass of the sample RC with respect to the reference RC is the integrated refraction;
- The ultra-small angle scattering is the sample RC standard deviation deconvolved from the reference RC;
- The ratio of the sample RC maximum with respect to the reference RC means here the maximum absorption ([36], [37]).

More advanced, a set of images can be obtained each containing information per pixel about the parameters just described: integrated absorption, integrated refraction and scattering images (and, if necessary, kurtosis and skewness images).

It is evident that accurate results can be obtained using very detailed analysis (pixel-by-pixel) of many more images than in Chapman's original approach. This can be useful especially for the investigation of the fine structure samples, since the resolution and the contrast of such objects in DEI is reduced due to the effect of multiple scattering. In MIR these problems are partially solved. But for medical applications the N/2 times (N is the number of desired different positions on the rocking curve) increase of the radiation dose cannot be ignored.

*Chapter 3*

# Experimental Investigations

## 3.1 THE DEI SET-UP AT ELETTRA

Experimental studies on multiple scattering have been carried out at the at the SYRMEP beamline of the synchrotron radiation facility ELETTRA in Trieste (Italy).

Elettra is a third generation synchrotron light source that can be operated at 2.0 or 2.4 GeV. The SYRMEP beamline radiation source is a bending magnet, and the beam is characterised by a vertical size of about 70-100 μm (FWHM). A schematic setup is shown in Figure 3.1.1. A detailed description of the DEI set-up used in the SYRMEP beamline can be found in references [38], [39].

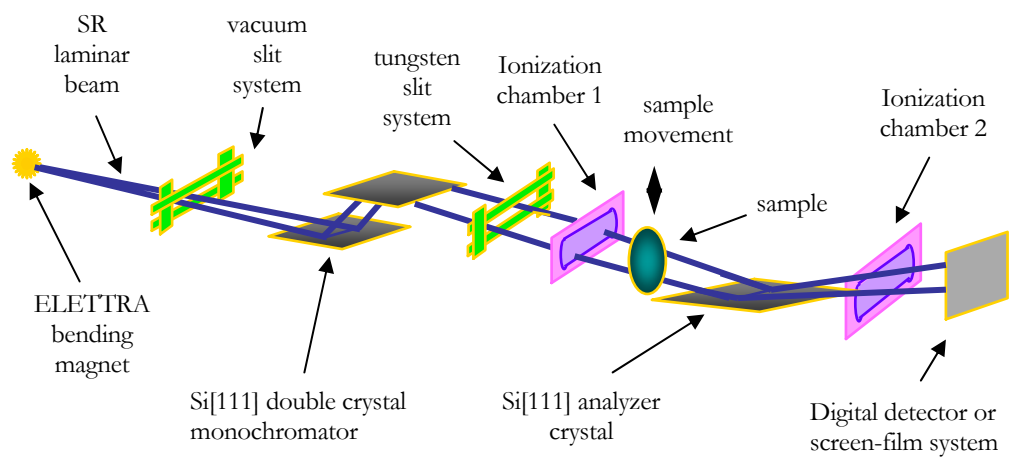

*Figure 3.1.1:* The DEI setup at Elettra

The monochromator is made of two Si(1,1,1) crystals (10mm thick, 140mm wide, 77 and 130mm long) utilised in the symmetric Bragg configuration. The monochromator is placed in Ultra High Vacuum (UHV) at a distance of 18 m from the source and is able to select an energy from 8.5 keV to 35 keV in the first harmonic. The sample, the analyser crystal and the detector are located at 23-25m from the source. The analyser is a single flat Si(1,1,1) perfect crystal (10mm thick, 130mm wide, 80mm long). Two ionisation chambers are placed before and after the analyser. From the ratio of the measured currents it is possible to determine the analyser's angle on the rocking curve. A CCD camera has been used as a detector. This camera is equipped with a 40μm thick intensifier screen (gadolinium oxysulphide scintillator) and has an active area of 29mm×29mm, subdivided into 2048×2048 pixels, having a minimum pixel size of 14×14 $\mu m^2$. Two kinds of optical systems before the CCD camera have been used. The area of the first taper is equal to the detector pixel area and gives a resolution of 14μm. The active area of the second taper is 1/16 of the original CCD area, shrinking correspondingly the pixel size and giving an effective resolution per pixel equal to 5μm.

### 3.2 SAMPLE DESCRIPTION

To study the effect of multiple scattering of X-rays, a phantom has been built simulating a simplified lung tissue, by filling a flat Plexiglas box with monodisperse PMMA (poly-methyl-methacrylate $(C_5H_8O_2)_n$) microspheres. The box was milled in the form of a stair, with five steps of different depth: 0.5, 1, 2, 3 and 5 mm, respectively. All stairs have the same width of 6 mm. Microspheres of three different sizes were used: 6, 30 and 100 μm in diameter. Each sample was exposed to energies of 17 keV, 25 keV and 30 keV. Plexiglas is a very stable polymer, from which it is possible to make very small objects (μm range) with high precision (in the order of 1 %) and nearly no absorption (in X-rays range) – qualities needed for our experiment. A photo of the phantom is shown in Figure 3.2.1 together with its simplified artistic schematic.

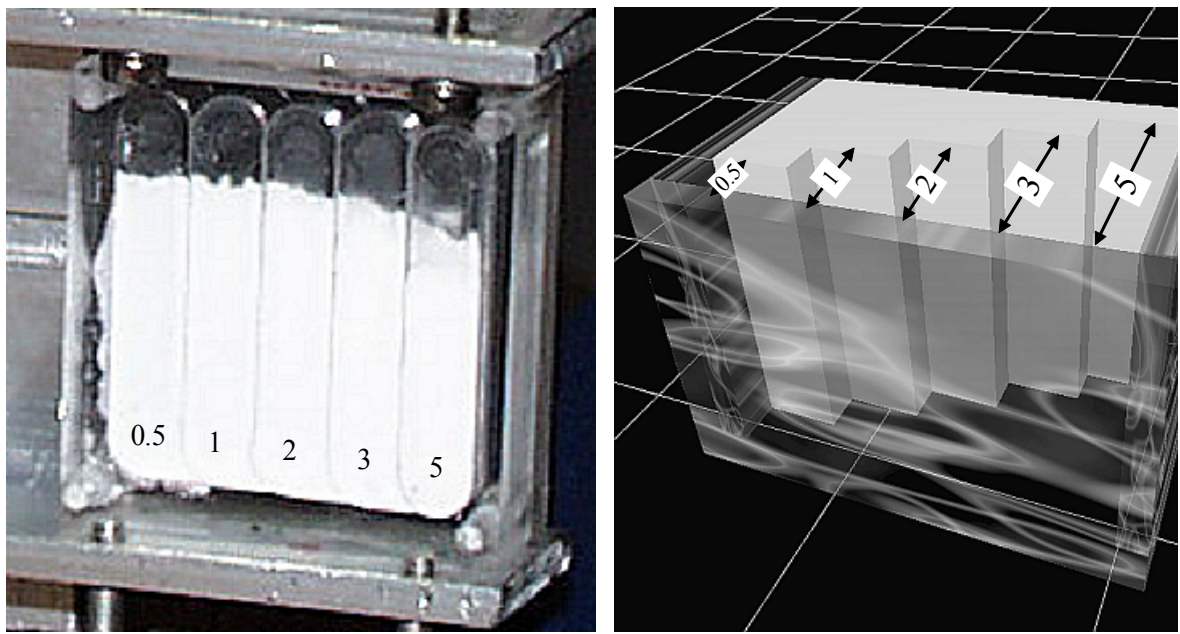

*Figure 3.2.1*: Plexiglas phantom filled with monodisperse PMMA microspheres

## 3.3 EXPERIMENTAL METHOD

Phantoms as described above were utilized in multiple image studies using the analyzer crystal. During the exposure time both the detector and the analyzer crystal are moved continuously. The beam traverses the sample through the stairs' depth. The analyzer crystal moves to scan N positions on the rocking curve: for each new position of the analyzer crystal we have a new position of the detector in order to catch the X-rays reflected from the analyzer crystal surface. An example of the resulting two-dimensional image for 100μm spheres at 17keV is presented in Figure 3.3.1.

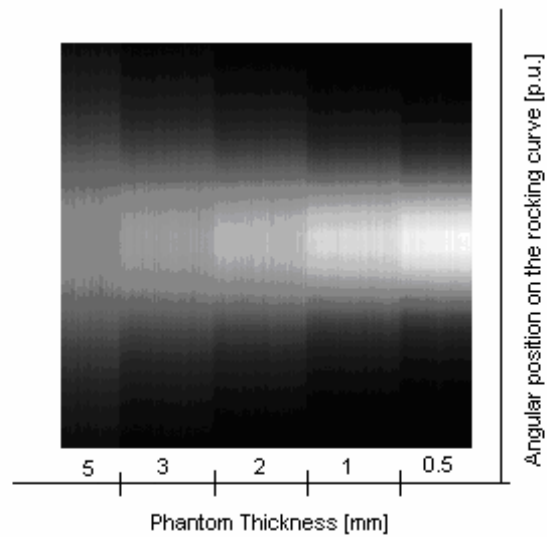

*Figure 3.3.1*: 100μm spheres at 17keV, SYRMEP ELETTRA (Trieste) from 04/2002

In Figure 3.3.1 the abscissa corresponds to the phantom stair thickness (mm) and the ordinate corresponds to the angular position on the rocking curve (arbitrary units). Analysis of the phantom scattered angle distribution with the analyzer crystal in a multi-image approach results in a broadening of the effective rocking curve. Visually the effect is noticeable in Figure 3.3.1 as the broadening of the white strip in the middle of the image. The effective rocking curve is a convolution of the intrinsic rocking curve of the analyzer crystal with the scattered (refraction) angle distribution of the X-rays passing through the sample (see equation 2.1.6 in Chapter 2). The crystal used for the analyzer is Si (1, 1, 1). Since in the PMMA wall region the absorption is dominant while scattering is negligible, we use this region to get the intrinsic rocking curve of the analyzer crystal. Thus, the deconvolution of the intrinsic rocking curve of the analyzer crystal from the effective rocking curve results in the measured scattered angle distribution. The width (standard deviation, σ) of this measured scattered angle distribution is equal to the quadratic subtraction of width of both distributions according to

$$\sigma_{sample} = \sqrt{\sigma^2_{measured} - \sigma^2_{RC}} \qquad (3.3.1)$$

where $\sigma_{RC}$ is the width of the intrinsic rocking curve and $\sigma_{measured}$ is the width of the measured effective rocking curve. For Si (1,1,1) the experimentally measured width are $\sigma_{RC} = (8.07 \pm 0.06)$ μradians for 17 keV, $\sigma_{RC} = (5.84 \pm 0.03)$ μradians for 25 keV and $\sigma_{RC} = (4.95 \pm 0.04)$ μradians for 30 keV.

## 3.4 EXPERIMENTAL RESULTS

In the experiment we measured the broadening of the rocking curve (or broadening of the refraction angle distribution) when the beam traverses the sample for the different sample thickness, described above, acquired at 17, 25 and 30 keV.

According to the Central Limit Theorem for relatively thick phantoms, when the number of the scatter centers is large and the individual refraction scatter processes are completely independent, the distribution of the projected (one-dimensional) scattering (refraction) angle is expected to be Gaussian in shape and the standard deviation $\sigma$ of these distributions is a square root function of the phantom thickness **x**:

$$\sigma = a \cdot \sqrt{x} \qquad (3.4.1)$$

As depicted in Figures 3.4.1-3.4.3, the standard deviation of the measured scattered angle distribution is shown as a function of the phantom thickness. Figures 3.4.1 shows the corresponding results for 6 μm diameter spheres at 17, 25 and 30 keV; Figures 3.4.2 – for 30 μm diameter spheres at 17, 25 and 30 keV and Figures 3.4.3 – for 100 μm diameter spheres at 17, 25 and 30 keV. For all samples the expected square root dependence can be clearly seen as shown by the fitted curves.

The related to figures data are presented in the Table 3.4.1.

From the experimental data two preliminary conclusions can be drawn:

- When the diameter of the spheres increases from 6 to 100 μm the standard deviation values of the refraction angle distribution decrease;

- When the energy is increased from 17 keV to 30 keV the standard deviation values of the refraction angle distribution decrease substantially. From equations (1.7) and (1.8) it is expected that this decrease occurs with a $1/E_\gamma^2$ law, where $E_\gamma$ is the energy of the X-ray photons.

Hence, for 17 keV and increasing the values of the spheres' diameter from 6 to 100 μm for 5mm phantom thickness changes the standard deviation from 40.07±0.91 μradians to 16.16±0.55 μradians; for 25 keV, correspondingly, from 18.28±0.50 μradians to 7.90±0.34 μradians and for 30 keV – from 13.23±0.40 μradians to 5.46±0.30 μradians (see Table 3.4.1).

On the graphs in Figures 3.4.1 – 3.4.3 the total errors are presented, i.e. statistical and systematic errors processed together. The statistical errors are in the range of 1-3%. The systematic errors, which are larger than the statistical errors by at least one order of magnitude, arise from the error on the rocking curve width and from the error on the box (phantom) thickness. The systematic error on the rocking curve width was considered to be equal to ±0.21 μradians and the systematic error on the phantom thickness due to the machining uncertainties – ±0.1 mm. Due to the systematic errors, the experimental values will have the largest errors at smaller layer thickness.

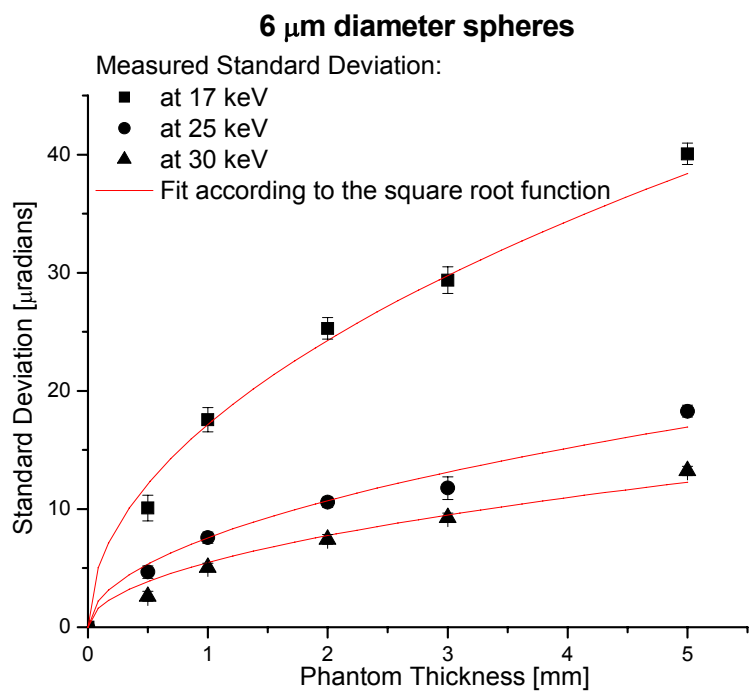

*Figure 3.4.1:* The comparison between the measured standard deviation values for 6 μm diameter spheres at different energies (17, 25 and 30 keV)

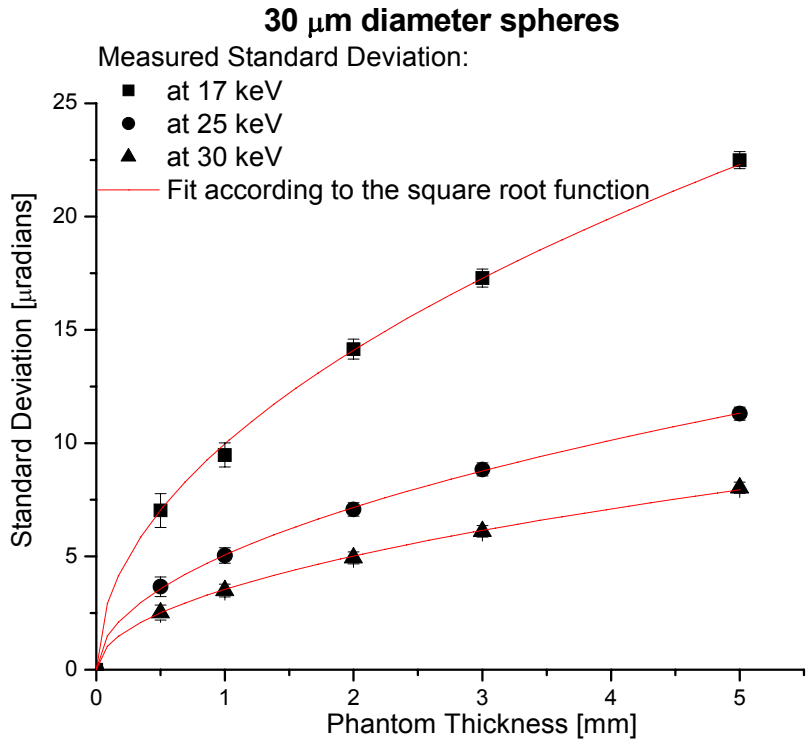

*Figure 3.4.2:* The comparison between the measured standard deviation values for 30 μm diameter spheres at different energies (17, 25 and 30 keV)

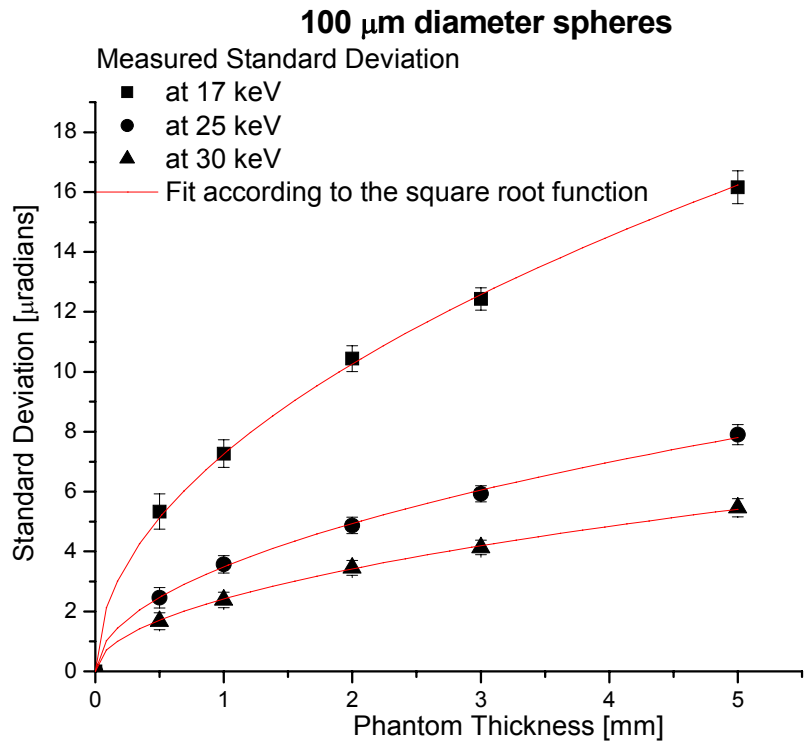

*Figure 3.4.3:* The comparison between the measured standard deviation values for 100 μm diameter spheres at different energies (17, 25 and 30 keV)

*Table 3.4.1* Measured standard deviation values for different samples at different energies

| Energy, [keV] | Spheres Diameter, [μm] | Measured standard deviation, [μradians] | | | | | Fit parameter according to equation (3.4.1) |
|---|---|---|---|---|---|---|---|
| | | Phantom Thickness [mm] | | | | | |
| | | 0.5 | 1 | 2 | 3 | 5 | |
| 17 | 100 | 5.33±±0.59 | 7.27±±0.46 | 10.44±±0.43 | 12.43±±0.38 | 16.16±±0.55 | a=7.29±0.04 |
| | 30 | 7.03±±0.74 | 9.48±±0.53 | 14.15±±0.44 | 17.29±±0.40 | 22.49±±0.38 | a=9.97±0.06 |
| | 6 | 10.09±±1.09 | 17.56±±1.03 | 25.30±±0.90 | 29.37±±1.12 | 40.07±±0.91 | a=17.47±0.36 |
| 25 | 100 | 2.46±±0.34 | 3.58±±0.29 | 4.87±±0.28 | 5.93±±0.27 | 7.90±±0.34 | a=3.49±0.02 |
| | 30 | 3.66±±0.43 | 5.05±±0.34 | 7.07±±0.30 | 8.84±±0.29 | 11.30±±0.29 | a=5.06±0.02 |
| | 6 | 4.67±±0.53 | 7.58±±0.47 | 10.59±±0.43 | 11.78±±0.96 | 18.28±±0.50 | a=7.58±0.27 |
| 30 | 100 | 1.68±±0.28 | 2.39±±0.27 | 3.45±±0.25 | 4.14±±0.25 | 5.46±±0.30 | a=2.42±0.01 |
| | 30 | 2.52±±0.33 | 3.50±±0.29 | 4.94±±0.26 | 6.11±±0.26 | 8.08±±0.25 | a=3.55±0.02 |
| | 6 | 2.62±±0.42 | 5.06±±0.35 | 7.45±±0.41 | 9.28±±0.38 | 13.23±±0.40 | a=5.49±0.22 |

For the experimental data the energy dependence of the standard deviation values fulfill the $1/E_\gamma^2$ law for all dimensions of the spheres: 6, 30 and 100 μm. This energy dependence is presented in Figure 3.4.4 for 6, 30 and 100 μm diameter spheres at a phantom thickness of about 3 mm and in Figure 3.4.5 for 6, 30 and 100 μm diameter spheres for all phantom thickness: 0.5, 1, 2, 3 and 5 mm. The good agreement with this law shows the reasonable reliability of the experiment.

To demonstrate that the behavior of the experimental data shown above can be theoretically understood, a detailed Monte Carlo simulation has been performed and is described in the next chapter.

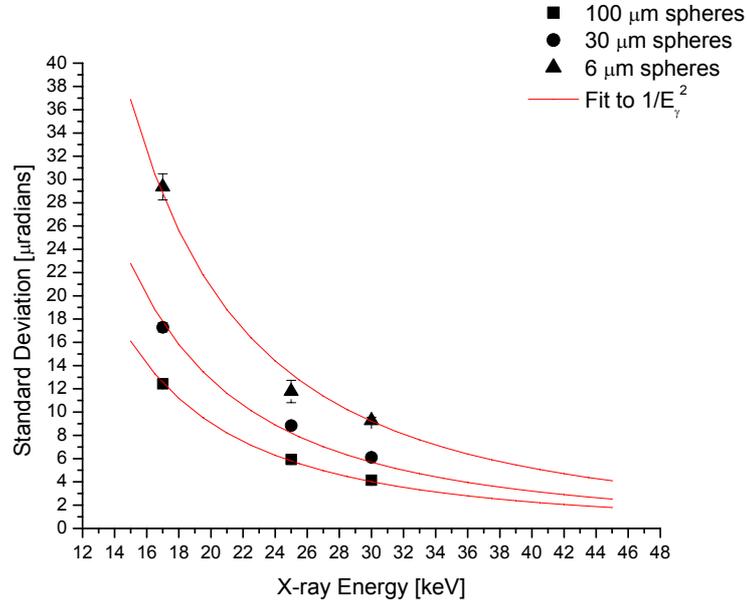

*Figure 3.4.4:* Measured standard deviation values of the refraction angle distribution as a function of the photon energy, Eγ, for 6, 30 and 100μm diameter spheres and a fixed phantom thickness of about 3 mm

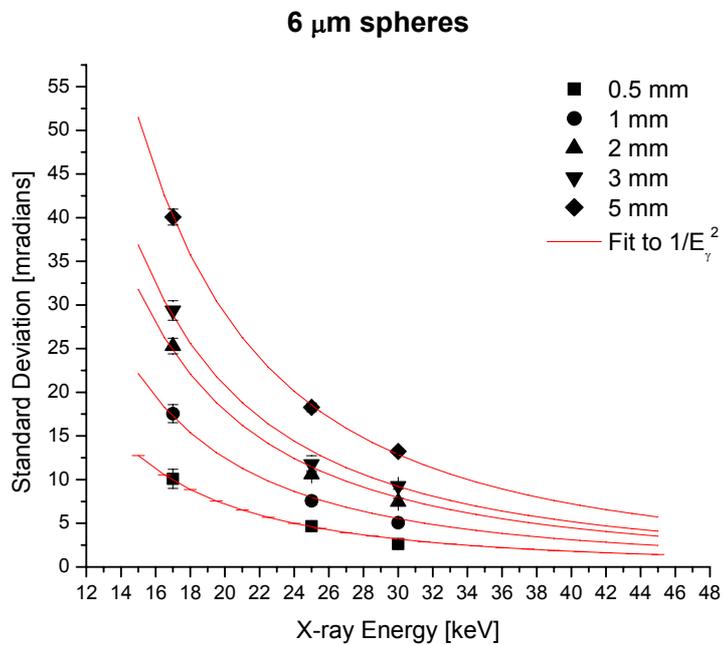

*Figure 3.4.5a:* Measured standard deviation values of the refraction angle distribution as a function of the photon energy, Eγ, for 6 μm diameter spheres for the phantom thickness of 0.5, 1, 2, 3 and 5 mm

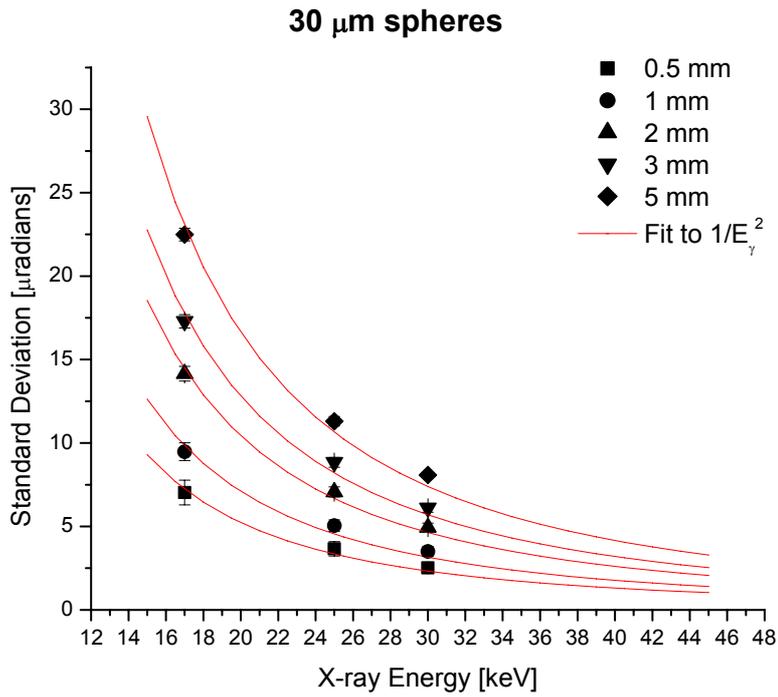

*Figure 3.4.5b:* Measured standard deviation values of the refraction angle distribution as a function of the photon energy, E$\gamma$, for 30 μm diameter spheres for the phantom thickness of 0.5, 1, 2, 3 and 5 mm

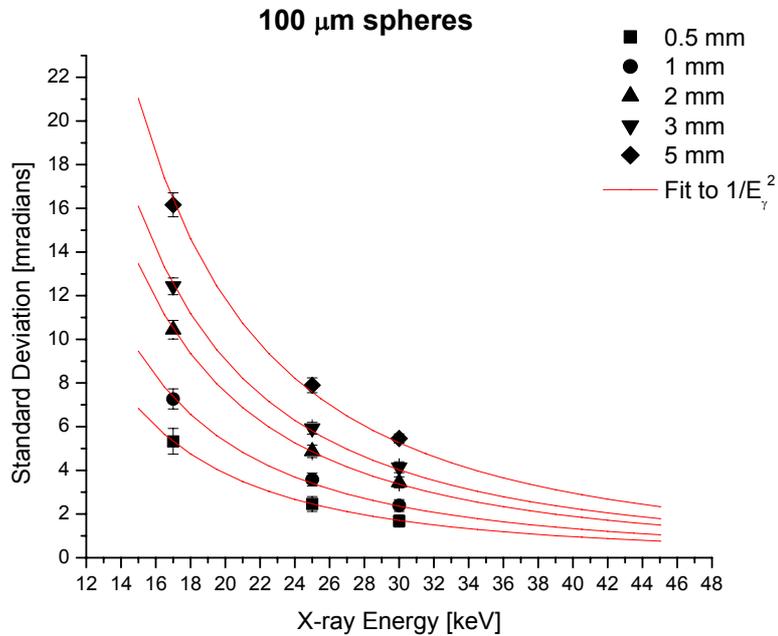

*Figure 3.4.5c:* Measured standard deviation values of the refraction angle distribution as a function of the photon energy, E$\gamma$, for 100 μm diameter spheres for the phantom thickness of 0.5, 1, 2, 3 and 5 mm

*Chapter 4*

# Monte Carlo Simulation of X-ray Multiple Refractive Scattering from Fine Structure Objects

## 4.1 INTRODUCTION

Monte Carlo (MC) calculations are by far the most successful method in simulating the stochastic process of particle transport in a scattering medium. This method was first proposed by Metropolis and Ulam [40] (actually MC has first been used by John von Neumann and Norbert Wiener in the Manhattan project (1943), but was kept secret at this time) and since then it has been broadly used to model different physical processes in particular in the radiological sciences. Later Koblinger and Zarand [41] (1973) used Monte Carlo simulations to determine the radiation dose to various organs during chest radiography. Sundararaman, Prasad and Vora [42] (1973) used Monte Carlo simulation to determine the bremsstrahlung spectra and intensities of the K characteristic lines for X-ray tubes. Reiss and Steinle [43] (1973) described Monte Carlo calculations of X-ray spectra in the diagnostic energy range for water phantoms. The first systematic radiotherapy study utilising Monte Carlo methods was made by Bruce and Johns [44] (1960). They computed scattered photon spectra for monoenergetic beams (50,100, 200, 500 and 1250 keV) incident upon low atomic number media. Their results were used to determine build-up factors and absorbed doses. Sidwell, Burlin and Wheatley [45] (1969) later calculated absorbed dose distributions for wide photon beams incident upon a water equivalent phantom corresponding to a human body. Sidwell and Burlin [46] (1973) later extended this work to the effect of lung inhomogeneties (cancer).

In this Chapter we will describe a Monte Carlo photon transport program for the simulation of multiple refractive scattering of X-ray photons on alveoli in lung tissue. This program is based on the refractive properties of X-rays in matter (absorption is neglected). For the ease of simulation and its experimental verification, lung tissue and, subsequently, the alveoli are represented by a three dimensional array of monodisperse PMMA microspheres. Multiple scattering is a quite general physical process that occurs when a wave or particle propagates in a medium with a high density of localized scatters like monodisperse PMMA microspheres. We applied the program to determine refraction angle distributions and their width (standard deviation values)for the same phantom described in the previous chapters at 17, 25 and 30 keV.

The MC results will be compared to the experimental measurements performed under similar conditions and taken with the DEI set-up of the SYRMEP beamline in ELETTRA (Trieste, Italy), described in the previous chapters.

## 4.2 SIMULATED PHANTOM

The phantom used in the experiment and described in Chapter 3 has been simulated using SIAMS S3D software [47]. This program builds a random close packing of spheres using an algorithm named 'drop and roll' or 'rolling algorithm', which yields a packing parameter of 0.60 independent of the sphere size. The packing density is given by the total volume of the PMMA spheres divided by the volume of the enclosing box.

The physical idea of this algorithm is formulated as following: the spheres' diameters are generated according to the chosen law of size distribution (in our case this is delta function, because all spheres have the same size) and are dropped inside the box either from one point, or from randomly chosen positions. As soon as the dropped sphere encounters an obstacle (the box wall or an already packed sphere), it sticks to it (without impact) and begins to slide on its surface in the direction of the minimum of the gravitational potential energy to the following obstacle. This direction is a projection of the free fall direction on the surface of the obstacle. The movement of the sphere stops, obviously, at the potential minimum that can be a point of intersection of three surfaces (three spheres, two spheres and one plane and etc.) or on a surface situated perpendicular to the direction of the free fall of the sphere (for example, on the bottom of the box). Electrostatic forces are not taken into account. Figure 4.2.1 shows the phantom simulated by the SIAMS 3D software with spheres of 30 $\mu$m in diameter. The output box dimensions for 500 of such spheres are 241.8 $\mu$m × 241.8 $\mu$m × 241.8 $\mu$m.

The data file, created by the SIAMS 3D software, contains the Cartesian coordinates of the centers of all spheres with the given radius and the dimensions of the box. This data file is the input of the Monte Carlo program. Two different sizes of microspheres were used: 30 and 100 $\mu$m. Each sample was simulated for three different energies: 17, 25 and 30 keV. The 6 $\mu$m sphere sample was not used in the simulations, since a very large number of spheres is needed to generate the required different thickness of such a sample. Since in the MC program the intersections of each X-ray with each sphere are investigated, rising the number of spheres makes the program very slow.

Visualization of the sample was made using IDL (Interactive Data Language). All codes in this thesis, including the visualization of the ray path inside the phantom (see Figure 4.3.4), are written in IDL software version 5.2.1 and are presented in the Appendix A.

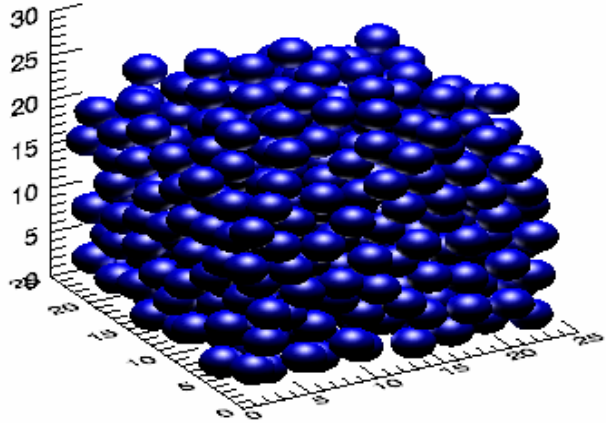

*Figure 4.2.1:* Simulated phantom from the experiment. Diameter of the spheres is 30μm, the box dimensions are 241.8μm ×241.8μm ×241.8μm (for 500 spheres)

### 4.3 MONTE CARLO PROGRAM ALGORITHM

The Monte Carlo simulation program is based on a three dimensional vector approach. The input parameters of the Monte Carlo program are the sphere diameter, the X-ray beam energy, the refractive indices of the sphere material (PMMA) and the embedding environment (air), the X-ray source (synchrotron radiation or X-ray tube radiation) and the thickness of the phantom containing the microspheres.

To investigate the path of each X-ray its interactions with the microspheres, refracted ray directions and, correspondingly, refraction angle distributions, based upon the refracted properties of X-rays in the media have been studied.

For X-ray photons above the K-edge, the refractive index is always smaller than unity. The refractive index for X-rays in air is approximately equal to unity ($n_1$ in the algorithm); while the refractive index for the X-rays in the PMMA microspheres is equal to $n_2 = 1 - \delta$, where δ is according to equation (1.4) expressed in terms of the wavelength λ of the X-rays, the classical radius $r_0$ and electron density $\rho_e$ and varies with the energy. For the PMMA at 17 keV δ is equal to $9.23 \cdot 10^{-7}$. The values for δ were obtained from the Henke data base [48] and in the range of 17 – 30 keV these values are presented in the Table A2.1.1 (Appendix A2).

We assume that each X-ray is defined by a starting position $\vec{S}$ and a direction vector $\vec{d}$. From the vector algebra we know that a direction vector is a vector that is invariant under translation, and a position vector is a vector whose starting point is always the origin. The parametric equation of a line is used to express the position vector $\vec{P}_{ray}$ of the generic X-ray:

$$\vec{P}_{ray} = \vec{S} + \alpha\vec{d}; \quad |\vec{d}| = 1 \tag{4.3.1}$$

The position vector of a generated sphere center is $\vec{C}$ and r is the radius of this sphere.

The problem now is to find the exact position of the intersection points of the X-ray with the sphere, if any. A new vector $\vec{q}$ will be taken from the center of the sphere to the point of intersection. This geometry is shown in Figure 4.3.1.

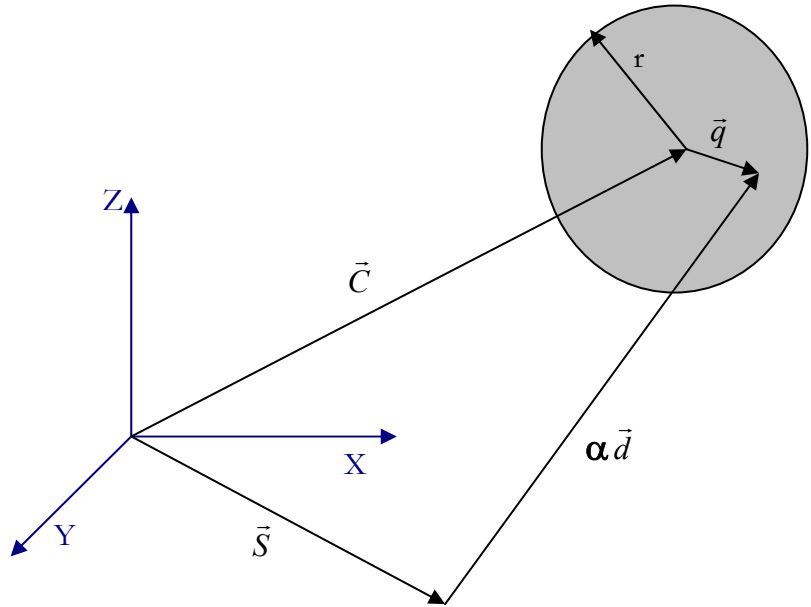

*Figure 4.3.1*: Vectors arrangement for the algorithm calculation

In this way we obtain standard approach:

$$\vec{S} + \alpha\vec{d} = \vec{C} + \vec{q} \tag{4.3.2}$$

$$\vec{q} \cdot \vec{q} = (\vec{S} - \vec{C}) \cdot (\vec{S} - \vec{C}) + 2\alpha\vec{d} \cdot (\vec{S} - \vec{C}) + \alpha^2 \tag{4.3.3}$$

The length of the vector $\vec{q}$ should be equal to the radius ($|\vec{q}| = r$) to obtain the desired points, which gives a quadratic equation in $\alpha$:

$$(\vec{S} - \vec{C}) \cdot (\vec{S} - \vec{C}) - r^2 + 2\alpha \vec{d} \cdot (\vec{S} - \vec{C}) + \alpha^2 = 0 \qquad (4.3.4)$$

The intersection points are given by the roots of the quadratic equation (4.3.4) as:

$$\vec{S} + \alpha_i \vec{d} \quad i = 1, 2 \qquad (4.3.5)$$

Since the ray is travelling in one direction, only positive values of $\alpha$ are used (if a negative value is returned for $\alpha$, then the object is behind the ray). Given that N is the number of spheres, we have N couples of roots for the quadratic equation (4.3.4). From these N couples we will choose only those, which have both roots positive. From the last chosen couples we look for the smallest positive value which, after substitution into the ray equation (4.3.1), determines the three Cartesian coordinates of the first intersection point (entrance point of the X-ray inside the sphere in the program). We also know the index of the couple, where the smallest positive $\alpha$ was found and, hence, the index of the sphere with which the ray has interacted. This information will be needed in the future in order to find the second intersection point (called exit point in the program), since we could not directly use the second root of the quadratic equation. The reason for this is the fact that the X-ray does not continue to propagate in the original direction inside the sphere but changes its direction because of the different refractive indices on the boundary of two media.

The corresponding unit surface normal in the intersection point is:

$$\vec{n} = \frac{\vec{q}}{|\vec{q}|} \qquad (4.3.6)$$

In order to find which direction the X-ray will take after it crosses the boundary between one medium and another, Snell's law has been used. Snell's law describes the refractive properties of the X-rays in the form:

$$n_1 \sin(\theta_1) = n_2 \sin(\theta_2) \qquad (4.3.7)$$

The angle between the incident ray direction and the surface normal is $\theta_1$; $\theta_2$ is the angle between the refracted ray direction and the surface normal; $n_1$ is the refractive index outside the sphere (for X-rays in air it is approximately unity); and $n_2$ is the refractive index inside the sphere (always less then unity). This physical model is shown in Figure 4.3.2. Snell's law can easily be derived from equations (1.4), (1.7) and (1.8). Its use is justified because the wavelength is much smaller than the sphere size and the product of sphere size and momentum transfer is still substantially larger than Planck's constant.

The direction vector for the refracted ray can be derived using a vector algebra approach. If the incident ray is characterized by the unit direction vector $\vec{d}$ and the refracted ray by unit direction vector $\vec{t}$, Snell's law can be rewritten as:

$$n_1(\vec{d} \times \vec{n}) = n_2(\vec{t} \times \vec{n}) \tag{4.3.8}$$

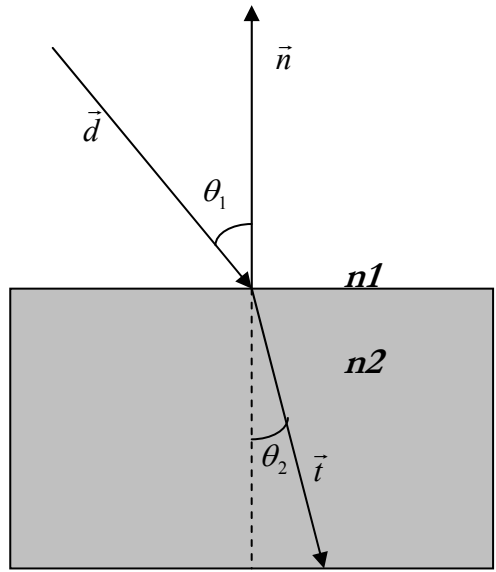

*Figure 4.3.2:* Snell's law demonstration

All three vectors are co-planar (lie in one plane). Direction vectors can be freely translated, thus it can be assumed that they all start at a fixed position in the plane. So, we can use the property of co-planar vectors which states:

$$\vec{t} = \alpha\vec{n} + \beta\vec{d} \tag{4.3.9}$$

Here α and β are scalar constants. Taking the cross-product of both sides of this equation with $\vec{n}$ gives:

$$(\vec{t} \times \vec{n}) = \beta(\vec{d} \times \vec{n}) \tag{4.3.10}$$

Equations (4.3.7) and (4.3.10) determine the value for the scalar constant β which is equal to:

$$\beta = \frac{n_1}{n_2} \tag{4.3.11}$$

Since $\vec{t}$ is a unit vector, it is also known that:

$$\vec{t} \cdot \vec{t} = 1 = (\alpha\vec{n} + \beta\vec{d}) \cdot (\alpha\vec{n} + \beta\vec{d}) = \alpha^2 + 2\alpha\beta(\vec{n} \cdot \vec{d}) + \beta^2 \tag{4.3.12}$$

This equation is quadratic in the scalar constant α, which can be calculated. The two solutions are:

$$\alpha = -\beta\vec{n} \cdot \vec{d} \pm \sqrt{\beta^2(\vec{n} \cdot \vec{d})^2 - \beta^2 + 1} \tag{4.3.13}$$

Only one of these two roots is correct. It is possible to show that the positive sign of the root is correct one for our purpose. If both refraction coefficients are equal, β is equal to unity (equation 4.3.11), the situation when the ray does not refract. If one substitutes this β into the equation (4.3.13), the scalar α will be obtained. If one substitutes then this α into equation (4.3.9), it comes out that the unit direction vector $\vec{t}$ is equal to the unit direction vector $\vec{d}$. Therefore, this X-ray did not change its direction, what contradicts the refraction expectation.

Substituting now the values of α and β into equation (4.3.9) gives:

$$\vec{t} = \frac{n_1}{n_2}\left(\left\{\sqrt{(\vec{n} \cdot \vec{d})^2 + \left(\frac{n_2}{n_1}\right)^2 - 1} - \vec{n} \cdot \vec{d}\right\}\vec{n} + \vec{d}\right) \tag{4.3.14}$$

There exists another possibility to represent the refracted direction $\vec{t}$. The scalar product between the normal to the surface $\vec{n}$ and the incident direction $\vec{d}$ can be expressed as a cosine of the angle between them and then Snell's law can be applied. The result is:

$$\cos(\theta_2) = \left[1 - \left(\frac{n_1}{n_2}\right)^2 (1 - \cos^2(\theta_1))\right]^{\frac{1}{2}} \tag{4.3.15}$$

$$\vec{t} = \frac{n_1}{n_2}\vec{d} - (\cos(\theta_2) - \frac{n_1}{n_2}\cos(\theta_1))\vec{n} \tag{4.3.16}$$

When a ray enters a sphere, this new direction is used, together with the entry point, to determine the second intersection with the sphere (that is the point where the ray leaves the sphere). The procedure to find the second intersection point is far simpler than to find the first intersection point, since the index of the sphere with which X-ray interacted is already known. The first intersection point then is taken as a new initial position of the X-ray and the refracted direction as a new initial direction, solving the quadratic equation (4.3.4) for the scalar α. There is no need to check again for intersections with all the spheres, so in equation (4.3.4) the center coordinates of the sphere where the first intersection occurred are used. After solving this equation, again two roots have been obtained (both positive). If one substitutes these $\alpha$ into the X-ray equation (4.3.1) with the new parameters, the already existing entrance point coordinates are obtained together with the exit point coordinates of the X-ray leaving the sphere. At this point, Snell's law is again applied to obtain a new direction vector for the X-ray with the refractive indices $n_1$ and $n_2$ interchanged.

This algorithm is repeated for each X-ray, propagating it till the end of the box. If the present X-ray leaves the box, the next X-ray is taken at the beginning of the box and is propagated in the same way. This procedure is repeated for all another photons falling onto the phantom. The angles between the refracted X-ray directions and the surface normal to the intersection points are called the space refraction angles. To obtain the projected refraction angles the refracted X-ray directions are projected onto the fundamental planes (zOx and yOx if the X-ray propagates in the Ox direction) and the output is made for the angles between the projected X-ray directions and the initial directions of the X-rays. The distributions of these angles are expected to follow an approximately Gaussian distribution. The refraction angles are small. If a ray in the energy range used enters at 45º with respect to the normal incidence, it is refracted by about 0.5 µradians in PMMA. Therefore, only a small fraction leaves the box sideways.

The flowdiagram of the program algorithm is shown in Figure 4.3.3 and the paths of the X-ray photons inside the box can be seen in Figures 4.3.4-4.3.5.

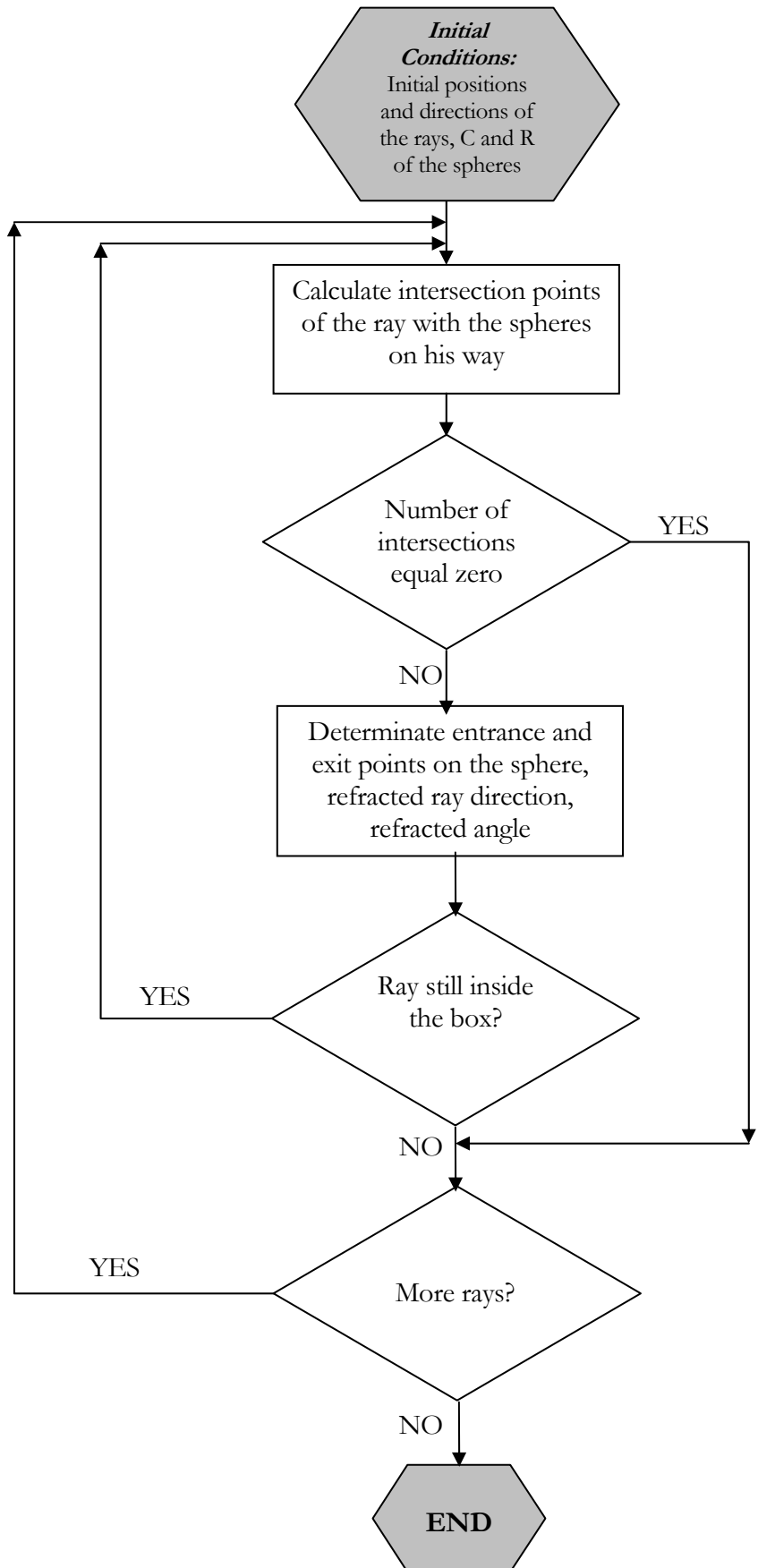

*Figure 4.3.3:* Flowdiagram of the program algorithm

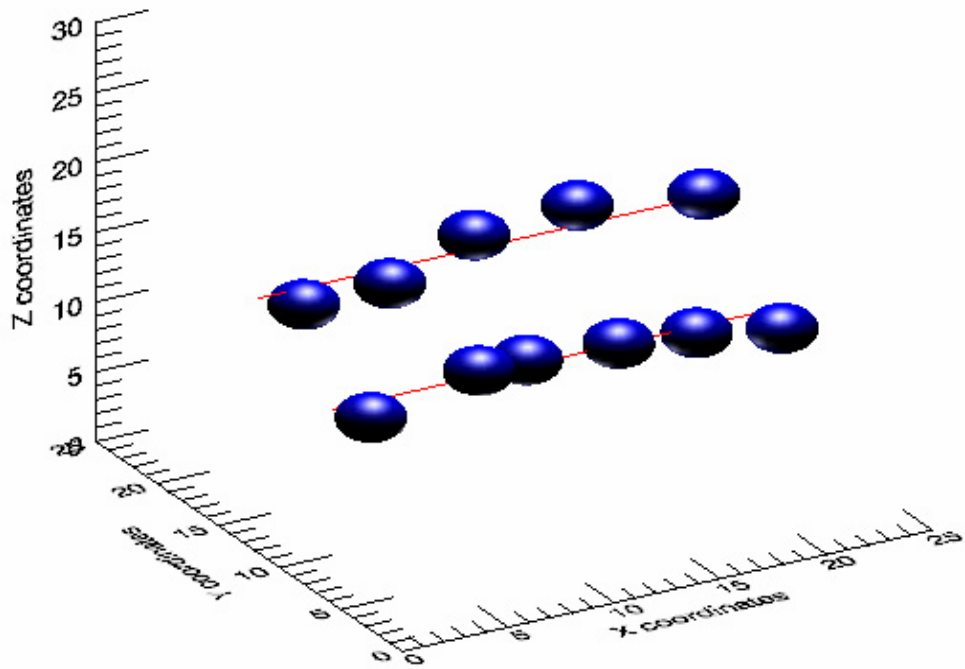

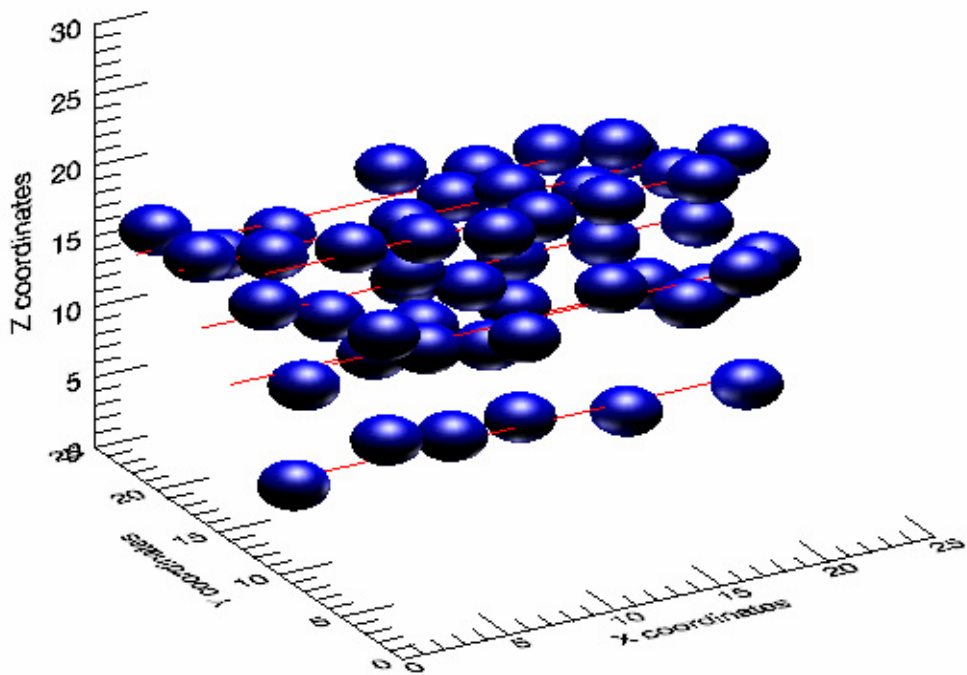

*Figure 4.3.4:* Two examples of the X-ray photons path' inside the box

*Chapter 5*

# Monte Carlo Simulation Results

## 5.1 PRELIMINARY TEST: SINGLE SPHERE SCATTERING

As a preliminary test of the proposed algorithm for the Monte Carlo program, the space angle distribution simulated for a single sphere has been compared to the algebraically calculated one. The simple schematic for a single sphere interaction is presented in Figure 5.1.1.

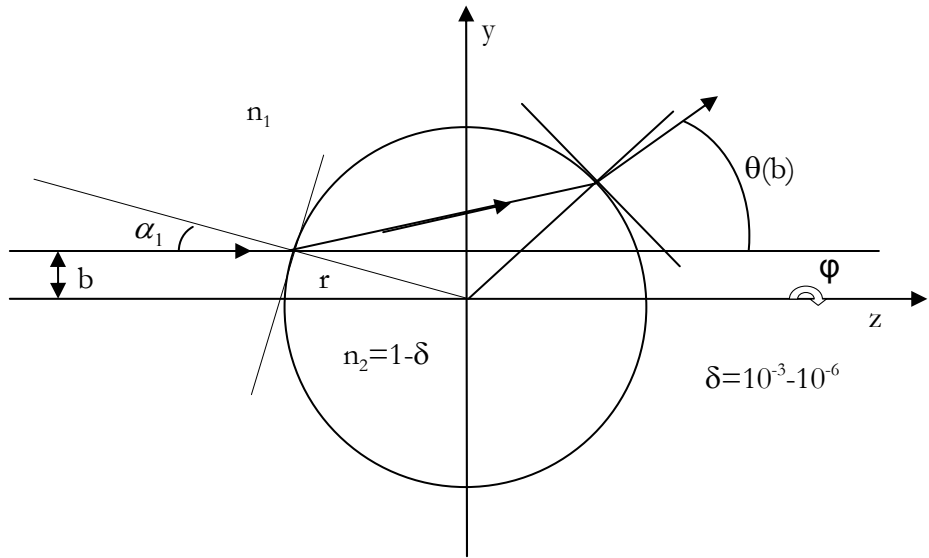

*Figure 5.1.1:* Single sphere interaction

The theoretical calculations of the space angle distribution are based on the following assumptions: the propagation direction of the incident beam changes due to refraction, when there is a density gradient in the object in the propagation direction. The deviation $\Delta\alpha$ at the interface, where the real part of the refractive index changes by $\Delta n = n_2 - n_1 = \delta$, can be calculated from the Snell's law using a Taylor expansion (see Figure 5.1.2). Hence, $\Delta\alpha = \dfrac{\delta}{n_1}\tan\alpha_1$, where $\alpha_1$ is the angle between the incident beam direction and the surface normal in the intersection point, $n_1 = 1$ for air. If one uses the impact parameter $b = r \cdot \sin(\alpha_1)$, where r is the radius of the sphere, then the deviation $\Delta\alpha$ at the interface can be expressed as $\Delta\alpha = \dfrac{\delta \cdot b}{\sqrt{r^2 - b^2}}$. Since the transverse displacement of the X-ray is very small, the refraction angle is very close to twice the deviation $\Delta\alpha$ at the interface:

$$\theta(b) = \dfrac{2 \cdot \delta \cdot b}{\sqrt{r^2 - b^2}} \qquad (5.1.1)$$

The probability density function $\dfrac{dN}{d\theta(b)}$ that the X-ray will scatter by the angle $\theta(b)$ is equal to:

$$\begin{aligned}\dfrac{dN}{d\theta(b)} &= \dfrac{dN}{db} \cdot \dfrac{db}{d\theta(b)} = \Psi \cdot \dfrac{\pi}{\delta} \cdot \dfrac{b \cdot \sqrt{r^2 - b^2}}{(1 + \dfrac{b^2}{r^2 - b^2})} = \\ &= \Psi \cdot \pi \cdot r^2 \cdot \dfrac{\theta(b)}{2\delta^2 + \theta^2(b) + \theta^4(b)/8\delta^2} \end{aligned} \qquad (5.1.2)$$

where $\Psi$ is a constant photon flux.

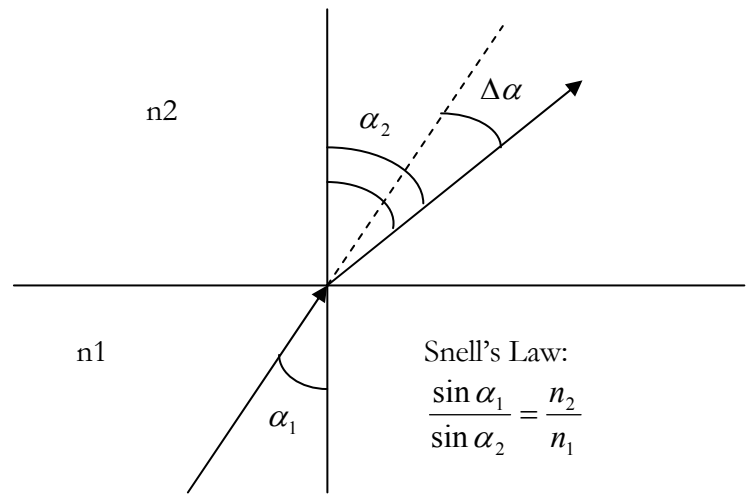

*Figure 5.1.2:* Snell's Law

Three kinds of test simulations have been done: with the X-rays falling on the sphere parallel to the OX axis; with the X-rays falling under some angle (e.g. under angle 11.47°), but still parallel between themselves; and with the X-rays falling on the sphere under random angles and, correspondingly, having random directions and random initial positions (see Figure 5.1.3 a), b) and c) correspondingly).

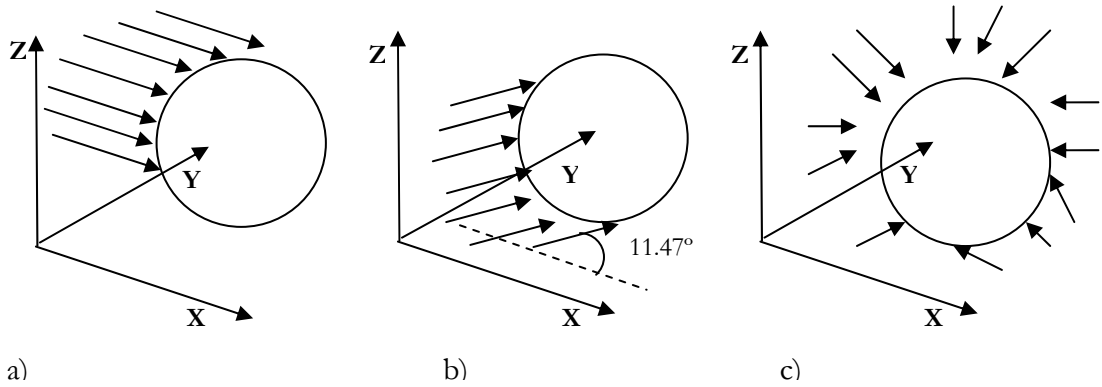

*Figure 5.1.3:* a) X-rays fall on the sphere parallel to OX axis b) X-rays fall on the sphere under angle 11.47° and parallel between themselves c) X-rays fall on the sphere under random angles having, correspondingly, random directions and random initial positions

In addition the expected theoretical distribution has been obtained according to the equation (5.1.2). The results of these tests are presented in Figures 5.1.4a-5.1.4c. Independently from the positions and from the directions of the X-rays, the space angle distributions should be identical. In the graphs CROSSES represent the Monte Carlo program output while the CONTINUOUS line corresponds to the theoretical expectation. It is possible to see from Figures 5.1.3a-5.1.3c that the simulated distributions are in good agreement with the theoretically predicted results and with each other, showing the correct idea of the algorithm.

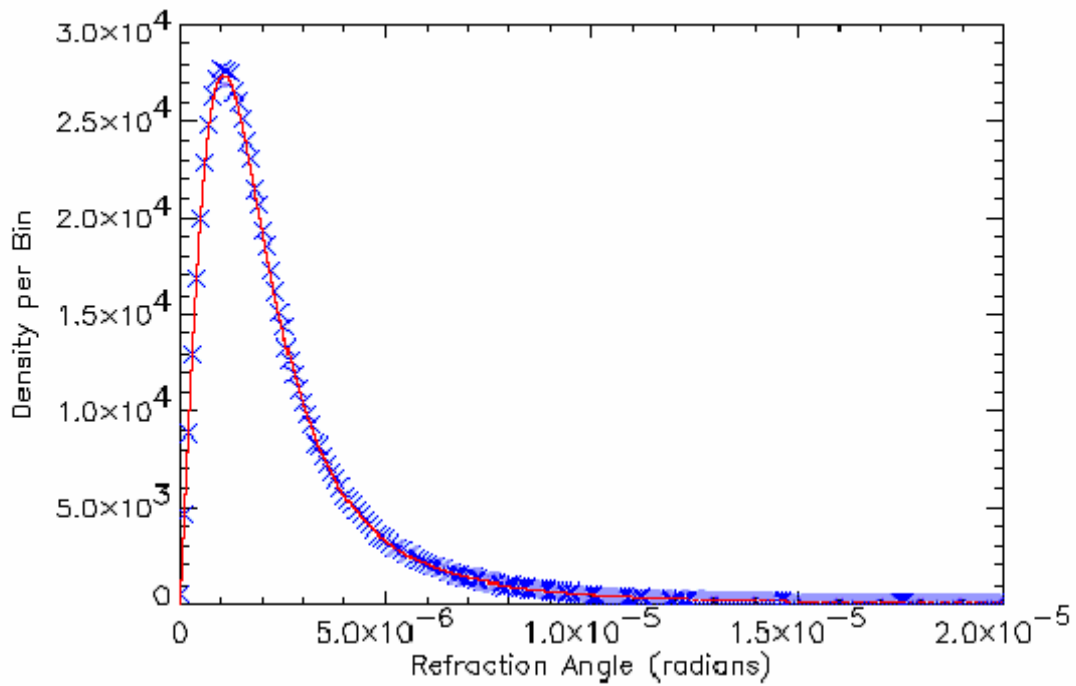

*Figure 5.1.4a:* Space angle distribution. Rays falling on the sphere are all parallel to the axis OX. CROSS corresponds to the simulated result, CONTINUOUS LINE corresponds to the theoretical result

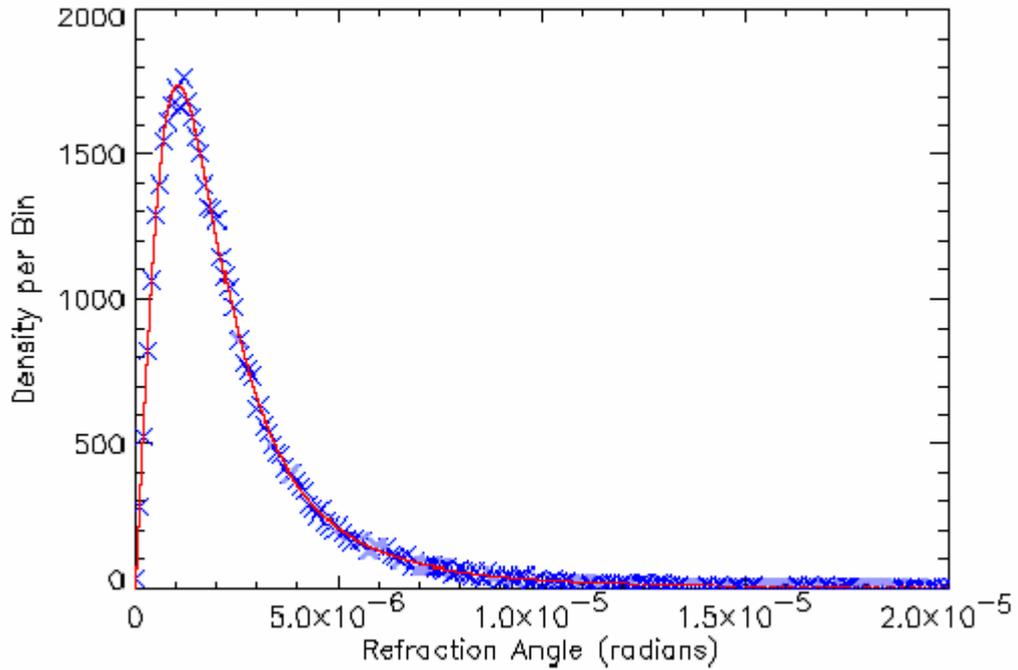

*Figure 5.1.4b:* Space angles distribution. Ray falling on the sphere under angle 11.47 degree, 1 million rays. CROSS corresponds to the simulated result, CONTINUOUS LINE corresponds to the theoretical result

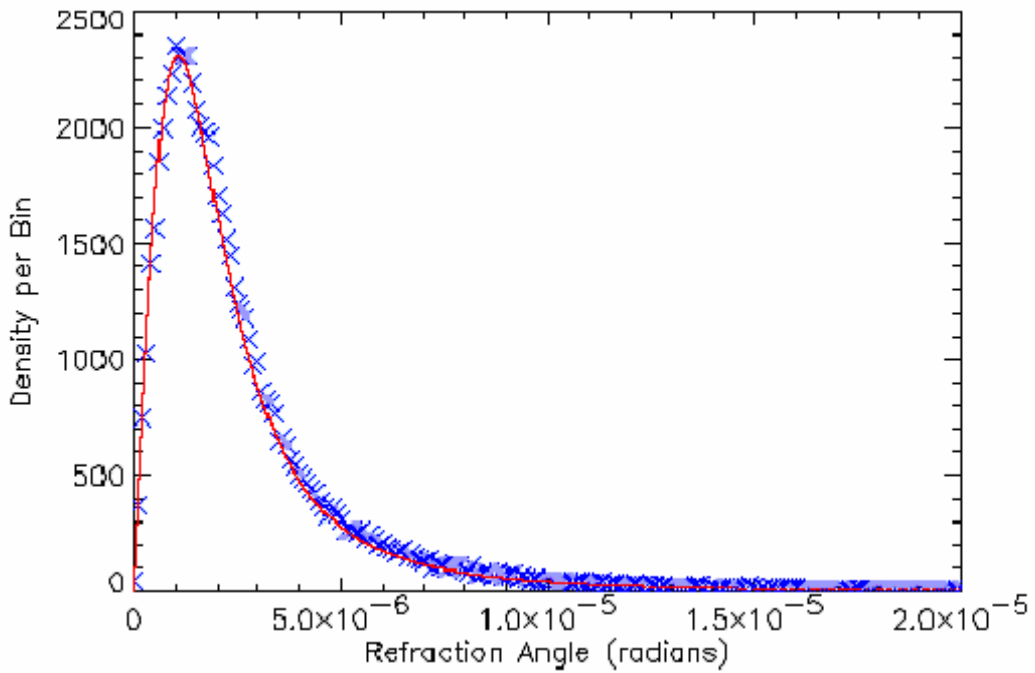

*Figure 5.1.4c:* Space angle distribution. Rays falling on the sphere under different angles, 5 million rays. CROSS corresponds to the simulated result, CONTINUOUS LINE corresponds to the theoretical result

The projected refraction angle distribution for 1 million rays falling on one 30 μm diameter sphere at 17 keV is presented in Figure 5.1.5. The DOTS correspond to the Monte

Carlo program output – projected refraction angles, and the CONTINUOUS LINE shows a Gaussian fit to the output results. The simulated distribution roughly resembles a Gaussian for small deflection angles, but at bigger angles it has larger tails than a Gaussian distribution. Nevertheless, Gaussian fit has been used in order to determine the standard deviation values of the simulated refraction angle distribution. In the case of 30μm diameter sphere in air embedded environment at 17keV the standard deviation of the fitted Gaussian is equal to 1.39±0.001 μradians.

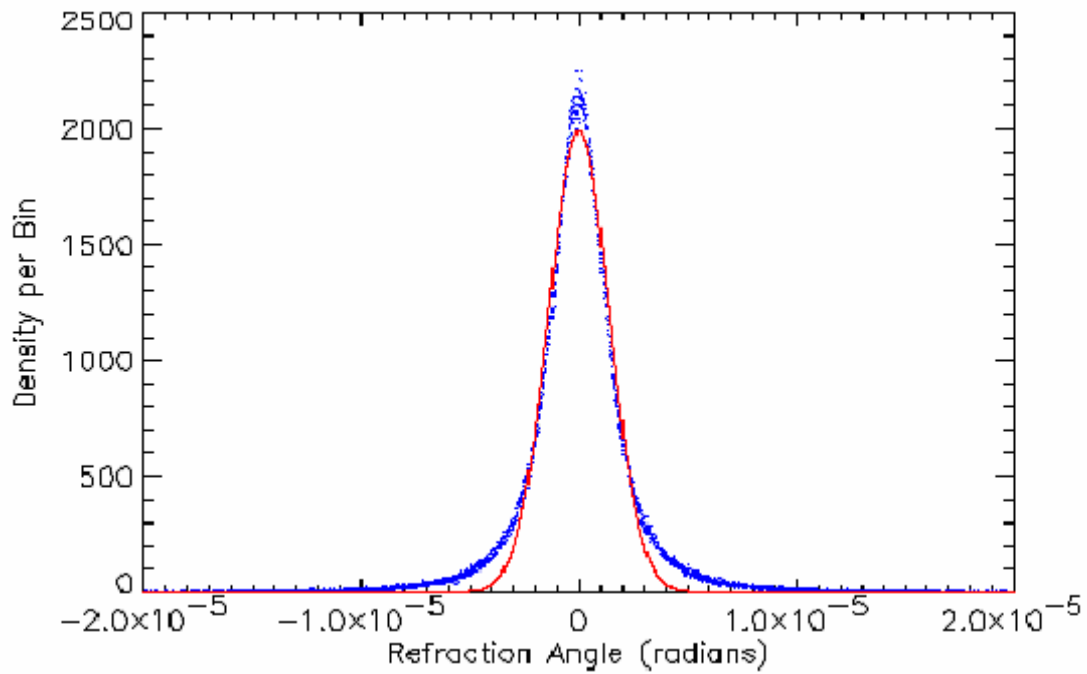

*Figure 5.1.5:* Projected angles distribution for 30 μm in diameter sphere at 17 keV. DOTS correspond to the simulated distribution, CONTINUOUS LINE correspond to a Gaussian fit

For 25 and 30 keV the standard deviation values for single sphere scattering are equal to 0.641±0.001 μradians and 0.446±0.001 μradians, respectively. No significant difference was found for the 30 and 100 μm microspheres.

## 5.2 MULTIPLE SCATTERING RESULTS

In this study 100000 X-ray photons have been generated for each of the five phantom thickness, for each of the three energies (17, 25, 30 keV) and two sphere diameters (30 and 100 µm).

To evaluate the Monte Carlo program output further the distribution of the number of scatters and the positions of the X-rays before the entrance into the phantom and after leaving it has been checked. According to the first, the shape of the distribution can represent the behavior of the X-rays inside the phantom. If most of the X-rays reaching the end of the box (phantom) without to leave it earlier, the distribution looks like it is presented in Figure 5.2.1a-5.2.1e for 30µm diameter spheres at 17keV. This distribution has an approximately Gaussian shape sometimes with a long tail toward the smallest bin number N, representing some low probability events (indicating that few X-rays leave the box sideway). The tail will be totally absent when none of the X-rays leaves the box sideways. The peak of the distribution gives the average value for the number of scatters N. This parameter depends strongly on the packing density and on the thickness of the phantom. As shown in Figure 5.2.1a for 30µm sphere at 17 keV for 0.48 mm phantom thickness the peak of the scatter center distribution is at N=12.23. The width of this distribution is equal to $N_\sigma = 1.66$. When the thickness of the phantom increases, the peak moves to the right, in the direction of increasing number of scatters. The distribution broadens. Hence, for 4.84 mm phantom thickness the peak is already at N=129.27 with a width equal to $N_\sigma = 5.84$.

For all other types of phantoms and the different energies that where generated a similar behavior was obtained.

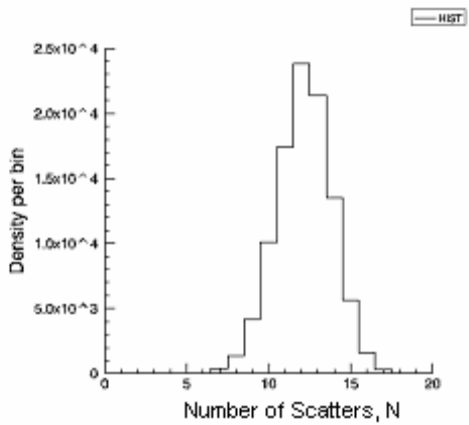
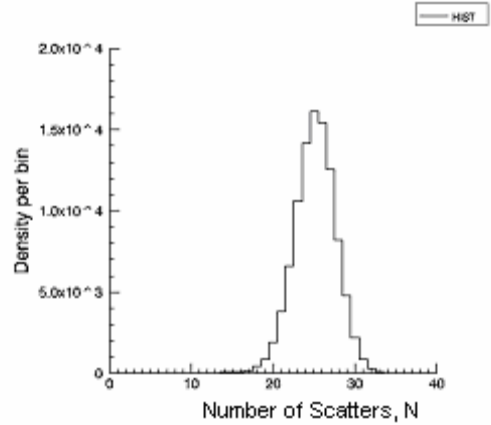

*Figure 5.2.1a*: 0.48mm phantom thickness (30μm in diameter sphere at 17 keV), peak at N=12.23

*Figure 5.2.1b:* 0.97mm phantom thickness (30μm in diameter sphere at 17 keV), peak at N=25.22

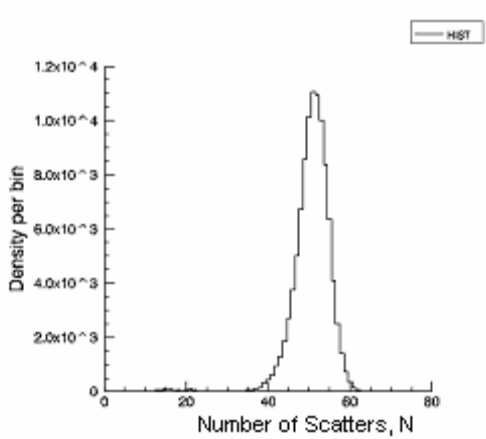
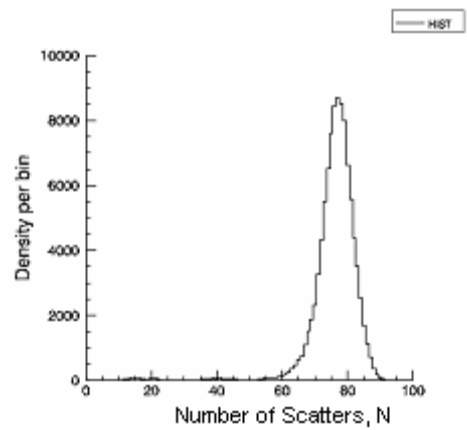

*Figure 5.2.1c*: 1.93mm phantom thickness (30μm in diameter sphere at 17 keV), peak at N=51.29

*Figure 5.2.1d*: 2.9mm phantom thickness (30μm in diameter sphere at 17 keV), peak at N=77.04

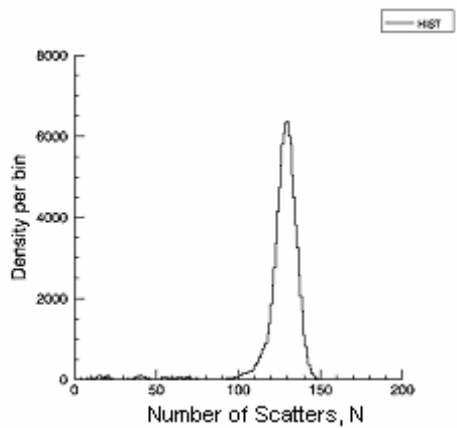

*Figure 5.2.1e:* 4.84mm phantom thickness (30μm in diameter sphere at 17 keV), peak at N=129.27

Another test distribution is the initial position of the X-rays. The initial positions are randomly distributed and follow a uniform distribution law. These distributions are presented in Figure 5.2.2a-5.2.2b for the positions before the entrance into the phantom and after leaving it. In these figures, as an example, the results are shown only for the 3 mm thickness phantom of the 30μm in diameter spheres at 17 keV. Clearly the distributions are uniform. In the Monte Carlo program the uniformity of distribution has been checked for each phantom thickness.

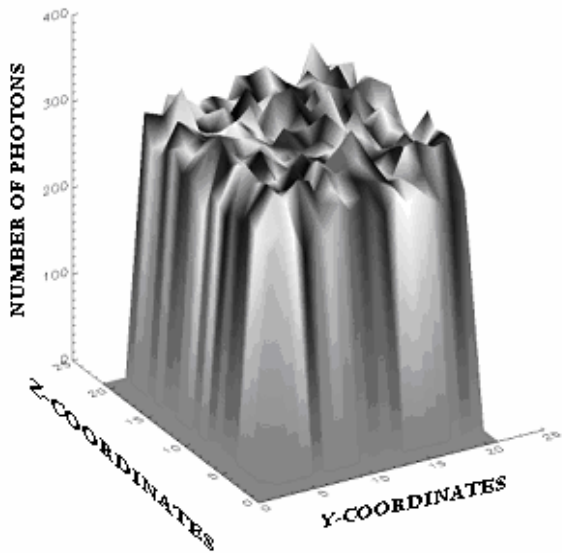
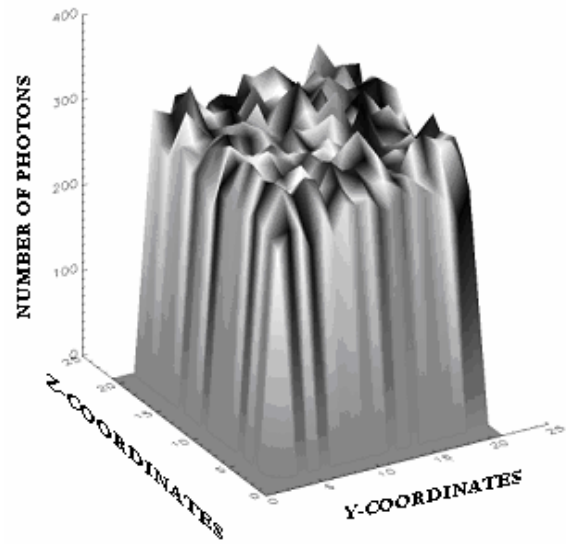

*Figure 5.2.2a:* X-rays initial position distribution (entrance into the phantom 3mm thickness)

*Figure 5.2.2b:* X-rays final position distribution (exit from the phantom 3mm thickness)

If the individual refraction scatter processes are completely independent, for large number of scatters the distribution of the projected one-dimensional scattering angle is expected to be Gaussian for purely statistical reasons (Central Limit Theorem and Convolution Theorem). Therefore, the standard deviation of the scattering (refraction) angle distribution should increase with the square root of the thickness of the sample [49].

The projected refraction angle distribution for the single sphere of 30μm diameter at 17 keV was presented in Figure 5.1.5. The change of the sphere diameter or energy will not change the general shape of the distribution.

Figures 5.2.3a-5.2.3f show the projected scattering angle distributions for 30μm diameter spheres at 17keV for six different phantom thickness; Figures 5.2.4a-5.2.4f – for 30μm diameter spheres at 25keV; Figures 5.2.5a-5.2.5f – for 100μm diameter spheres at 17keV for six different phantom thickness and Figures 5.2.6a-5.2.6e – for 100μm spheres at 25keV. The CROSSES correspond to the Monte Carlo simulated refraction angle distribution and the CONTINUOUS LINE is a Gaussian fit to this distribution.

As it can be seen from these Figures, the refraction angle distribution approaches, as expected, a Gaussian shape. Clearly visible is also the broadening of the scattering distribution, while increasing the thickness of the phantom and, thus, the number of scatters.

The standard deviation values from the Gaussian fits of the refraction angle distributions as a function of the phantom thickness at 17, 25 and 30 keV have been plotted for both 30 and 100 µm in diameter spheres, generated with the inherent packing density 0.6. The results are shown in Figures 5.2.7-5.2.12 for 30 µm spheres at 17 keV, for 30 µm spheres at 25 keV, for 30 µm spheres at 30 keV, for 100 µm spheres at 17 keV, for 100 µm spheres at 25 keV and for 100 µm spheres at 30 keV, respectively. The symbols indicate the simulated values while the line represents the fit results described below.

The simulated standard deviation values related to the graphs are also presented in the Tables 5.2.1-5.2.6. The statistical error for the sample standard deviation is obtained from the formula 5.2.1, where n is the number of X-ray photons falling onto the phantom [49]:

$$\Delta\sigma = \frac{\sigma}{\sqrt{2\cdot(n-1)}} \tag{5.2.1}$$

Since the error values are on the percent and permille level (see Table 5.2.1-5.2.6) it is not possible to see the errors bar in the correspondent graphs. Nevertheless, they are present.

The conditions of statistical independence and large number of scatters are not completely fulfilled over the full range of phantom thickness, in particular not at small phantom thickness. Therefore deviations from a pure square root law are expected. Indeed, pure square root fits to the Monte Carlo results do not give satisfactory fits.

Obviously, for thin layers, the number of scatters is not large, and statistical independence is only guaranteed if the positions of the sphere centers are chosen completely random. Due to the gravitational force (and possible contributions from electrostatic force on the PMMA spheres) during the filling of the experimental sample and in the construction of the Monte Carlo sample, the spheres will get some degree of order, resulting in position correlations, and therefore correlations in the scattering angles.

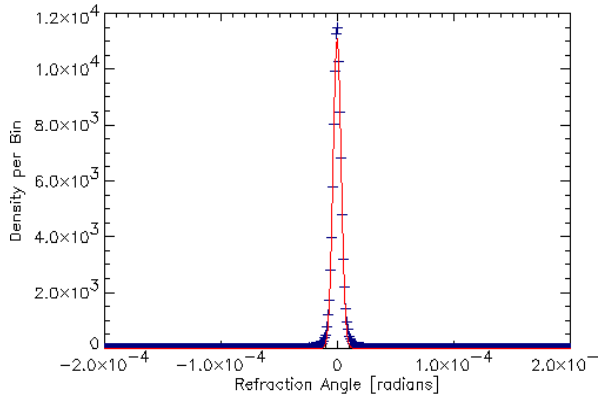

*Figure 5.2.3a*: 30μm spheres at 17keV, 0.24mm phantom thickness (500 spheres), 100000 rays

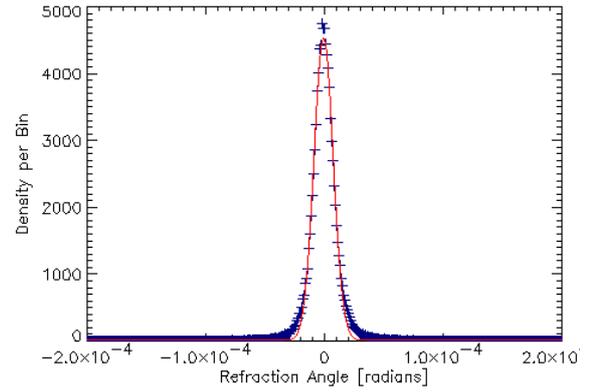

*Figure 5.2.3b*: 30μm spheres at 17keV, 0.48mm phantom thickness (1000 spheres), 100000 rays

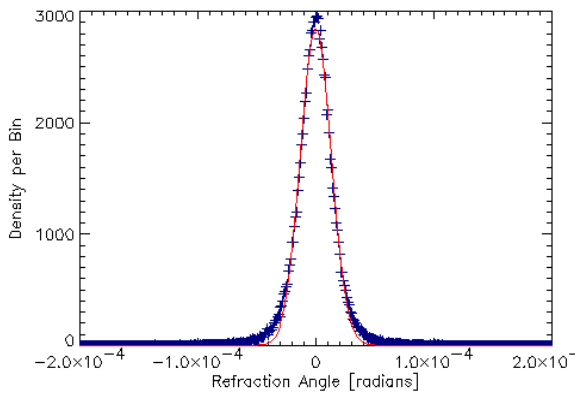

*Figure 5.2.3c*: 30μm spheres at 17keV, 0.97mm phantom thickness (2000 spheres), 100000 rays

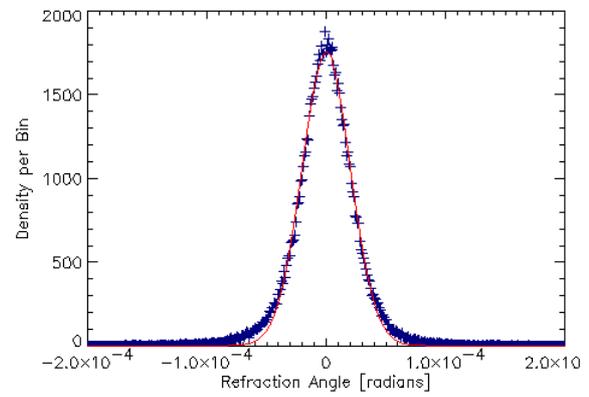

*Figure 5.2.3d*: 30μm spheres at 17keV, 1.93mm phantom thickness (4000 spheres), 100000 rays

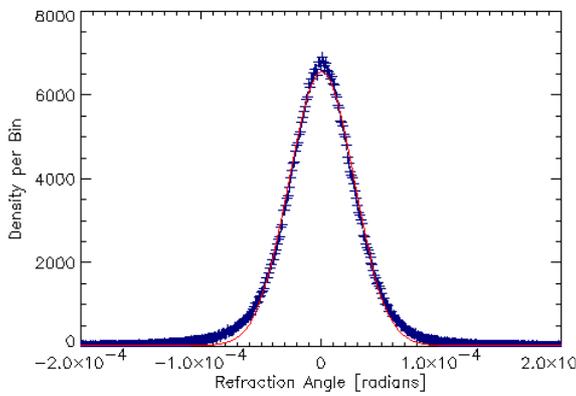

*Figure 5.2.3e*: 30μm spheres at 17keV, 2.9mm phantom thickness (6000 spheres), 500000 rays

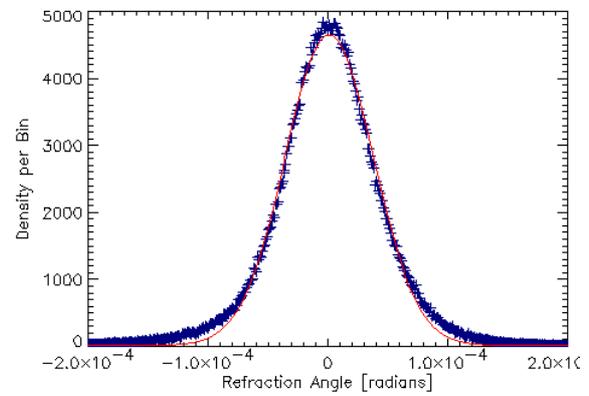

*Figure 5.2.3f*: 30μm spheres at 17keV, 4.84mm phantom thickness (10000 spheres), 500000 rays

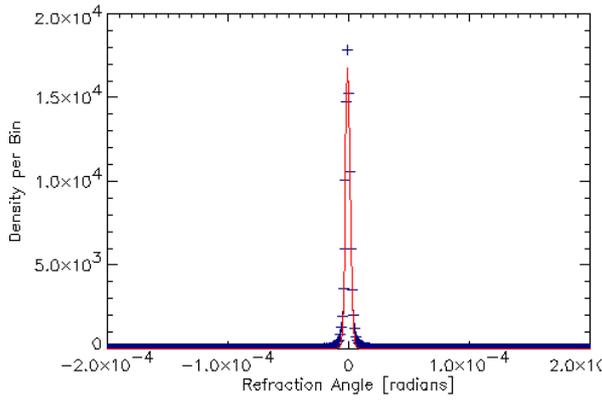

*Figure 5.2.4a*: 30μm spheres at 25 keV, 0.24mm phantom thickness (500 spheres), 100000 rays

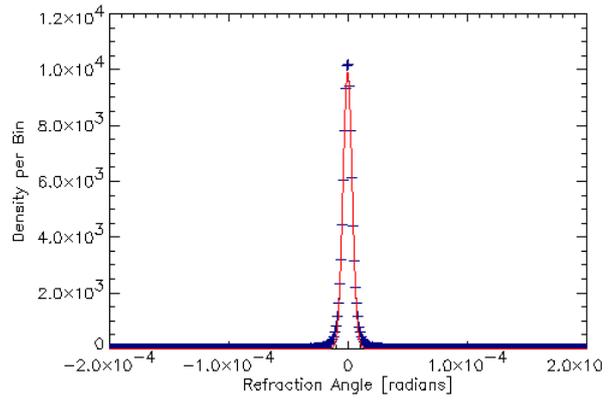

*Figure 5.2.4b*: 30μm spheres at 25keV, 0.48mm phantom thickness (1000 spheres), 100000 rays

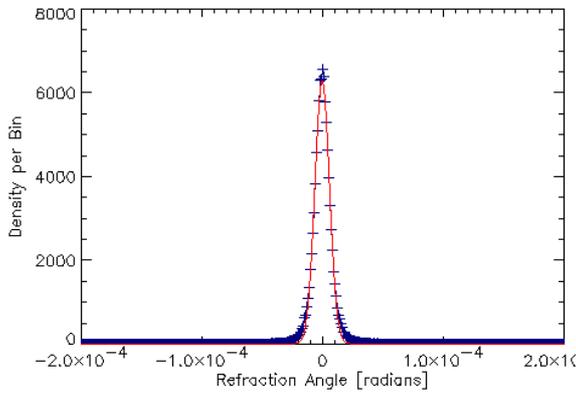

*Figure 5.2.4c*: 30μm spheres at 25keV, 0.97mm phantom thickness (2000 spheres),100000 rays

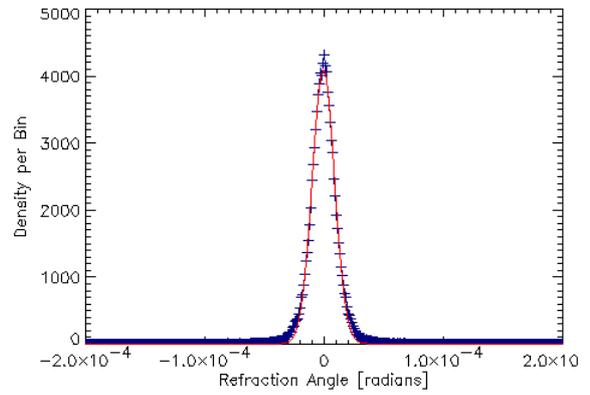

*Figure 5.2.4d*: 30μm spheres at 25keV, 1.93mm phantom thickness (4000 spheres), 100000 rays

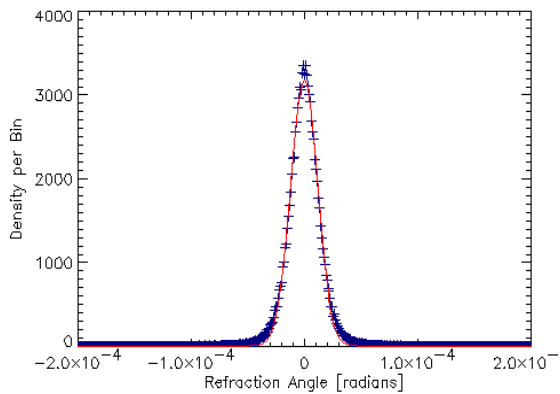

*Figure 5.2.4e*: 30μm spheres at 25keV, 2.9mm phantom thickness (6000 spheres), 100000 rays

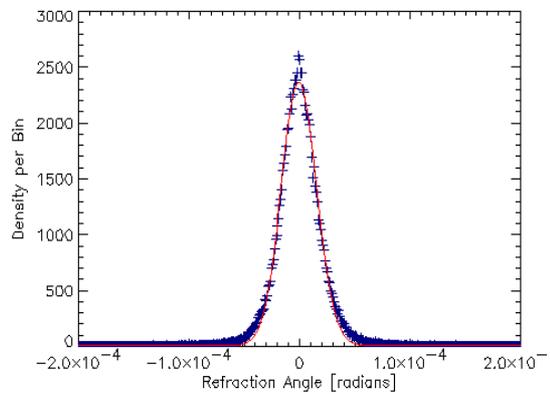

*Figure 5.2.4f*: 30μm spheres at 25keV, 4.84mm phantom thickness (10000 spheres), 100000 rays

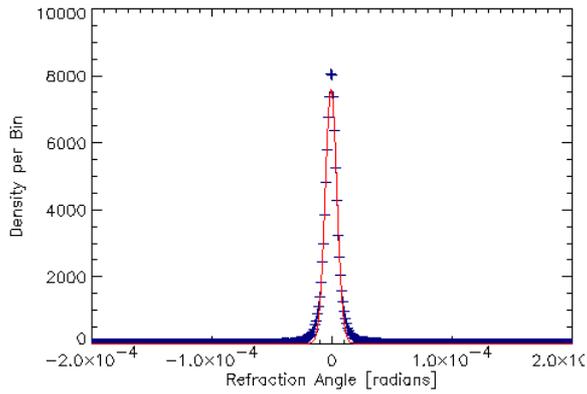

*Figure 5.2.5a*: 100μm spheres at 17keV, 0.81mm phantom thickness (500 spheres), 100000 rays

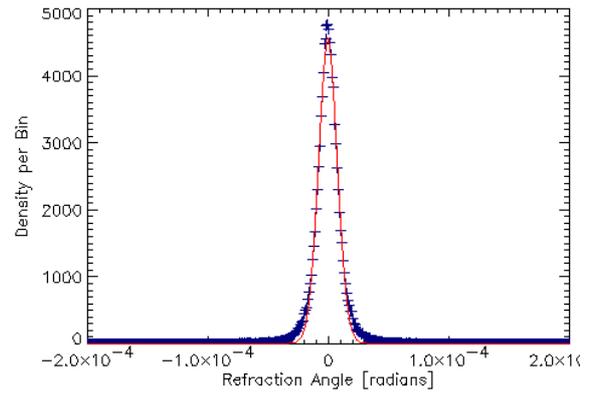

*Figure 5.2.5b*: 100μm spheres at 17keV, 1.61mm phantom thickness (1000 spheres), 100000 rays

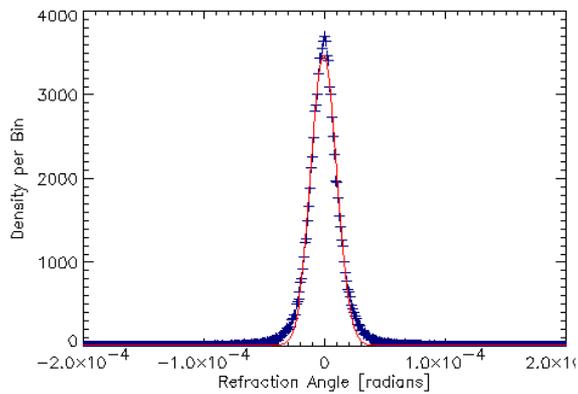

*Figure 5.2.5c*: 100μm spheres at 17keV, 2.42mm phantom thickness (1500 spheres), 100000 rays

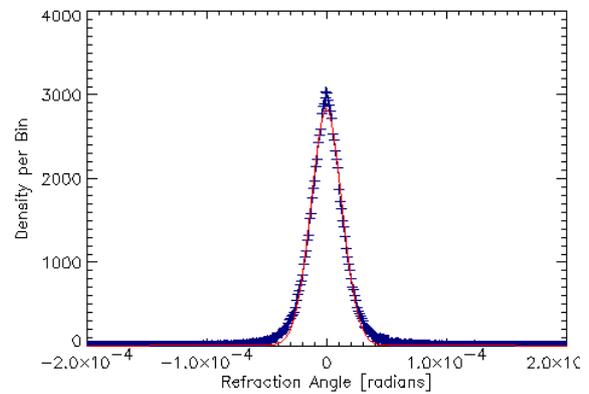

*Figure 5.2.5d*: 100μm spheres at 17keV, 3.22mm phantom thickness (2000 spheres), 100000 rays

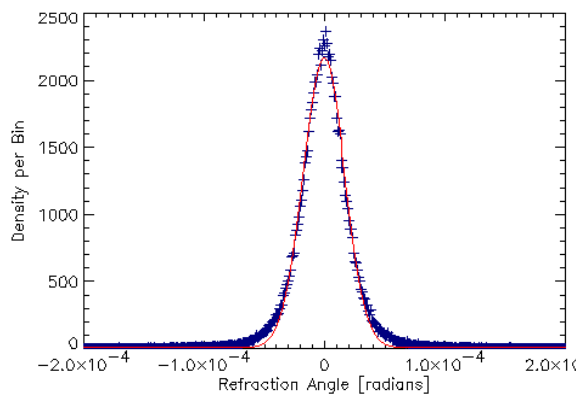

*Figure 5.2.5e*: 100μm spheres at 17keV, 4.84mm phantom thickness (3000 spheres), 100000 rays

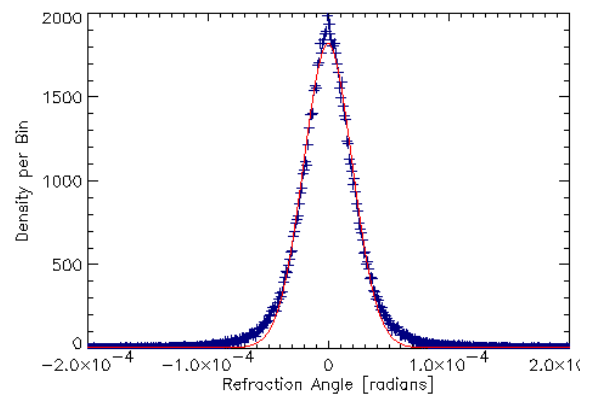

*Figure 5.2.5f*: 100μm spheres at 17keV, 6.45mm phantom thickness (4000 spheres), 100000 rays

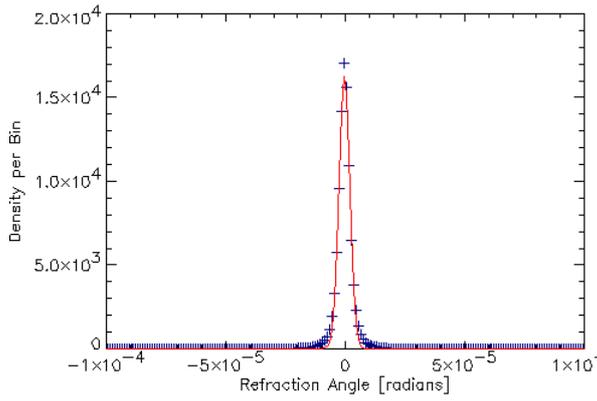
*Figure 5.2.6a*: 100µm spheres at 25keV, 0.81mm phantom thickness (500 spheres), 100000 rays

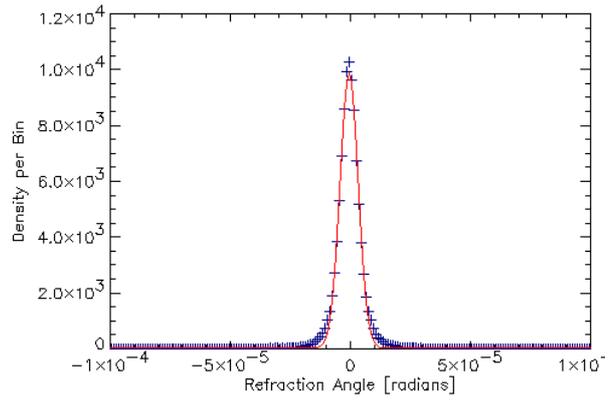
*Figure 5.2.6b*: 100µm spheres at 25keV, 1.61mm phantom thickness (1000 spheres), 100000 rays

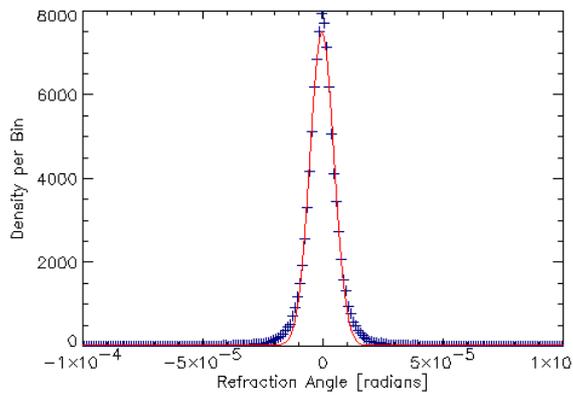
*Figure 5.2.6c*: 100µm spheres at 25keV, 2.42mm phantom thickness (1500 spheres), 100000 rays

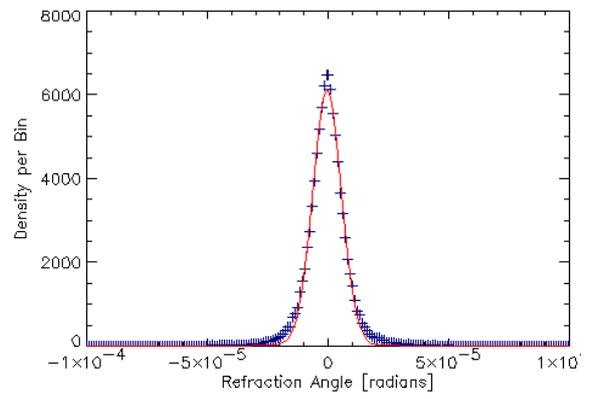
*Figure 5.2.6d*: 100µm spheres at 25keV, 3.22mm phantom thickness (2000 spheres), 100000 rays

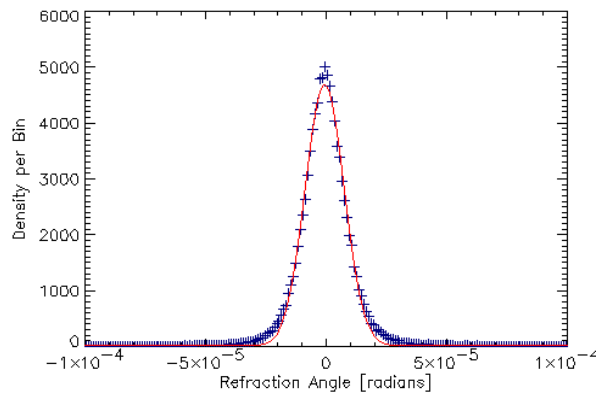
*Figure 5.2.6e*: 100µm spheres at 25keV, 4.84mm phantom thickness (3000 spheres), 100000 rays

Keeping in mind that the transverse displacement of the X-rays, even after scattering in a phantom thickness of 5 mm, is typically less than 100 nanometers compared to the sphere diameter of 30 or 100 µm, the above mentioned correlations may be responsible for an additional linear term in the dependence of the standard deviation values of the scattering angle versus layer thickness. An example is an aligned linear array of spheres where the incident beam is parallel to the axis.

A convolution of a linear and a square root term is certainly unphysical, because in this case the linear term would dominate the behavior at large thickness, contradicting the statistical nature of the process. There is also no reason for the very long range correlations required for such a behavior, and short range correlations would not prevent a square root behavior at large thickness.

To take into account these arguments, the Monte Carlo results of the standard deviation values for the scattering angle distributions versus the layer thickness have been fitted to an horizontal parabola according to (5.2.2):

$$Y = \sqrt{a \cdot X + b^2} - b \qquad (5.2.2)$$

where Y is the standard deviation of the scattering angle and X is the phantom thickness. This is a two parameter fit, having a finite slope (linear term) close to the origin, forcing the fit through it (retaining the constraint: no scatterer → no scattering angle) and showing a square root behavior at large thickness.

As shown in Figures 5.2.7-5.2.8 there is perfect agreement of the fits with the simulated results for all sphere sizes and energies. Moreover, as expected, an increase of the sphere diameter results in a decrease of the standard deviation of the refraction angle distribution. The MC results shown in Figures 5.2.7-5.2.8 together with the fit parameters *a* and *b* according to the equation (5.2.2) are also presented in the Tables 5.2.1 – 5.2.6.

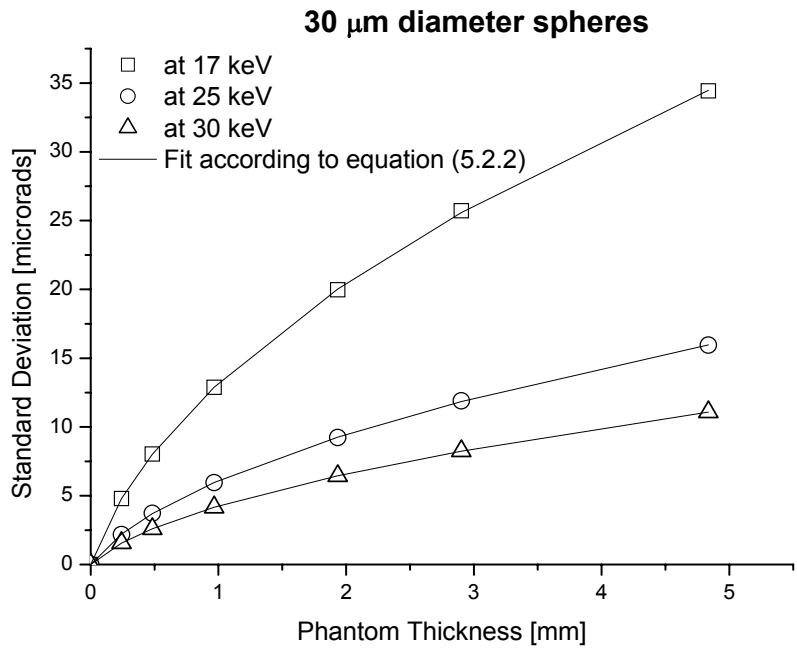

*Figure 5.2.7:* The comparison between the simulated at different energies (17keV, 25keV and 30keV) standard deviation values for 30μm in diameter spheres

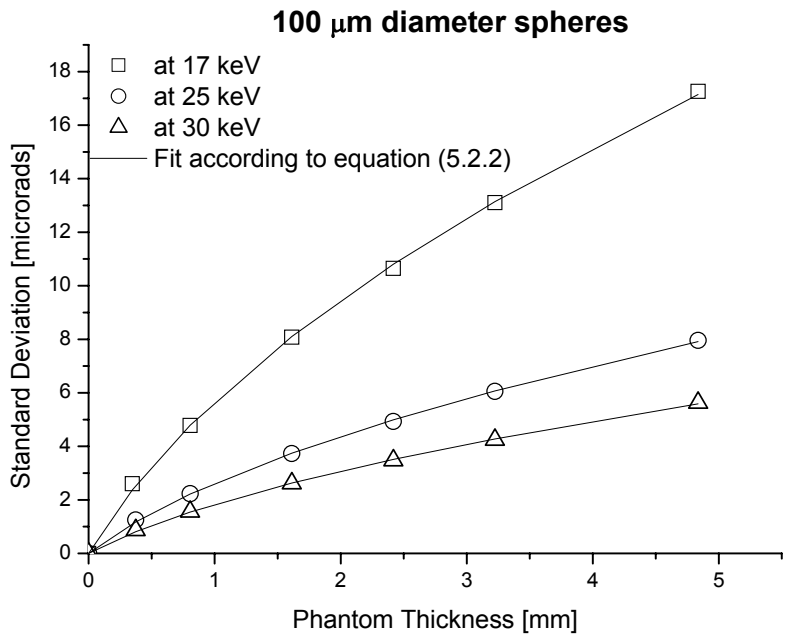

*Figure 5.2.8:* The comparison between the simulated at different energies (17keV, 25keV and 30keV) standard deviation values for 100μm in diameter spheres

| Table 5.2.1: Results for the 30μm diameter spheres at 17 keV |||
|---|---|---|
| Phantom Thickness, [mm] | Simulated Standard Deviation, [μradians] | Fit parameters according to equation (5.2.2) |
| 0.03 | 1.39±0.001 | a=328.26±9.82<br>b=5.81±0.51 |
| 0.24 | 4.79±0.01 | |
| 0.48 | 8.03±0.02 | |
| 0.97 | 12.88±0.03 | |
| 1.93 | 19.97±0.05 | |
| 2.90 | 25.71±0.06 | |
| 4.84 | 34.43±0.08 | |

| Table 5.2.2 Results for the 30μm diameter spheres at 25keV |||
|---|---|---|
| Phantom Thickness, [mm] | Simulated Standard Deviation, [μradians] | Fit parameters according to equation (5.2.2) |
| 0.03 | 0.641±0.001 | a=70.57±2.12<br>b=2.72±0.24 |
| 0.24 | 2.174±0.005 | |
| 0.48 | 3.726±0.008 | |
| 0.97 | 5.96±0.01 | |
| 1.93 | 9.23±0.02 | |
| 2.90 | 11.90±0.03 | |
| 4.84 | 15.95±0.04 | |

| Table 5.2.3 Results for the 30μm diameter spheres at 30keV |||
|---|---|---|
| Phantom Thickness, [mm] | Simulated Standard Deviation, [μradians] | Fit parameters according to equation (5.2.2) |
| 0.03 | 0.446±0.001 | a=33.56±0.93<br>b=1.79±0.15 |
| 0.24 | 1.575±0.004 | |
| 0.48 | 2.596±0.006 | |
| 0.97 | 4.159±0.009 | |
| 1.93 | 6.44±0.01 | |
| 2.90 | 8.25±0.02 | |
| 4.84 | 11.08±0.02 | |

| Table 5.2.4 Results for the 100μm diameter spheres at 17keV |||
|---|---|---|
| Phantom Thickness, [mm] | Simulated Standard Deviation, [μradians] | Fit parameters according to equation (5.2.2) |
| 0.37 | 2.606±0.006 | a=117.88±3.93<br>b=7.70±0.46 |
| 0.81 | 4.78±0.01 | |
| 1.61 | 8.08±0.02 | |
| 2.42 | 10.65±0.02 | |
| 3.22 | 13.10±0.03 | |
| 4.84 | 17.26±0.04 | |

| Table 5.2.5 Results for the 100μm diameter spheres at 25keV |||
|---|---|---|
| Phantom Thickness, [mm] | Simulated Standard Deviation, [μradians] | Fit parameters according to equation (5.2.2) |
| 0.37 | 1.25±0.003 | |
| 0.81 | 2.231±0.005 | |
| 1.61 | 3.732±0.008 | a=23.003±0.852 |
| 2.42 | 4.93±0.01 | b=3.07±0.22 |
| 3.22 | 6.05±0.01 | |
| 4.84 | 7.96±0.02 | |

| Table 5.2.6 Results for the 100μm diameter spheres at 30keV |||
|---|---|---|
| Phantom Thickness, [mm] | Simulated Standard Deviation, [μradians] | Fit parameters according to equation (5.2.2) |
| 0.37 | 0.872±0.002 | |
| 0.81 | 1.559±0.003 | |
| 1.61 | 2.617±0.006 | a=11.74±0.49 |
| 2.42 | 3.475±0.008 | b=2.29±0.18 |
| 3.22 | 4.244±0.009 | |
| 4.84 | 5.63±0.01 | |

In the Monte Carlo simulation the energy dependence of the standard deviation values fulfills the $1/E_\gamma^2$ law, as expected. If the energy is increased from 10 keV to 30 keV, the standard deviation values of the refraction angle distribution decrease for both 30μm and 100μm spheres exactly with the square of the photon energy, Eγ, as required for energies well above the K-edges involved due to classical electrodynamics for quasi-free electrons [50], [51] (see equations 1.7 and 1.8). This energy dependence is presented in Figure 5.2.9a for 30 μm and 100 μm sphere diameter at a fixed phantom thickness of about 3 mm as an example for the correct behavior of the Monte Carlo program. For comparison Figure 5.2.9b shows the energy dependence for a single 30 μm diameter sphere. For a single 100 μm diameter sphere the standard deviation values are equal to those for the 30 μm diameter sphere since the standard deviation values of the refraction angle distribution do not depend on the sphere's size.

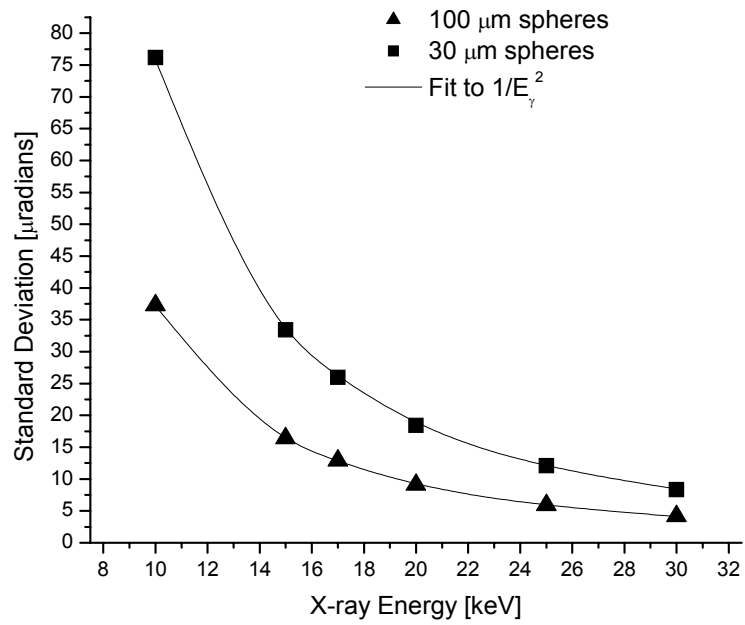

*Figure 5.2.9a*: Simulated standard deviation values of the refraction angle distribution as a function of the photon energy, Eγ, for 30μm and 100μm sphere diameter and a fixed phantom thickness of about 3 mm

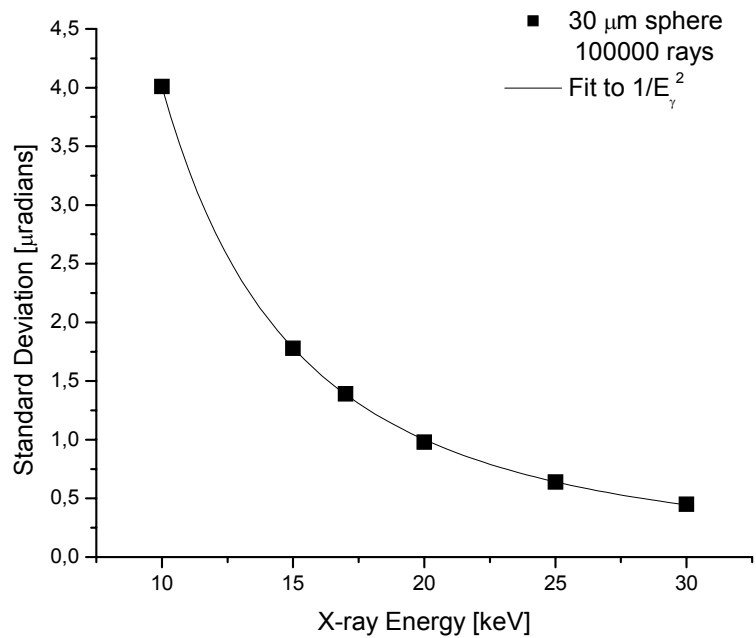

*Figure 5.2.9b*: Simulated standard deviation values of the refraction angle distribution as a function of the photon energy, Eγ, for 30μm single sphere

Obviously, and trivially, the standard deviation values are strongly dependent on the phantom packing density. The less the packing density of the sample, the smaller are the values of the width of the refraction angle distribution. Larger spheres (e.g. 100 µm) are usually packed closer to each other due to the influence of the gravitational force, which is dominant compared to the electrostatic one and occurs during the filling procedure of the PMMA spheres in the Plexiglas containers. Smaller spheres, smaller water vapour intake and higher resistivity increase the relative importance of electrostatic forces with respect to gravitational force. It is known that for these reasons polystyrene spheres cannot be handled in air. Therefore, PMMA spheres have been chosen to perform the experiments. Due to electrostatic forces the smaller spheres will not arrive at the minimum gravitational potential position as modeled in the Monte Carlo generation for the spheres position. This effect will decrease the effective packing density. Hence, for smaller spheres the electrostatic forces become dominant, which results in a smaller packing density.

It seems extremely difficult to include the effect of the electrostatic forces between the spheres in the Monte Carlo program. To get an idea about the electrostatic effect, it was included in a coarse way in the Monte Carlo simulation by generating with the SIAMS 3D software spheres of larger sphere diameter, e.g. 40 µm with the inherent packing density of 0.6. In the Monte Carlo simulation the centers of the spheres from this SIAMS 3D data file have then been used in the ray tracing of spheres with smaller diameter, e.g. 30 µm, thus changing the packing density parameter from 0.6 to 0.26. The same method was used to generate different packing densities for 100 µm in diameter spheres. For example, to obtain a packing density equal to 0.52, by SIAMS 3D software a sample with sphere diameter 105 µm was generated and then the centers of the spheres in the created data file have been used in the ray tracing of spheres with smaller diameter, e.g. 100 µm. The Figures 5.2.10-5.2.12 show the simulated standard deviation values of the scattering angle distribution versus the layer thickness for packing densities of 0.6, 0.26 and of 0.15, for 30 µm diameter spheres at 17, 25 and 30 keV, respectively, together with a fit according to (5.2.2). Figure 5.2.13 shows the simulated standard deviation values of the scattering angle distribution versus the layer thickness for packing densities of 0.6, 0.52 and of 0.45, for 100 µm diameter spheres at 17 keV together with a fit according to (5.2.2). The expected decrease with decreasing packing density is statistically significant and clearly visible.

The results related to the graphs are also presented in the Tables 5.2.7 – 5.2.10.

Systematic errors in the simulation results arise from the way the simulated phantom was created: the required phantom (box) thickness was obtained by artificially connecting a suitable number of generated smaller thickness boxes (this is due to the SIAMS 3D evaluation program limitations); moreover, the box was filled in one case with the spheres in the direction of X-rays propagation and, alternatively, filled with the spheres in the direction of the gravitational force (normal to the X-ray direction). For the 30 µm diameter spheres at 17 keV the systematic error is calculated to be equal to 2.5 %.

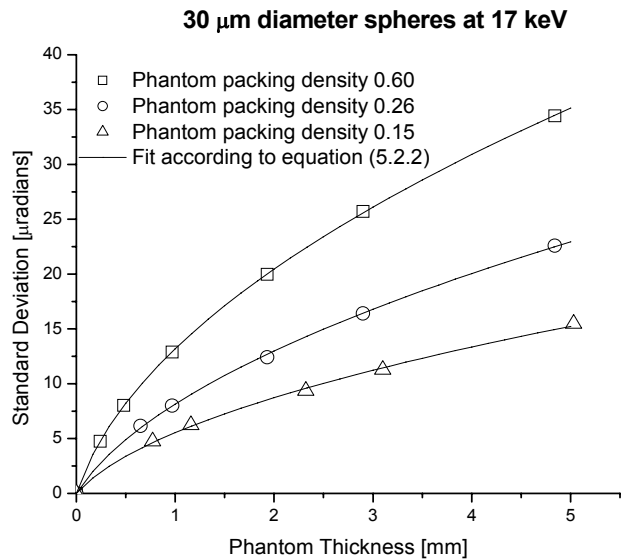

*Figure 5.2.10*: Standard deviation values of the refraction angle distribution as a function of the phantom thickness for three different packing densities at 17 keV and 30 µm sphere diameter

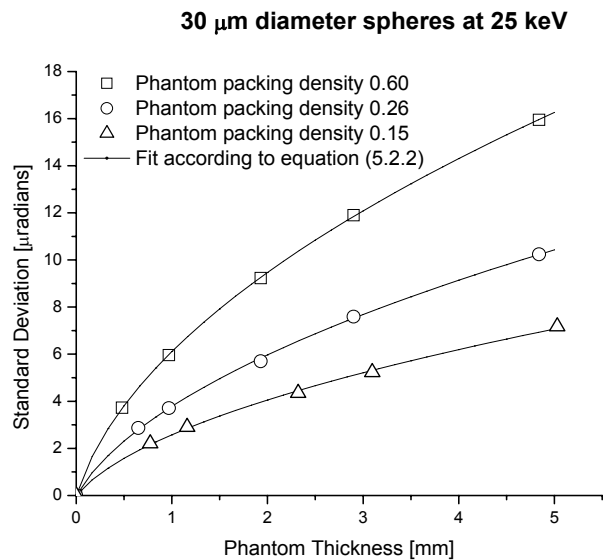

*Figure 5.2.11*: Standard deviation values of the refraction angle distribution as a function of the phantom thickness for three different packing densities at 25 keV and 30 µm sphere diameter

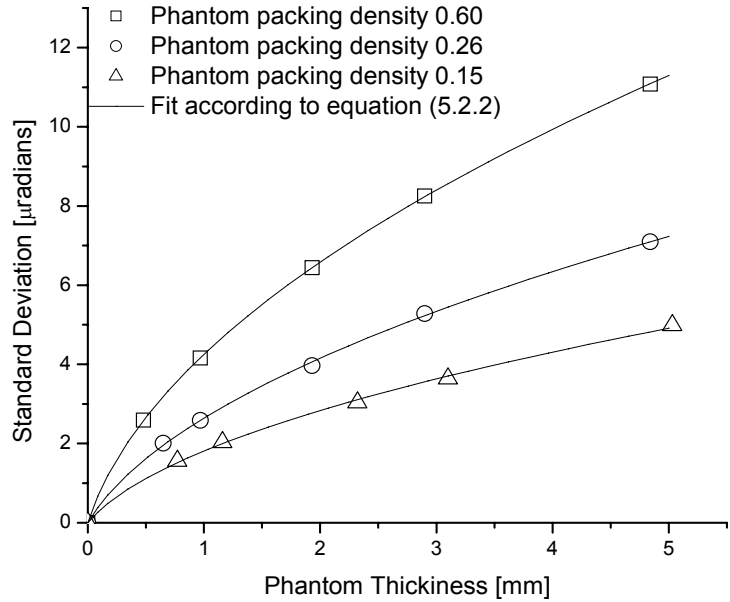

*Figure 5.2.12*: Standard deviation values of the refraction angle distribution as a function of the phantom thickness for three different packing densities at 30 keV and 30 μm sphere diameter

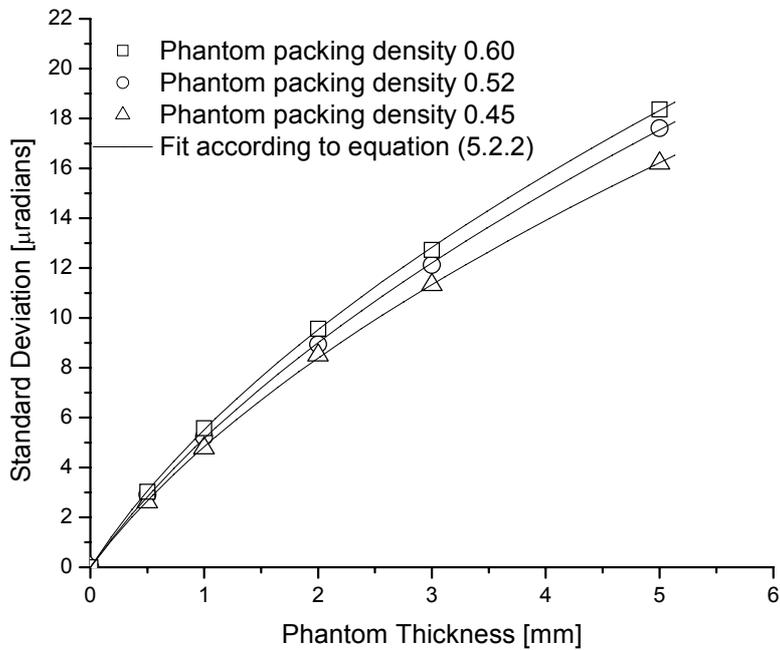

*Figure 5.2.13*: Standard deviation values of the refraction angle distribution as a function of the phantom thickness for three different packing densities at 17 keV and 100 μm sphere diameter

For 100 μm diameter spheres the systematic error of the standard deviation values has been evaluated to be equal to 1.3 %. In Figure 5.2.8 and in the Table 5.2.4 the results are presented for the phantom in which the required thickness were obtained by artificially connecting an integer number of smaller thickness boxes. The same values will be used for comparison with the experimental data in Chapter 6. In Figure 5.2.13 and in the Table 5.2.10 the phantom was directly obtained by filling the box with the necessary thickness by the PMMA spheres for all packing densities.

| colspan="6" | *Table 5.2.7:* Simulated standard deviation values for 30 μm diameter spheres at 17 keV |
|---|---|---|---|---|---|
| Phantom Thickness, [mm] | Simulated standard deviation, [μradians], phantom packing density 0.6 | Phantom Thickness, [mm] | Simulated standard deviation, [μradians], phantom packing density 0.26 | Phantom Thickness, [mm] | Simulated standard deviation, [μradians], phantom packing density 0.15 |
| 0.48 | 8.03±0.02 | 0.65 | 6.16±0.01 | 0.77 | 4.74±0.01 |
| 0.97 | 12.88±0.03 | 0.97 | 8.01±0.02 | 1.16 | 6.23±0.01 |
| 1.93 | 19.97±0.05 | 1.93 | 12.41±0.03 | 2.32 | 9.37±0.02 |
| 2.90 | 25.71±0.06 | 2.90 | 16.41±0.04 | 3.10 | 11.27±0.03 |
| 4.84 | 34.43±0.08 | 4.84 | 22.60±0.05 | 5.03 | 15.47±0.03 |

| colspan="6" | *Table 5.2.8:* Simulated standard deviation values for 30 μm diameter spheres at 25 keV |
|---|---|---|---|---|---|
| Phantom Thickness, [mm] | Simulated standard deviation, [μradians], phantom packing density 0.6 | Phantom Thickness, [mm] | Simulated standard deviation, [μradians], phantom packing density 0.26 | Phantom Thickness, [mm] | Simulated standard deviation, [μradians], phantom packing density 0.15 |
| 0.48 | 3.726±0.008 | 0.65 | 2.864±0.006 | 0.77 | 2.210±0.005 |
| 0.97 | 5.96±0.01 | 0.97 | 3.708±0.008 | 1.16 | 2.904±0.006 |
| 1.93 | 9.23±0.02 | 1.93 | 5.70±0.01 | 2.32 | 4.35±0.01 |
| 2.90 | 11.90±0.03 | 2.90 | 7.59±0.02 | 3.10 | 5.23±0.01 |
| 4.84 | 15.95±0.04 | 4.84 | 10.24±0.02 | 5.03 | 7.17±0.02 |

| Table 5.2.9: Simulated standard deviation values for 30 μm diameter spheres at 30 keV |||||||
|---|---|---|---|---|---|
| Phantom Thickness, [mm] | Simulated standard deviation, [μradians], phantom packing density 0.6 | Phantom Thickness, [mm] | Simulated standard deviation, [μradians], phantom packing density 0.26 | Phantom Thickness, [mm] | Simulated standard deviation, [μradians], phantom packing density 0.15 |
| 0.48 | 2.596±0.006 | 0.65 | 2.011±0.004 | 0.77 | 1.558±0.003 |
| 0.97 | 4.159±0.009 | 0.97 | 2.588±0.006 | 1.16 | 2.033±0.006 |
| 1.93 | 6.44±0.01 | 1.93 | 3.965±0.009 | 2.32 | 3.033±0.007 |
| 2.90 | 8.25±0.02 | 2.90 | 5.28±0.01 | 3.10 | 3.643±0.008 |
| 4.84 | 11.08±0.02 | 4.84 | 7.10±0.02 | 5.03 | 4.99±0.01 |

| Table 5.2.10: Simulated standard deviation values for 100 μm diameter spheres at 17 keV ||||
|---|---|---|---|
| Phantom Thickness, [mm] | Simulated standard deviation, [μradians], phantom packing density 0.6 | Simulated standard deviation, [μradians], phantom packing density 0.52 | Simulated standard deviation, [μradians], phantom packing density 0.45 |
| 0.5 | 3.031±0.008 | 2.914±0.007 | 2.61±0.006 |
| 1 | 5.57±0.01 | 5.22±0.01 | 4.77±0.01 |
| 2 | 9.56±0.02 | 8.94±0.02 | 8.50±0.02 |
| 3 | 12.72±0.03 | 12.12±0.03 | 11.32±0.03 |
| 5 | 18.37±0.04 | 17.61±0.04 | 16.20±0.04 |

# Chapter 6

# Comparison of the simulated results with the experimental data

To compare the simulated results with the experimental data the standard deviation values of the simulated refraction angle distributions have been plotted as a function of the phantom thickness together with the measured standard deviation values for both 30 and 100 μm spheres diameter at the three energies of 17, 25 and 30 keV, respectively. For both the simulated results and the experimental data a fit according to equation (5.2.2) has been performed.

The measured standard deviation values for 30 μm spheres at 17 keV are compared to the Monte Carlo simulated standard deviation values for phantom packing densities of 0.6, 0.26 and 0.15. How the packing parameter was varied has been already discussed in the previous Chapter 5. The results are shown in Figure 6.1.1.

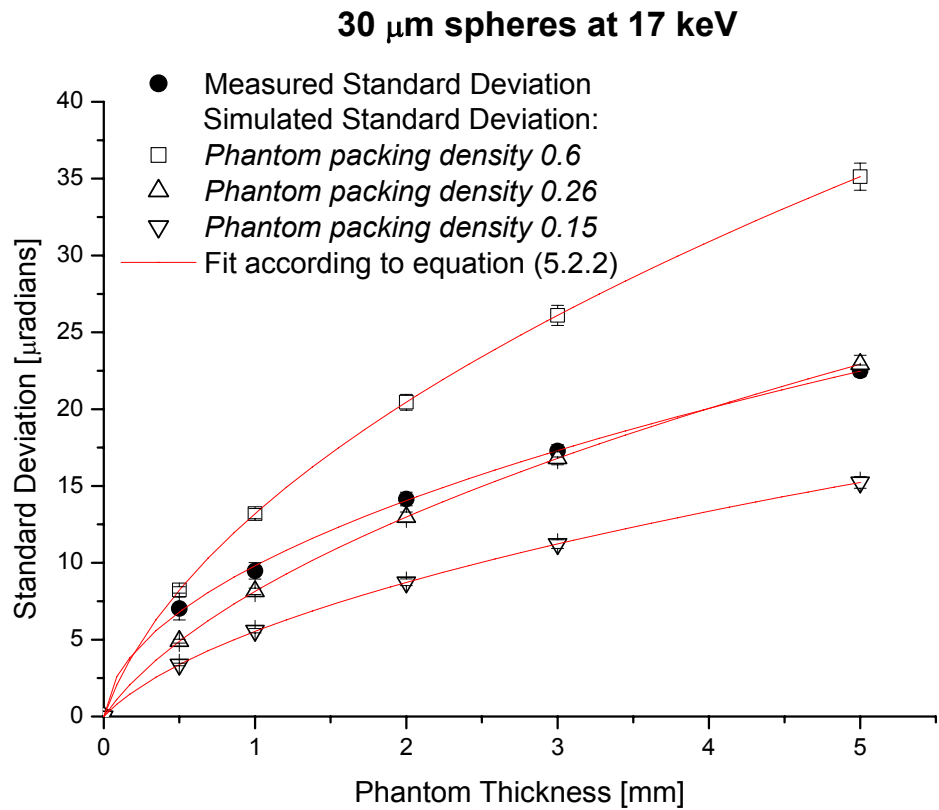

*Figure 6.1.1:* Comparison between measured and simulated standard deviation values as a function of the phantom thickness: 30 μm diameter spheres at 17 keV. For the simulated results an inherent packing densities of 0.6, 0.26 and 0.15 have been used

As can be seen from Figure 6.1.1, the simulated standard deviation values for the phantom with the packing density 0.6 (originally generated by the SIAMS 3D software) are far higher compared to the measured values. The simulated standard deviation values for the phantom with the packing density 0.15 are far lower than the measured standard deviation values. The best agreement between the simulated results and the measured data are obtained for the simulated phantom packing density 0.26. Hence, in order to obtain a reasonable agreement between simulated and measured standard deviation values, it makes sense to adjust the simulated packing density to obtain simulated results as close as possible to the measured ones. The standard deviation depends on the packing density as a square root function. Moreover, an approximate packing density in the experimental phantom can be derived if one plots the standard deviation values for a certain fixed phantom thickness versus the packing density. The intersection point between the curve obtained in this way and the line parallel to the abscissa (representing the measured value for the standard deviation at the same fixed phantom thickness) gives the approximate number for the experimental packing density or, the same, adjusted simulated packing density. The new simulated standard deviation values can be obtained by equation:

$$\sigma_2 = \sigma_1 \cdot \sqrt{\frac{k_2}{k_1}} \tag{6.1.1}$$

where $k_2$ is the adjusted packing density (which is more close to the packing density of the experimental phantom), $k_1$ is the packing density of a specific simulated phantom (usually correspondent to the simulated standard deviation values closest to the experimental one) and $\sigma_1$ is the standard deviation value for the phantom with the packing density value $k_1$.

For the simulated results compared to the experimental data for 30 μm spheres at 17 keV, the best agreement was found for the simulated phantom with the adjusted packing density equal to 0.30. Figure 6.1.2 shows the correspondent comparison.

The related to Figure 6.1.1 and Figure 6.1.2 numerical values are presented in the Tables 6.1.1 - 6.1.2.

Since the agreement between the fits according to equation (5.2.2) and the simulated results is perfect, for convenience of comparison with the experimental data the simulated standard deviation values for all the phantoms were found exactly for the phantom thickness of 0.5, 1, 2, 3 and 5 mm.

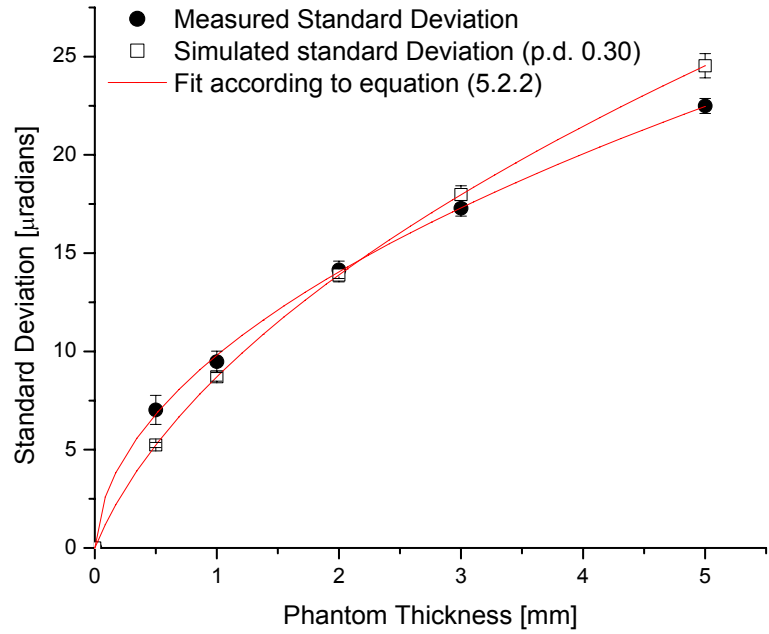

*Figure 6.1.2:* Comparison between measured and simulated standard deviation values as a function of the phantom thickness: 30 µm diameter spheres at 17 keV. For the simulated results the adjusted packing density of 0.30 has been used

| Phantom Thickness, [mm] | *Table 6.1.1:* Simulated results for 30µm diameter spheres at 17 keV | | |
|---|---|---|---|
| | Simulated standard deviation, [µradians], phantom packing density 0.60 | Simulated standard deviation, [µradians], phantom packing density 0.26 | Simulated standard deviation, [µradians], phantom packing density 0.15 |
| 0.5 | 8.24±0.21 | 4.90±0.12 | 3.38±0.09 |
| 1 | 13.20±0.33 | 8.14±0.20 | 5.58±0.14 |
| 2 | 20.45±0.51 | 12.98±033 | 8.73±0.22 |
| 3 | 26.10±0.65 | 16.80±0.42 | 11.22±0.28 |
| 5 | 35.13±0.88 | 22.93±0.58 | 15.23±0.38 |

| Phantom Thickness, [mm] | *Table 6.1.2*: Comparison between the measured standard deviation values and the simulated standard deviation values with the adjusted packing density of 0.30, for 30µm diameter spheres at 17 keV | | |
|---|---|---|---|
| | Measured standard deviation, [µradians] | Phantom Thickness, [mm] | Simulated standard deviation, [µradians], phantom packing density 0.30 |
| 0.5 | 7.03±0.74 | 0.5 | 5.24±0.13 |
| 1 | 9.48±0.53 | 1 | 8.71±0.22 |
| 2 | 14.15±0.44 | 2 | 13.89±0.35 |
| 3 | 17.29±0.40 | 3 | 17.98±0.45 |
| 5 | 22.49±0.38 | 5 | 24.54±0.62 |

In Figure 6.1.3 the experimental data for 30 μm spheres at 25 keV are compared to the Monte Carlo results for the phantom packing densities of 0.6, 0.26 and 0.15. The experimental data are close to the simulated results for the phantom packing density of 0.26, but not really in good agreement with them. In order to achieve a reasonable coincidence, the simulated phantom packing density is needed to be adjusted using the equation 6.1.1. The adjusted phantom packing density is equal to 0.36. The comparison between the experimental standard deviation values and the simulated standard deviation values of the phantom with the adjusted packing density of 0.36 is shown in Figure 6.1.4. The related numerical values are presented in the Tables 6.1.3 – 6.1.4.

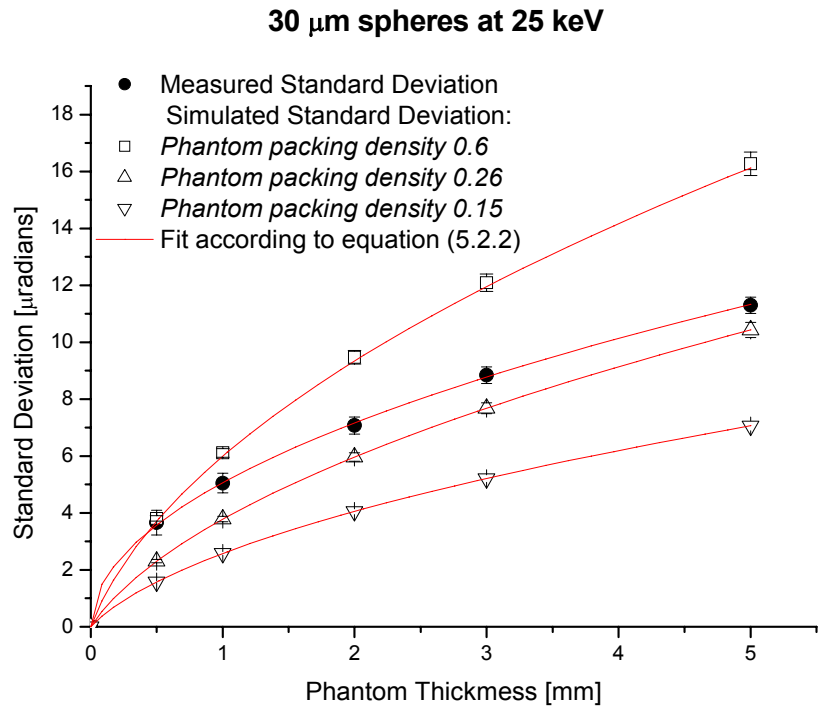

*Figure 6.1.3:* Comparison between measured and simulated standard deviation values as a function of the phantom thickness: 30 μm diameter spheres at 25 keV. For the simulated results the inherent packing densities of 0.6, 0.26 and 0.15 have been used

| *Table 6.1.3:* Simulated standard deviation values for 30μm diameter spheres at 25 keV | | | |
|---|---|---|---|
| Phantom Thickness, [mm] | Simulated standard deviation, [μradians], phantom packing density 0.60 | Simulated standard deviation, [μradians], phantom packing density 0.26 | Simulated standard deviation, [μradians], phantom packing density 0.15 |
| 0.5 | 3.81±0.10 | 2.31±0.06 | 1.58±0.04 |
| 1 | 6.11±0.15 | 3.78±0.09 | 2.58±0.06 |
| 2 | 9.47±0.24 | 5.97±0.15 | 4.05±0.10 |
| 3 | 12.09±0.30 | 7.68±0.19 | 5.21±0.13 |
| 5 | 16.27±0.41 | 10.43±0.26 | 7.06±0.18 |

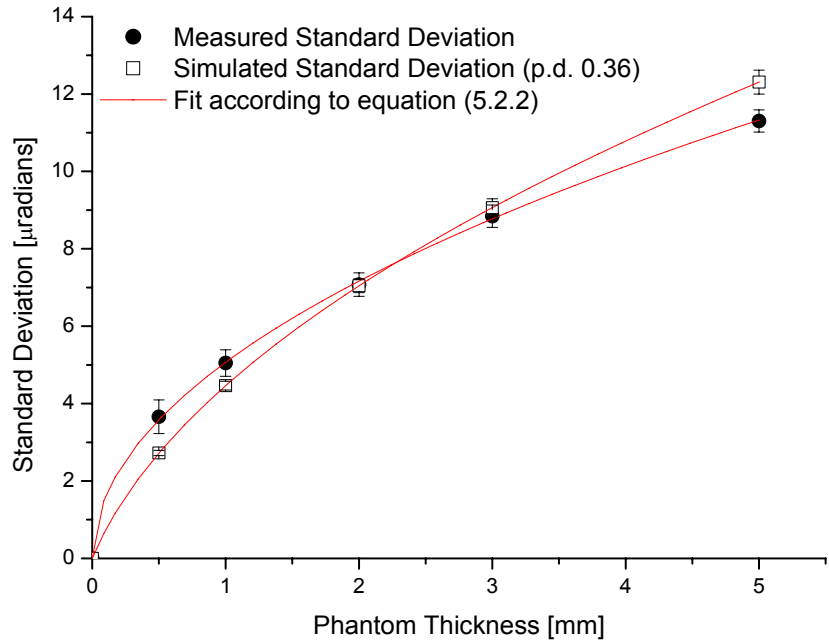

*Figure 6.1.4:* Comparison between measured and simulated standard deviation values as a function of the phantom thickness: 30 μm diameter spheres at 25 keV. For the simulated results the adjusted packing density of 0.36 has been used

| Phantom Thickness, [mm] | Measured standard deviation, [μradians] | Phantom Thickness, [mm] | Simulated standard deviation, [μradians], phantom packing density 0.36 |
|---|---|---|---|
| 0.5 | 3.66±0.43 | 0.5 | 2.72 ±0.07 |
| 1 | 5.05±0.34 | 1 | 4.46±0.11 |
| 2 | 7.07±0.30 | 2 | 7.04±0.18 |
| 3 | 8.84±0.29 | 3 | 9.06±0.23 |
| 5 | 11.30±0.29 | 5 | 12.31±0.31 |

*Table 6.1.4*: Comparison between the measured standard deviation values and the simulated standard deviation values with adjusted packing density of 0.36, for 30 μm diameter spheres at 25 keV

In Figure 6.1.5 the measured data for 30 μm spheres at 30 keV are compared to the simulated results for the phantom packing densities of 0.6, 0.26 and 0.15. The behavior is similar to that for the same spheres dimension at 17 and 25 keV. Hence, the adjusted phantom packing density is found and it is equal approximately to 0.36 (the value equal to that found for the phantom at 25 keV). The comparison between the experimental standard deviation values and the simulated standard deviation values for the phantom with the adjusted packing density of 0.36 is shown in Figure 6.1.6. The related data are presented in the Tables 6.1.5 – 6.1.6.

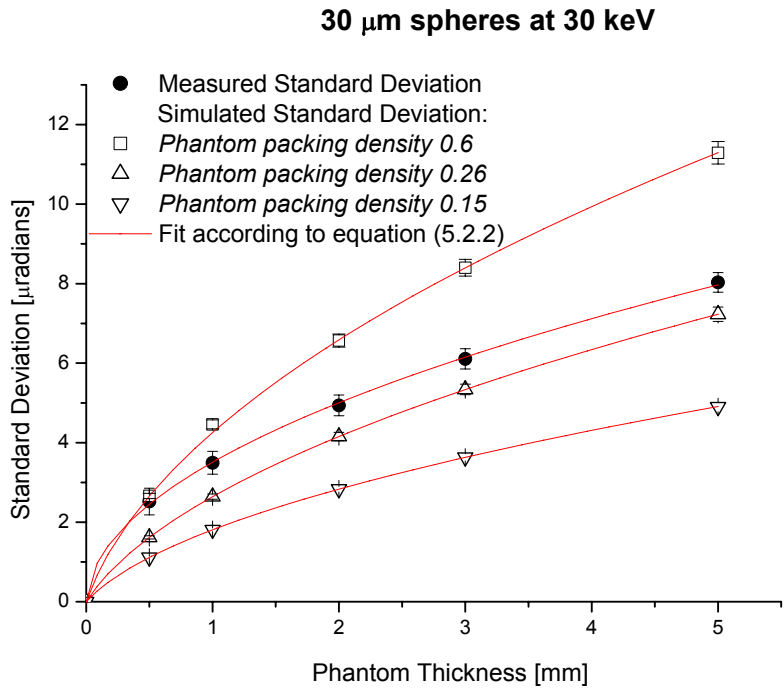

*Figure 6.1.5:* Comparison between measured and simulated standard deviation values as a function of the phantom thickness: 30 µm diameter spheres at 30 keV. For the simulated results the inherent packing density of 0.6, 0.26 and 0.15 have been used

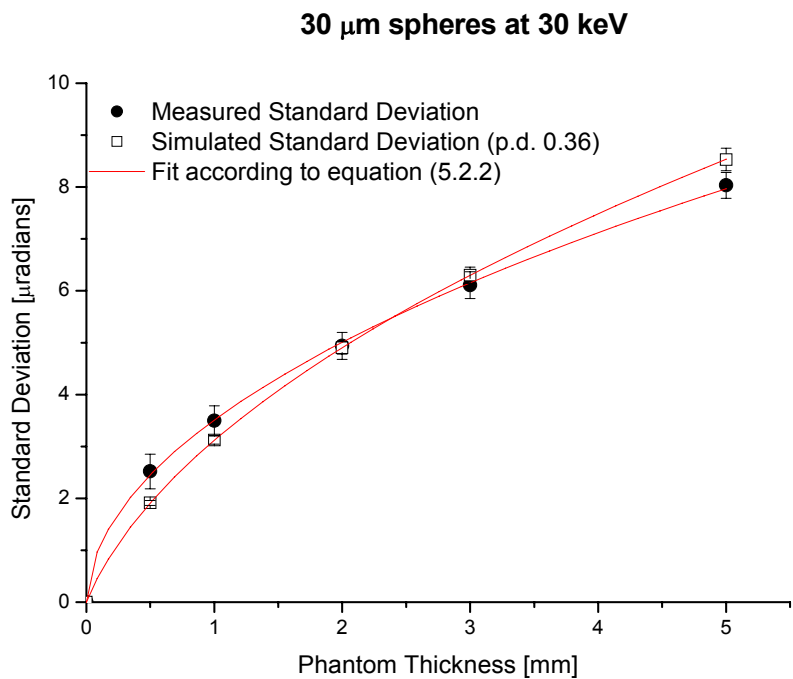

*Figure 6.1.6:* Comparison between measured and simulated standard deviation values as a function of the phantom thickness: 30 µm diameter spheres at 30 keV. For the simulated results the adjusted packing density of 0.36 has been used

| Table 6.1.5: Simulated standard deviation values for 30μm diameter spheres at 30 keV ||||
|---|---|---|---|
| Phantom Thickness, [mm] | Simulated standard deviation, [μradians], phantom packing density 0.60 | Simulated standard deviation, [μradians], phantom packing density 0.26 | Simulated standard deviation, [μradians], phantom packing density 0.15 |
| 0.5 | 2.66±0.07 | 1.62±0.04 | 1.12±0.03 |
| 1 | 4.46±0.11 | 2.64±0.07 | 1.81±0.05 |
| 2 | 6.57±0.16 | 4.16±0.10 | 2.83±0.07 |
| 3 | 8.40±0.21 | 5.34±0.13 | 3.63±0.09 |
| 5 | 11.29±0.28 | 7.23±0.18 | 4.90±0.12 |

| Table 6.1.6: Comparison between the measured standard deviation values and the simulated standard deviation values with adjusted packing density of 0.36, for 30 μm diameter spheres at 30 keV ||||
|---|---|---|---|
| Phantom Thickness, [mm] | Measured standard deviation, [μradians] | Phantom Thickness, [mm] | Simulated standard deviation, [μradians], phantom packing density 0.36 |
| 0.5 | 2.52±0.33 | 0.5 | 1.91±0.05 |
| 1 | 3.50±0.29 | 1 | 3.12±0.08 |
| 2 | 4.94±0.26 | 2 | 4.90±0.12 |
| 3 | 6.11±0.26 | 3 | 6.30±0.16 |
| 5 | 8.03±0.25 | 5 | 8.53±0.21 |

In Figures 6.1.7 – 6.1.9 the simulated standard deviation values for the standard deviation of the refraction angle distributions are compared to the measured data for 100 μm diameter spheres for 17, 25 and 30 keV, respectively. The packing density in all simulated phantoms is 0.6. As can be seen from these figures, the agreement between measured and simulated standard deviation values for this inherent simulated phantom packing density is sufficient. The related data are presented in the Tables 6.1.7 – 6.1.9. The reason why the packing density for 100 μm spheres at all energies is larger than for 30 μm spheres was discussed in the Chapter 5: the larger the sphere size the influence of the gravitational force become more decisive compared to the electrostatic forces, which are dominant at small sphere sizes. Hence, the spheres are packed more close to each other, which results in a higher packing density. As can be seen from the figures and from the tables, the expectations discussed in Chapter 5 are completely fulfilled.

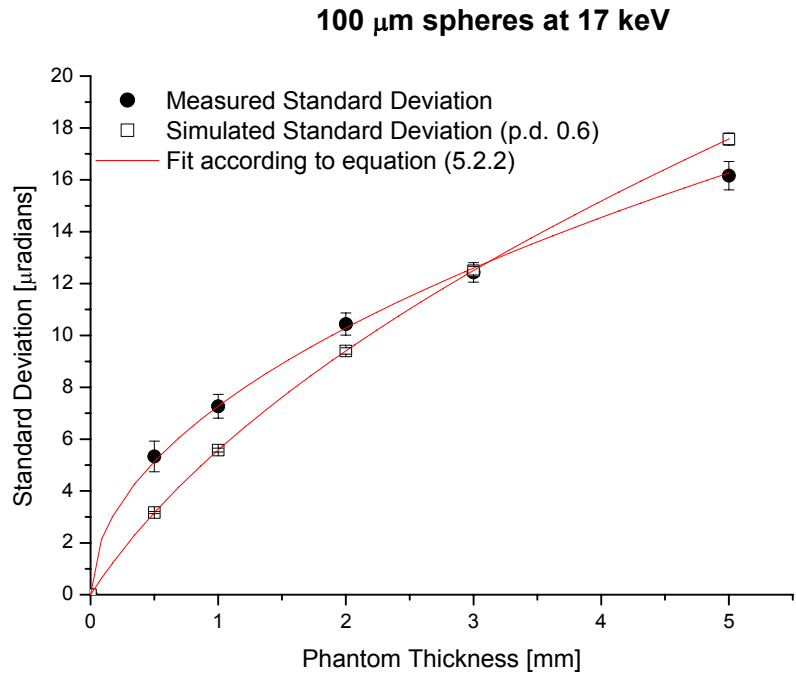

*Figure 6.1.7:* Comparison between measured and simulated standard deviation values as a function of the phantom thickness: 100 μm diameter spheres at 17 keV. For the simulated results the inherent packing density of 0.6 has been used

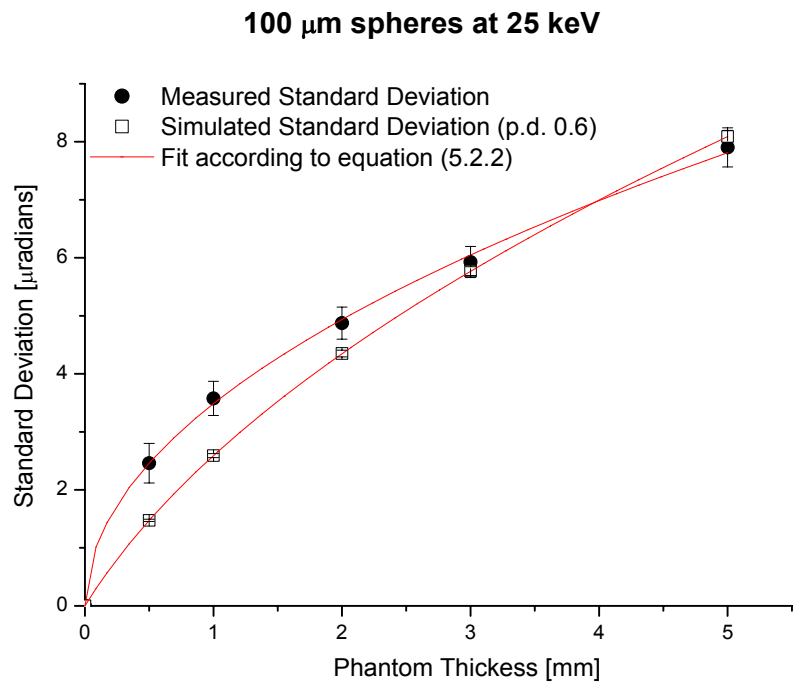

*Figure 6.1.8:* Comparison between measured and simulated standard deviation values as a function of the phantom thickness: 100 μm diameter spheres at 25 keV. For the simulated results the inherent packing density of 0.6 has been used

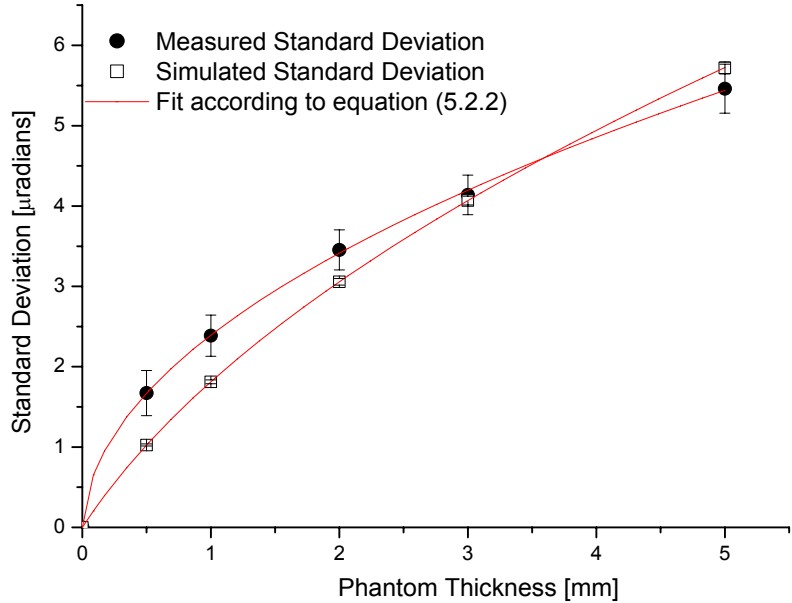

*Figure 6.1.9:* Comparison between measured and simulated standard deviation values as a function of the phantom thickness: 100 µm diameter spheres at 30 keV. For the simulated results the inherent packing density of 0.6 has been used

| *Table 6.1.7*: Comparison between the measured standard deviation values and the simulated standard deviation values with inherent packing density of 0.60, for 100 µm diameter spheres at 17 keV | | | |
|---|---|---|---|
| Phantom Thickness, [mm] | Measured standard deviation, [µradians] | Phantom Thickness, [mm] | Simulated standard deviation, [µradians], phantom packing density 0.6 |
| 0.5 | 5.33±0.59 | 0.5 | 3.17±0.04 |
| 1 | 7.27±0.46 | 1 | 5.58±0.07 |
| 2 | 10.44±0.43 | 2 | 9.40±0.12 |
| 3 | 12.43±0.38 | 3 | 12.50±0.17 |
| 5 | 16.16±0.55 | 5 | 17.57±0.23 |

| *Table 6.1.8*: Comparison between the measured standard deviation values and the simulated standard deviation values with inherent packing density of 0.60, for 100 µm diameter spheres at 25 keV | | | |
|---|---|---|---|
| Phantom Thickness, [mm] | Measured standard deviation, [µradians] | Phantom Thickness, [mm] | Simulated standard deviation [µradians], phantom packing density 0.6 |
| 0.5 | 2.46±0.34 | 0.5 | 1.47±0.02 |
| 1 | 3.58±0.29 | 1 | 2.59±0.03 |
| 2 | 4.87±0.28 | 2 | 4.35±0.06 |
| 3 | 5.93±0.27 | 3 | 5.77±0.08 |
| 5 | 7.90±0.34 | 5 | 8.09±0.11 |

| Phantom Thickness, [mm] | Measured standard deviation, [µradians] | Phantom Thickness, [mm] | Simulated standard deviation, [µradians], phantom packing density 0.6 |
|---|---|---|---|
| 0.5 | 1.67±0.28 | 0.5 | 1.026±0.01 |
| 1 | 2.39±0.27 | 1 | 1.81±0.02 |
| 2 | 3.45±0.25 | 2 | 3.06±0.04 |
| 3 | 4.14±0.25 | 3 | 4.09±0.05 |
| 5 | 5.46±0.30 | 5 | 5.72±0.08 |

*Table 6.1.9*: Comparison between the measured standard deviation values and the simulated standard deviation values with inherent packing density of 0.60, for 100 µm diameter spheres at 30 keV

From the comparison of the measured standard deviation values of the refraction angle distributions to the results of the Monte Carlo program the following conclusions can be drawn:

A reasonable agreement between the simulated results and the measured data is obtained after the simulated phantom packing densities have been adjusted to 0.30, 0.36 and 0.36 for the 30 µm diameter spheres at 17, 25 and 30 keV, respectively, and left unadjusted at 0.6 for 100 µm diameter spheres at all energies.

The agreement is reasonable but not really good. At small phantom thickness it is possible to notice the large discrepancies between the measured and the simulated standard deviation values. As it was discussed in Chapter 3 (Experimental Investigations), the experimental values at small phantom thickness have the largest errors, mainly because the experimental width of the rocking curve is only slightly increased by the scattering in the sample with resulting uncertainties in the deconvolution. The big systematic error on the small thickness arises also from the phantom thickness machining. According to the simulated standard deviation values, for thin layers the number of scatters is not large. Moreover, the condition of statistical independence is not completely fulfilled (see Chapter 5), since it is possible only in the case if the positions of the spheres' centers are chosen completely random. In our case the spheres have some degree of order due to the gravitational and electrostatic forces, resulting the scattering angle correlations. These facts can be responsible for the additional linear term in the dependence of the refraction angle distributions' standard deviation values for the phantom thickness.

In view of the presence of the deviations from the ideal conditions in both the experimental data and the simulated results, the agreement is considered sufficient and quite interesting to use the Monte Carlo simulation program for further more complicated studies in lung alveoli and bronchi imaging.

# *Chapter 7*
# Summary

A Monte Carlo program based on a three dimensional vector approach was developed to model multiple refractive scattering of X-ray photons in objects with a fine structure. A particular interest was paid to the investigation of lung tissue. Alveoli are low contrast and low absorbing structures. Hence, they are not visible in the conventional radiography which is based on the changes in the absorption arising from density differences and from variation in the thickness and composition of the object. Another possibility to image fine structure objects is to use the phase imaging techniques. As known, the phase change constant $\delta$ at low energies (15-30 keV) is 1000 times larger than the absorption constant $\beta$. The Diffraction Enhance Imaging (DEI) technique is one of the recent phase sensitive techniques based on the use of an analyzer crystal placed between the sample and the detector. With such a setup it became possible to improve the visibility of structures having characteristics similar to that of the alveoli: the bronchi, micro-calcifications in the breast as possible indicator of the cancer etc. Nevertheless, in case of overlapping microstructures, such as alveoli in lung, arose the problems of multiple scattering, which is typical for objects with big amount of small structures. Multiple scattering in DEI, like any scattered radiation independently of its origin, reduces the spatial resolution and, correspondingly image contrast. Hence, deeper investigations were essential to understand in detail the phenomenon of multiple scatter. For this purpose the Monte Carlo program has been developed. The output of this Monte Carlo program is a refraction (scattering) angle distributions and the standard deviations of these distributions. Simulated results were experimentally verified by measuring with a DEI setup at the SYRMEP beamline at ELETTRA (Italy) the broadening of the analyzer rocking curve caused by highly scattering phantoms. Phantoms, simulating a simplified lung tissue, have been obtained by filling a Plexiglas box in the form of a stair with monodisperse PMMA microspheres. The stairs had a depth of 0.5, 1, 2, 3 and 5 mm and same width of 6 mm. Two diameters of spheres were used: 30 and 100 µm exposed to energies of 17, 25 and 30 keV. For the ease of the simulation the alveoli in the Monte Carlo program are modelled as multiple layers of these monodisperse PMMA microspheres.

According to the Central Limit Theorem and Convolution Theorem, for large number of scatters in the phantom and in case of complete independence of the individual refraction scatter processes the refraction (scattering) angle distribution is expected to be Gaussian.

Moreover, the standard deviation of this distribution should increase with the square root of the phantom thickness. Indeed, from the Monte Carlo program the refraction angle distributions output approach a Gaussian shape. More difficult is to evaluate the agreement of the simulation output with the expected theoretical square root dependence. The standard deviation values of the measured refraction angle distributions as a function of the phantom thickness had been compared to the correspondent simulated results for 17, 25 and 30 keV X-ray photon energy and 30 and 100 μm in diameter spheres. The agreement is reasonable but not really good. The largest discrepancies occur at small phantom thickness. In this case the experimental data have the largest errors. Moreover, the condition of statistical independence and large number of scatters in the simulation are not completely fulfilled over the full range of phantom thickness. Therefore deviation from a pure square root law can be expected and a new fit function was proposed.

The standard deviation values are strongly dependent on the phantom packing density. The larger spheres (e.g. 100 μm) are usually packed close to each other due to the influence of gravitational force, which is dominant compared to the electrostatic one and occurs when filling the PMMA spheres in the Plexiglas box. For the smaller spheres (e.g. 30 μm) increase the importance of the electrostatic forces with respect to the gravitational one. This forces decrease the effective packing density of the small spheres. Hence, the agreement between the measured data and the Monte Carlo simulations is reasonable if the packing density of the simulated phantom is adjusted to 0.30, 0.36 and 0.36 for the 30 μm spheres at 17, 25 and 30 keV, respectively, and left unadjusted at 0.6 for 100 μm at all. Such packing densities seem realistic considering the preparation of the phantom in the experiment. The goal of this study has been to verify the working principle on simple phantoms. In the next step more complex biological samples, such as lung tissues, will be imitated and examined quantitatively. In addition, the reduction in normal DEI signals caused by multiple scattering in the object can be studied to find out which system parameters are optimal to use with such a specific kind of sample like lung tissues.

# APPENDIX A

## A1. Monte Carlo photon transport program

pro sphere_generation90_debug, rays

; NAME:

;      sphere_generation90_debug.pro

; PURPOSE:

;      **Simulation of multiple refractive scattering in DEI technique**

; CALLING SEQUENCE:

;      sphere_generation90_debug, rays

; INPUTS:

;      **rays** - number of X-rays falling on the phantom

;      **n1** - refraction coefficient outside the spheres

;      **n2** - refraction coefficient inside the spheres

;      **delta** - real part of the refraction coefficient

;      **xsize, ysize, zsize** - dimensions of the box

;      **N** - number of the neighboring spheres (to check for the

;      interactions with the ray)

;      **direction** - initial direction of the ray

;      (all rays parallel between each other and to the axe X)

;      **x_source[0:2,*]** - initial position of the rays,

;      here random variable

;      **C, C1**-arrays of the centers of the spheres

;      **R, R1**-arrays of the radius of the spheres

; OPTIONAL KEYWORDS:

;      NONE

; COMMON BLOCKS:

;      share1 - coordinates of the centers of the spheres and their radius

;      need to be accessed by another procedure:roots60_last.pro

; PROCEDURE:

;     other procedures to work with:roots60_last.pro

;

; OUTPUTS:

;     **angle** – *projected onto XY plane refraction angle*

;     **angle**1 - *projected onto XZ plane refraction angle*

;     **ent_point** - *the entrance point of the x-ray into the sphere (first*

;  *intersection point)*

;     **exit_point** - *the exit point of the x-ray out of the sphere*

;     *(second intersection point)*

;     **index[m1]**-*index of the sphere with which ray has been interacted*

;   RESTRICTIONS:

; HISTORY:

;     *written by A.Khromova, March 2004*

   common share1, coords,R1

  ; *The DEVICE procedure provides device-dependent control over the current graphics device. To get an 8-bit image to display in color, or to use a color table tool, you will have to turn color decomposition OFF:*

   device,decomposed=0

; *Open a file to read a data (Cartesian coordinates of the spheres'*
; *centers and spheres' radius)*
OPENR, 1, FILEPATH('sphp1.dat', SUBDIR = ['procedures']), ERROR = err

; *To check if file opens without the errors*
   IF (err NE 0) then PRINTF, -2,!ERR_STRING

; *To open files to write the output data*

    OPENW,2, 'ent_point.dat'

    OPENW,3, 'exit_point.dat'

    OPENW,4, 'sphere_number.dat'

    OPENW,5, 'refraction_angles_30_1layer_xy_17keV_debug_500tho.dat'

    OPENW,7, 'refraction_angles_30_1layer_xz_17keV_debug_500tho.dat'

    OPENW,6,'x_source.dat'

    openw,8,'number_of_scatters_30_1layer_17keV_debug_500tho.dat'

; *Establish error handler. When errors occur, the index of the error is*

; *returned in the variable Error_status.*

; *Initially, this argument is set to zero.*

    CATCH, Error_status

  IF Error_status NE 0 THEN BEGIN; This statement begins the error ;handler.

    PRINT, 'Error index:', Error_status

    PRINT, 'Error message:', !ERR_STRING

    stop

  ENDIF

  ;  CONSTANTS ---------------------------------------------

;Energy=17;keV

n1=1.0D; *refraction coefficient outside spheres*

delta=9.22829372D-07; PMMA at 17keV

;delta=4.25990919D-07; PMMA at 25 keV

n2=1-delta; *refraction coefficient inside spheres*

N=300L

xsize=24.18D; dimensions of the box

```
ysize=24.18D
zsize=24.18D

direction=[1.0D,0.0D,0.0D]
seed=-1L;A variable or constant used to initialize the random sequence ; on input, and in which the state of the random number generator is saved ; on output.

; VARIABLES ----------------------------------------------
coords=dblarr(4,500); array contains Cartesian coordinates of the
;centers and radius of the spheres from the data file
coords_sorted=dblarr(4,500);array contains sorted in ascending order ;by x-coordinates data from the previous array(coords[0:3,*])
values=dblarr(2,N);array contains roots of the quadratic equation to ;find the first intersection point of the ray with the sphere
;values1=dblarr(2,N);array contains roots of the quadratic equation to find the second intersection point of the ray with the sphere
position=dblarr(3,N); array contains initial position of the ray
C=dblarr(3,800); array contains Cartesian coordinates of the spheres'
        ; centers (whole number of spheres in the box)
R=dblarr(1,800); array contains radius of the spheres (whole number of
        ; the spheres in the box)
C1=dblarr(3,N); array contains Cartesian coordinates of the spheres'
        ; centers (close neighboring spheres)
R1=dblarr(N);array contains radius of the spheres (close neighboring
; spheres)
x_source=dblarr(3,rays); array contains random initial positions of
; the rays
t=dblarr(3); array contains refracted direction of the ray inside the
; sphere
t1=dblarr(3);array contains refracted direction of the ray which goes ; out of the sphere
```

```
d=dblarr(3);initial direction of the ray inside or outside the sphere
ni=dblarr(3); array contains Cartesian coordinates of the normals to
; the intersection points(normalized vector)
n=dblarr(3);array contains Cartesian coordinates of the normals to the
         ;intersection points
cross_prod=dblarr(3);array contains cross product of two unit
            ;directions of the ray
normal=dblarr(3); array contains Cartesian coordinates of the normals
                ;to the intersection points(not normalized vector)
number=0L; number of scatter centers

; INITIALIZATION ------------------------------------------

; Read data from the sphp1.dat into the variable coords
readf,1,coords
coords[0,*]=(coords[0,*]+12.09)
coords[1,*]=(coords[1,*]+12.09)
coords[2,*]=coords[2,*]
position[*,*]=0

;Sorting of the spheres' centers coordinates in ascending order by the
; X-coordinate
index=sort(coords[0,*])
coords_sorted=coords[*,index]
x_source[0,*]=0
x_source[1,*]=randomu(seed,rays)*18.18+3.0
x_source[2,*]=randomu(seed,rays)*18.18+3.0
```

```
C[0:2,*]=0

R[0,*]=0

C[0:2,0:499]=coords_sorted(0:2,*)

R[0,0:499]=coords_sorted(3,*)

p=0L

m1=0L

C1=C[0:2,0:299]; close neighboring array

R1=R[*,0:299]
```

; CYCLE BEGIN  ------------------------------------

```
while p lt rays do begin

while (x_source[0,p] lt xsize) and (x_source[1,p] lt ysize) and (x_source[2,p] lt zsize) do begin

    ; Output of the initial positions of the rays

    printf,6,x_source[0:2,p]

    ;print,x_source[0:2,p]

    ; Initial position of the rays

    position[0:2,*]=rebin(x_source[0:2,p],3,300)
```

;Output of roots from the alfa quadratic equation(see parametric ray ;equation, alfa – displacement of the ray from the initial position ;according to the initial direction)

```
        values[0:1,*]=roots60_last(C1,position,direction)
```

; Here we choose only positive couples of the roots and roots where both

; values not equal zero

```
j=where((values[0,*] ne 0.D) and (values[1,*] ne 0.D) and (values[0,*] gt 0.D) and (values[1,*] gt 0.D),count)
```

```
    ; If there is no such a numbers this is mean that there are no
    ; intersection points of the ray with the sphere and we trace the next ray
        if count eq 0 then goto, skip_it

    ; If we have needed couples, we choose from the values the minimum one.
    ; This value is responsible for the first intersection point
        min_displ=min(values[0:1,j])

    ; k-this is the index of the sphere where the intersection has occurred
    k=where((values[0,*] eq min_displ) or (values[1,*] eq min_displ),count1)

    ; Now we can find the entrance point of the ray into the sphere
    ent_point=position+[min_displ*direction[0],min_displ*direction[1],min_displ*direction[2]]
    ; Since we were using close neighboring array, now we should find the ;index of the sphere in the whole (real) array
        m1=k[0]+m1
        normal=ent_point-C[0:2,m1]
        ni=-normal/norm(normal);normal to the unit surface in the entrance ;point
        d=direction;initial direction of the ray
        n=ni

; To find the angle between normal and initial direction we will use a
; vector product...
        cross_prod[0]=d[1]*n[2]-d[2]*n[1]
        cross_prod[1]=d[2]*n[0]-d[0]*n[2]
        cross_prod[2]=d[0]*n[1]-d[1]*n[0]
        num=sqrt(cross_prod[0]^2+cross_prod[1]^2+cross_prod[2]^2)

        incident_angle=asin(num/sqrt(total(d^2,1)*total(n^2,1)))*180/!pi
        ;incident_angle=asin(d[1]/sqrt(d[0]*d[0]+d[1]*d[1]))
```

;... *Or scalar product*

    dot_product=total(n*d,1)

    ;possible variant, *the same----------------*

    ;scalar=d[0]*n[0]+d[1]*n[1]+d[2]*n[2]

    ;incident_angle=(!pi-acos(scalar/(norm(d)*norm(n))))

    ;delta_theta=delta/k1*tan(incident_angle)

    ;refraction_angle1=incident_angle-delta_theta

    ; *From the Snell's law the refraction angle then will be:*

refraction_angle1=asin(n1*sin(incident_angle)/k2)

; *Here we will find the direction of the refracted ray after he entered the ;sphere*

    t[0]=n1/n2*( (sqrt(dot_product^2+(n2/n1)^2-1)-dot_product)*n[0]+d[0])

    t[1]=n1/n2*( (sqrt(dot_product^2+(n2/n1)^2-1)-dot_product)*n[1]+d[1])

    t[2]=n1/n2*( (sqrt(dot_product^2+(n2/n1)^2-1)-dot_product)*n[2]+d[2])

    t=t/norm(t)

;*Equivalent variant*

            ;t[0]=n1/n2*d[0]-(cos(refraction_angle1)-n1/n2*cos(incident_angle))*n[0]

            ;t[1]=n1/n2*d[1]-(cos(refraction_angle1)-n1/n2*cos(incident_angle))*n[1]

            ;t[2]=n1/n2*d[2]-(cos(refraction_angle1)-n1/n2*cos(incident_angle))*n[2]

 ; *Now the initial position of the ray will change and will be equal to the first intersection point(ent_point)*
; *Initial direction will be equal to the first refraction direction inside the sphere*
; *Algorithm to find the first intersection point will be applied again in ; order to find the 2d intersection point. But since we already know the ; index of the sphere with which the ray interacted, we don't need to check ; for the intersections with all the spheres*

```
x_source[*,p]=ent_point

y=x_source[*,p]-C[0:2,m1]

A=total(y*y,1)-R[0,m1]*R[0,m1]

B=2*t ## transpose(y)

C3d=t[0]*t[0]+t[1]*t[1]+t[2]*t[2]

discriminant=B^2-4*A*C3d

x1=(-B+sqrt(discriminant))/(2*C3d)

x2=(-B-sqrt(discriminant))/(2*C3d)

exit_point=ent_point+[x1*t[0],x1*t[1],x1*t[2]]

normal=exit_point-C[0:2,m1]

ni=normal/norm(normal);normal to the surface in the 2d intersection ; point

d=t; new initial direction of the ray

n=ni

cross_prod[0]=d[1]*n[2]-d[2]*n[1]

cross_prod[1]=d[2]*n[0]-d[0]*n[2]

cross_prod[2]=d[0]*n[1]-d[1]*n[0]

num=sqrt(cross_prod[0]^2+cross_prod[1]^2+cross_prod[2]^2)

incident_angle2=asin(num/sqrt(total(d^2,1)*total(n^2,1)))

dot_product=total(n*d,1)

refraction_angle2=asin(n2*sin(incident_angle2))

; Now we will find the direction of the refracted ray when it goes out
; from the sphere

            t1[0]=n2/n1*( (sqrt(dot_product^2+(n1/n2)^2-1)-dot_product)*n[0]+d[0])
```
81

$$t1[1]=n2/n1*(\;(sqrt(dot\_product^2+(n1/n2)^2-1)-dot\_product)*n[1]+d[1])$$

$$t1[2]=n2/n1*(\;(sqrt(dot\_product^2+(n1/n2)^2-1)-dot\_product)*n[2]+d[2])$$

;t1[0]=n2/n1*d[0]-(cos(refraction_angle2)-n2/n1*cos(incident_angle2))*n[0]

;t1[1]=n2/n1*d[1]-(cos(refraction_angle2)-n2/n1*cos(incident_angle2))*n[1]

;t1[2]=n2/n1*d[2]-(cos(refraction_angle2)-n2/n1*cos(incident_angle2))*n[2]

t1=t1/norm(t1)

*; Here we will find projected refraction angles (angle between projected*

*; direction of the refracted ray and initial direction OX in our case,*

*; since all rays have initial direction parallel to axe X)*

angle=asin(t1[1]/sqrt(t1[0]*t1[0]+t1[1]*t1[1])); *onto the plane XY*

angle1=asin(t1[2]/sqrt(t1[0]*t1[0]+t1[2]*t1[2]));*onto the plane XZ*

*; When ray goes out from the first sphere we are looking for the next*

*; sphere with which it will interact.*

*; Ray will have a new initial position (equal to the exit point now) and*

*; new direction (equal to the 2d refracted direction)*

x_source[*,p]=exit_point

direction=t1

;*Since spheres are sorted by the X-coordinates, and we know the size of* ;*the box in the x-direction, we took before 300 sphere to check for the* ;*intersections (first sphere was at 0 x-coordinate). Now we will find the* ;*index of the sphere from which we will start to check for the new* ;*intersections of the ray*

m2=m1+299

;*Output of the intersection points and index of the intersected spheres*

;printf,2,ent_point

```
    ;printf,3,exit_point

    ;printf,4,k

    ;print, index[m1]

    ;printf,4,index[m1]

; New segment of the whole array to check for intersections

    C1=C[0:2,m1:m2[0]]

    R1=R[0,m1:m2[0]]

    number=number+1; the number of the scatters centers

  endwhile

  skip_it:

  print,number

  printf,8,number

; Next ray

  p=p+1

; Initial parameters for each ray should be equal

  m1=0L

  C1=C[0:2,0:299]

  R1=R[*,0:299]

  direction=[1.0D,0.0D,0.0D]

  number=0L
```

; We interested only in the last refraction angle(when ray goes out of the ;box)

   printf,5,angle

   print,angle

   printf,7,angle1

   print,angle1

   print,p

   print,'--------------------'

   endwhile

;we have to close all the file we opened before

 close,1,2,3,4,5,6,7,8

end

## A2. PMMA refractive index for different energies

| Table A2.1.1: PMMA refractive index for different energies | | | | | |
|---|---|---|---|---|---|
| Energy, [eV] | Delta $\delta$ | Beta $\beta$ | Energy, [eV] | Delta $\delta$ | Beta $\beta$ |
| 17000 | 9.22829E-7 | 4.84683E-10 | 21457.77 | 5.79005E-07 | 2.3096E-10 |
| 17096.83 | 9.12395E-07 | 4.75256E-10 | 21579.99 | 5.72461E-07 | 2.27179E-10 |
| 17194.21 | 9.02079E-07 | 4.66046E-10 | 21702.91 | 5.6599E-07 | 2.23479E-10 |
| 17292.15 | 8.9188E-07 | 4.57047E-10 | 21826.53 | 5.59593E-07 | 2.19861E-10 |
| 17390.65 | 8.81796E-07 | 4.48254E-10 | 21950.86 | 5.53268E-07 | 2.16321E-10 |
| 17489.71 | 8.71826E-07 | 4.39666E-10 | 22075.89 | 5.47014E-07 | 2.12857E-10 |
| 17589.33 | 8.61969E-07 | 4.31274E-10 | 22201.63 | 5.40832E-07 | 2.09469E-10 |
| 17689.52 | 8.52224E-07 | 4.23073E-10 | 22328.09 | 5.34719E-07 | 2.06155E-10 |
| 17790.28 | 8.42589E-07 | 4.15058E-10 | 22455.28 | 5.28675E-07 | 2.02911E-10 |
| 17891.61 | 8.33063E-07 | 4.07226E-10 | 22583.18 | 5.227E-07 | 1.99738E-10 |
| 17993.52 | 8.23645E-07 | 3.99575E-10 | 22711.81 | 5.16793E-07 | 1.96632E-10 |
| 18096.01 | 8.14333E-07 | 3.92098E-10 | 22841.18 | 5.10952E-07 | 1.93593E-10 |
| 18199.09 | 8.05126E-07 | 3.84791E-10 | 22971.28 | 5.05177E-07 | 1.9062E-10 |
| 18302.75 | 7.96024E-07 | 3.77651E-10 | 23102.13 | 4.99467E-07 | 1.87709E-10 |
| 18407 | 7.87024E-07 | 3.70674E-10 | 23233.72 | 4.93822E-07 | 1.8486E-10 |
| 18511.85 | 7.78127E-07 | 3.63855E-10 | 23366.06 | 4.88241E-07 | 1.82072E-10 |
| 18617.29 | 7.6933E-07 | 3.57191E-10 | 23499.15 | 4.82723E-07 | 1.79343E-10 |
| 18723.34 | 7.60633E-07 | 3.50678E-10 | 23633 | 4.77267E-07 | 1.76671E-10 |
| 18829.98 | 7.52034E-07 | 3.44312E-10 | 23767.62 | 4.71873E-07 | 1.74056E-10 |
| 18937.24 | 7.43532E-07 | 3.38092E-10 | 23903 | 4.6654E-07 | 1.71495E-10 |
| 19045.11 | 7.35127E-07 | 3.32012E-10 | 24039.15 | 4.61267E-07 | 1.68988E-10 |
| 19153.59 | 7.26816E-07 | 3.26069E-10 | 24176.07 | 4.56054E-07 | 1.66533E-10 |
| 19262.69 | 7.186E-07 | 3.20262E-10 | 24313.78 | 4.509E-07 | 1.64129E-10 |
| 19372.41 | 7.10476E-07 | 3.14584E-10 | 24452.27 | 4.45805E-07 | 1.61775E-10 |
| 19482.75 | 7.02445E-07 | 3.09035E-10 | 24591.55 | 4.40766E-07 | 1.5947E-10 |
| 19593.72 | 6.94504E-07 | 3.03612E-10 | 24731.63 | 4.35785E-07 | 1.57212E-10 |
| 19705.33 | 6.86653E-07 | 2.98311E-10 | 24872.5 | 4.3086E-07 | 1.55001E-10 |
| 19817.57 | 6.78891E-07 | 2.93209E-10 | 25014.17 | 4.25991E-07 | 1.52835E-10 |
| 19930.45 | 6.71217E-07 | 2.88316E-10 | 25156.65 | 4.21177E-07 | 1.50714E-10 |
| 20043.98 | 6.6363E-07 | 2.83509E-10 | 25299.94 | 4.16417E-07 | 1.48635E-10 |
| 20158.15 | 6.56128E-07 | 2.78643E-10 | 25444.05 | 4.11711E-07 | 1.46599E-10 |
| 20272.97 | 6.48712E-07 | 2.73805E-10 | 25588.98 | 4.07058E-07 | 1.44604E-10 |
| 20388.44 | 6.41379E-07 | 2.69074E-10 | 25734.74 | 4.02458E-07 | 1.42649E-10 |
| 20504.58 | 6.34129E-07 | 2.6445E-10 | 25881.32 | 3.9791E-07 | 1.40734E-10 |
| 20621.37 | 6.26961E-07 | 2.59928E-10 | 26028.74 | 3.93413E-07 | 1.38857E-10 |
| 20738.83 | 6.19874E-07 | 2.55506E-10 | 26177 | 3.88967E-07 | 1.37017E-10 |
| 20856.96 | 6.12868E-07 | 2.51184E-10 | 26326.11 | 3.84572E-07 | 1.35214E-10 |
| 20975.76 | 6.0594E-07 | 2.46957E-10 | 26476.06 | 3.80226E-07 | 1.33447E-10 |
| 21095.24 | 5.99091E-07 | 2.42823E-10 | 26626.87 | 3.75929E-07 | 1.31715E-10 |
| 21215.39 | 5.92319E-07 | 2.38781E-10 | 26778.53 | 3.71681E-07 | 1.30017E-10 |
| 21336.24 | 5.85624E-07 | 2.34827E-10 | 26931.06 | 3.67481E-07 | 1.28352E-10 |

| Energy, [eV] | Delta $\delta$ | Beta $\beta$ |
|---|---|---|
| 27084.46 | 3.63328E-07 | 1.2672E-10 |
| 27238.74 | 3.59222E-07 | 1.2512E-10 |
| 27393.89 | 3.55163E-07 | 1.23551E-10 |
| 27549.92 | 3.51149E-07 | 1.22012E-10 |
| 27706.85 | 3.47181E-07 | 1.20503E-10 |
| 27864.67 | 3.43258E-07 | 1.19022E-10 |
| 28023.38 | 3.39379E-07 | 1.17571E-10 |
| 28183 | 3.35544E-07 | 1.16146E-10 |
| 28343.54 | 3.31753E-07 | 1.14749E-10 |
| 28504.98 | 3.28004E-07 | 1.13379E-10 |
| 28667.34 | 3.24297E-07 | 1.12034E-10 |
| 28830.63 | 3.20633E-07 | 1.10714E-10 |
| 28994.85 | 3.1701E-07 | 1.09419E-10 |
| 29160.01 | 3.13428E-07 | 1.08148E-10 |
| 29326.1 | 3.09886E-07 | 1.069E-10 |
| 29493.14 | 3.06384E-07 | 1.05676E-10 |
| 29661.14 | 3.02922E-07 | 1.04474E-10 |
| 29830.09 | 2.99499E-07 | 1.03294E-10 |
| 30000 | 2.96115E-07 | 1.02135E-10 |

# APPENDIX B

## B1. X-ray tube spectrum simulation

### *B1.1. Silver as a target material*

The choice of the target material (and hence the atomic number) of an X-ray tube determines for a given electron energy the intensity of radiation (quantity) and the radiation energy (quality) of the characteristic radiation. The primary requirements of the target material are a high melting point, good thermal conductivity and a high heat capacity (because of the predominant production of heat in the generation of x-rays). Tungsten emerges as an outstanding anode material for both short and long load times. Nevertheless, in our X-ray tube tungsten was exchanged with silver. The reason for this is that we need to operate in the mammographic range (17-25 keV), but tungsten has characteristic lines shifted to higher energies ($K_{\alpha 1} = 59.3 keV$, $K_{\alpha 2} = 57.9 keV$, $K_{\beta 1} = 67.2 keV$, $K_{\beta 2} = 69 keV$). Silver, on the contrary, has characteristic lines in the needed range ($K_{\alpha 1} = 22.163 keV$, $K_{\alpha 2} = 21.990 keV$, $K_{\beta 1} = 24.942 keV$, $K_{\beta 2} = 24.912 keV$), a high melting point (832°C) and very high thermal conductivity (4.18 W/cm K).

### *B1.2. Bremsstrahlung and characteristic radiation*

The X-ray tube spectrum consists of a Bremsstrahlung (so cold 'white radiation') with sharp peaks of a characteristic radiation. When high-energy electrons, produced by a cathode, penetrate the surface of the anode, their kinetic energy is lost to heat and radiative processes. The most probable process is ionization of the target atoms, which is going on without high energy photon production and with great loss to heat.

Another process occurs when an electron approaches very close to the nucleus of the target and undergoes a radiation loss. The electron decelerates in the Coulomb field of the nucleus because of the loss of its energy due to receding from the interaction. The loss in energy will appear as a photon with energy hν and the primary electron will recede with energy E-hν. This process is called *Bremsstrahlung*. The energy transformation usually occurs in several steps because the electron orbits lie at varying distances from the nucleus. The energies of the quanta from all the multi-stage braking processes therefore take arbitrary values below the maximum energy. Thus, x-ray quanta of every possible frequency are generated. They form a *continuous spectrum*.

Let an electron have enough energy in order to kick a strongly bound electron in the nucleus out of its orbit. For example, in silver 25.514keV is required to eject the K electron from the atom. If this voltage is supplied, all K lines will appear. Once the K electron is ejected, the space may be filled with an electron from the L, M, or N shells. In this process the energy difference is liberated and is emitted in the form of a photon of characteristic frequency $\nu$ ($h\nu = E_K - E_L$, for example). The x-rays have a characteristic wavelength corresponding to the difference in energy of the energy levels between which the transition occurred. This is called *characteristic radiation*. For example, when an electron moves from the $L_{III}$ to the K shell, energy 22.514-3.351=22.163keV is radiated. We refer to this as a $KL_{III}$ transition. The photon emitted is called the $K_{\alpha 1}$ line.

### B1.3. X-ray tube simulation

For more details about the theoretical description of the tube spectrum we refer to reference [52]. Here will be described only the main features of the spectrum, which were used for simulation.

**a) Bremsstrahlung simulation**

Kramers in his work [53] proposed the spectral distribution of the number dN of pseudo monochromatic photons with photon energies from E to E+dE or wavelengths from a $\lambda + d\lambda$ in the way shown below:

$$dN = \Omega \cdot i \cdot \sigma_{K,E} \cdot const \cdot Z \cdot \left(\frac{E_0}{E} - 1\right) dE \qquad (B1.3.1)$$

$$dN = \Omega \cdot i \cdot \sigma_{K,\lambda} \cdot const \cdot hc_0 \cdot Z \cdot \left(\frac{\lambda}{\lambda_0} - 1\right) \cdot \frac{1}{\lambda^2} d\lambda$$

where $E_0 = eV[keV]$, $\lambda_0 = \frac{hc_o}{E_0}$, $\lambda = \frac{hc_0}{E}$, $\Omega$ [sr] – solid angle of photon flux from point source, $i$ [mA] – x-ray tube current, V[kV] – x-ray tube voltage, h=6.6260755×10$^{-34}$ Js is the Planck's constant, $c_0 = 2.99792458 \times 10^8\, ms^{-1}$ – speed of light in vacuum, $e = 1.60217733 \times 10^{-19} C$ — charge of the electron, $\sigma_{K,E} [s^{-1} sr^{-1} mA^{-1} keV^{-1}]$ and $\sigma_{K,\lambda}$ — Kramers' cross-sections, $[dN] = s^{-1}$, $[const] = s^{-1} sr^{-1} mA^{-1} keV^{-1}$, [E]=keV.

Smith and Reed [54] modified Kramers' cross-section: they proposed exponents x and n for a better fit of the experimental data compared to theory.

$$\sigma_{SR,E} = const \cdot Z^n \cdot \left(\frac{E_0}{E} - 1\right)^x \tag{B1.3.2}$$

$\sigma_{SR,E} = s^{-1} sr^{-1} mA^{-1} keV^{-1}$, [n] and [x] are dimensionless.

When electrons decelerate in the target with production of X-rays, the latter can be absorbed on their path from the point of origin to the surface. This gives rise to the so called depth distribution function (DDF). In Love and Scott's concept [55], the DDF is assumed to be constant from the surface to depth $2\overline{\rho z}$:

$$\overline{\rho z} = \rho z_m \frac{0.49269 - 1.0987\eta + 0.78557\eta^2}{0.70256 - 1.09865\eta + 1.0046\eta^2 + \ln U_0} \ln U_0 \tag{B1.3.3}$$

$$\rho z_m = \frac{A}{Z}\left(0.787 \times 10^{-5} \sqrt{J} E_0^{\frac{3}{2}} + 0.735 \times 10^{-6} E_0^2\right)$$

$$\eta = E_0^m \left[0.1904 - 0.2236 \ln Z + 0.1292(\ln Z)^2 - 0.0149(\ln Z)^3\right]$$

where $U_0 = \frac{E_0}{E}$, J=0.0135Z, $m = 0.1382 - \left(\frac{0.9211}{\sqrt{Z}}\right)$, ρ — density in g $cm^{-3}$, z — linear dimension of depth in cm, ρz — depth expresses in mass per unit area (g $cm^{-2}$) and A — relative atomic weight.

Considering the photoelectric absorption of photons of energy E in the target of element j and using the equations B1.3.1-B1.3.3 the number of photons will be the following:

$$dN = \Omega \cdot i \cdot \sigma_{SR,E} \cdot dE \frac{1 - \exp(-\tau_{E,j} \cdot 2\overline{\rho z} \sin\varphi / \sin\varepsilon)}{\tau_{E,j} \cdot 2\overline{\rho z} \sin\varphi / \sin\varepsilon} \tag{B1.3.4}$$

$$= \Omega \cdot i \cdot const \cdot Z \cdot \left(\frac{E_0}{E} - 1\right)^x \frac{1 - \exp(-\tau_{E,j} \cdot 2\overline{\rho z} \sin\varphi / \sin\varepsilon)}{-\tau_{E,j} \cdot 2\overline{\rho z} \sin\varphi / \sin\varepsilon} dE$$

Here φ is an incident angle of the electrons with regard to the target surface, $\tau_{E,j}$ is the mass absorption coefficient for photons of energy E in the element j (cm $g^{-1}$), and ε is a take-off angle of x-rays with regard to the target surface.

The value of *const* is $1.36 \times 10^9 \, sr^{-1} mA^{-1} keV^{-1} s^{-1}$ and x=1.109-0.00435Z+0.00175$E_0$ by Schossmann [56]

**b) Results and Discussions**

Figures B1.3.1 – B1.3.3 demonstrate the result of the Bremsstrahlung simulations. The program can be found in Appendix B2. All figures were computed for an incident angle of the electrons equal to 40°, the take-off angle of the x-rays equal to 20° and the current 3mA (parameters of X-rays tube we used) for voltages of 30kV, 40kV and 60kV. The solid angle Ω is easy to calculate: if the area of the detector is 2.5cm×5cm=12.5 $cm^2$, the solid angle is Ω=12.5 $cm^2$/10000 $cm^2$=0.00125sr. All spectra are calculated for an Ag target without filtration.

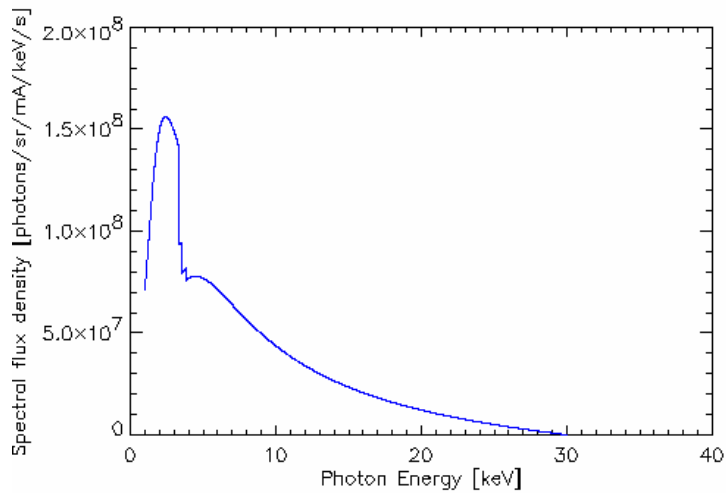

*Figure B1.3.1: Ag target at 30kV*

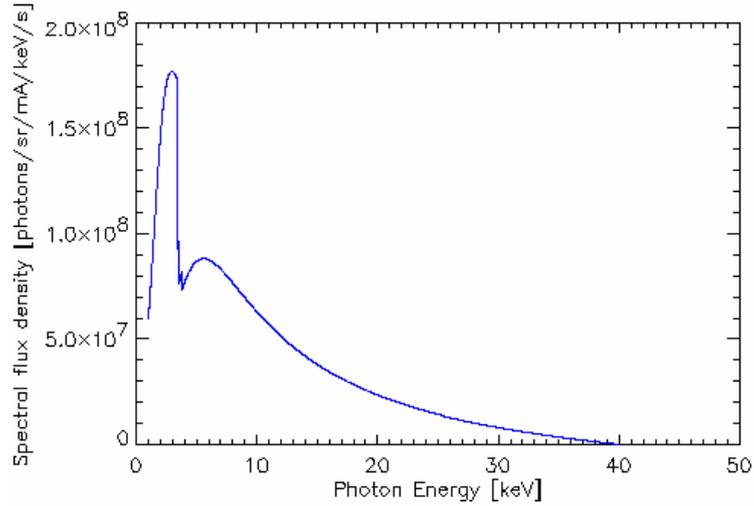

*Figure B1.3.2: Ag target at 40kV*

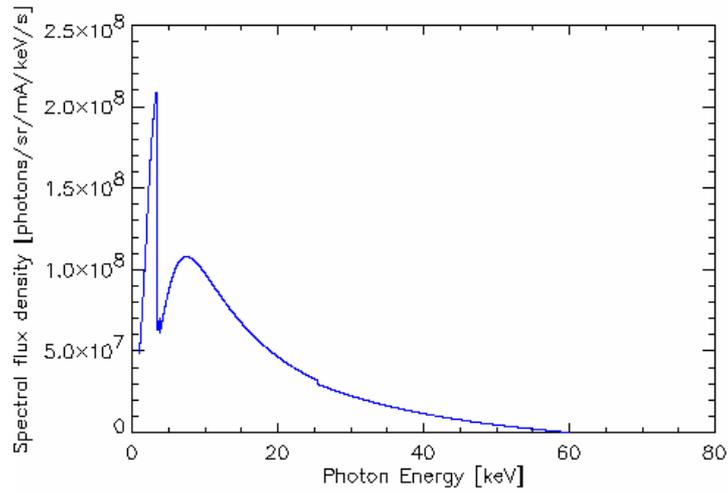

*Figure B1.3.3:* Ag target at 60 keV

With the developed program it is possible to simulate Bremsstrahling for all elements from the Mendeleev table for different electron energy.

**c) Characteristic radiation**

Characteristic count-rates $N_{jkl}$ are given by [52]:

$$N_{jkl} = \Omega \cdot i \cdot Const_{kl} \cdot \frac{1}{S_{jk}} \cdot R \cdot \omega_{jk} p_{jkl} f(\chi_{jkl}) \tag{B1.3.5}$$

Here j defines the element, k the ionized atomic level and l the level from where the k vacancy is filled. $Const_{kl} = 6 \times 10^{13} \, s^{-1} sr^{-1} mA^{-1}$. The stopping power factor $\frac{1}{S_{jk}}$ is following:

$$\frac{1}{S_{jk}} = \frac{z_k b_k}{Z}(U_0 \ln U_0 + 1 - U_0)\left[1 + 16.05\sqrt{\frac{J}{E_{jk}}}\frac{\sqrt{U_0} \ln U_0 + 2(1-\sqrt{U_0})}{U_0 \ln U_0 + 1 - U_0}\right] \quad (B1.3.6)$$

Here $z_K = 2$, $z_L = 8$, $b_k = 0.35$, $b_L = 0.25$, $U_0 = \frac{E_0}{E_{jk}}$, J=0.0135Z with energy $E_{jk}$ of the absorption edge of the corresponding characteristic radiation; $z_K$ and $b_k$ are valid for K-shells and $z_L$ and $b_L$ for L-shells.

R is a backscattering factor and by Myklebust [57] is equal to:

$$R = 1 - 0.008517Z + 3.613\times 10^{-5}Z^2 + 0.009583Z e^{-U_0} + 0.001141 E_0 \quad (B1.3.7)$$

Fluorescence $\varpi_{jk}$ for silver is equal 0.842 and the transition probabilities are 2.961, 1.571, 0.502, 0.0827 for $K_{\alpha 1}$, $K_{\alpha 2}$, $K_{\beta 1}$, $K_{\beta 2}$, respectively. The last term in the equation (B1.3.5) is responsible for the absorption of characteristic x-rays:

$$\chi_{jkl} = \frac{\tau}{\sin \varepsilon}$$
$$f(\chi_{jkl}) = \frac{1 - \exp(-\tau \cdot 2\overline{\rho z} \sin \varphi / \sin \varepsilon)}{\tau \cdot 2\overline{\rho z} \sin \varphi / \sin \varepsilon} \quad (B1.3.8)$$

Here absorption coefficient $\tau$ is for the energy $E_{jkl}$ of characteristic photons in the element j.

### d) Results and discussion

Figures B1.3.4-B1.3.6 show the Bremsstrahlung together with the Characteristic lines for 30kV, 40kV and 60kV for the Ag target calculated as described above. On the ordinate axis the logarithm of the number of photons dN is plotted for convenience of the presentation of the graphs.

In Figure B1.3.7 and B1.3.8 the zoom of the characteristic lines (for 30kV just like example) is presented. In this way it is easier to see the correctness of the characteristic line values.

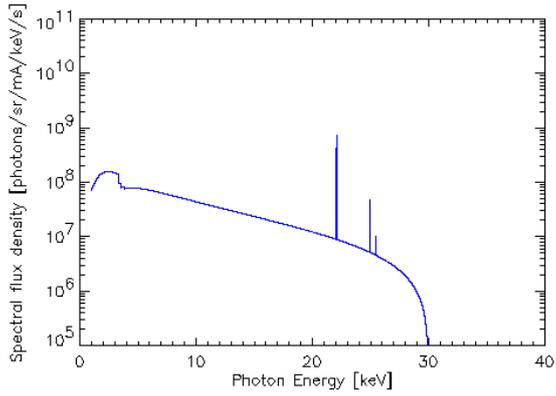
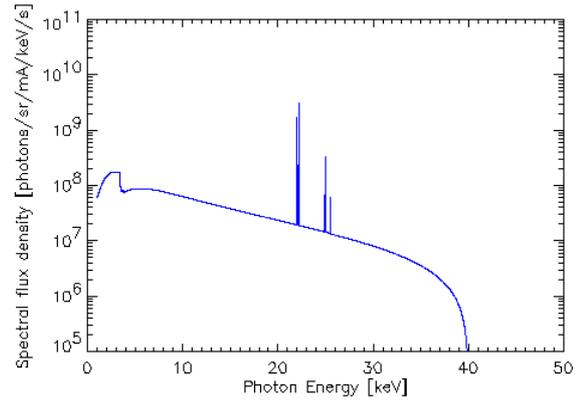

*Figure B1.3.4*: Ag target: Bremsstrahlung with characteristic lines of $K_{\alpha 1}$, $K_{\alpha 2}$, $K_{\beta 1}$, $K_{\beta 2}$ at 30kV. Logarithmic graphic.

*Figure B1.3.5*: Ag target: Bremsstrahlung with characteristic lines of $K_{\alpha 1}$, $K_{\alpha 2}$, $K_{\beta 1}$, $K_{\beta 2}$ at 40kV. Logarithmic graphic.

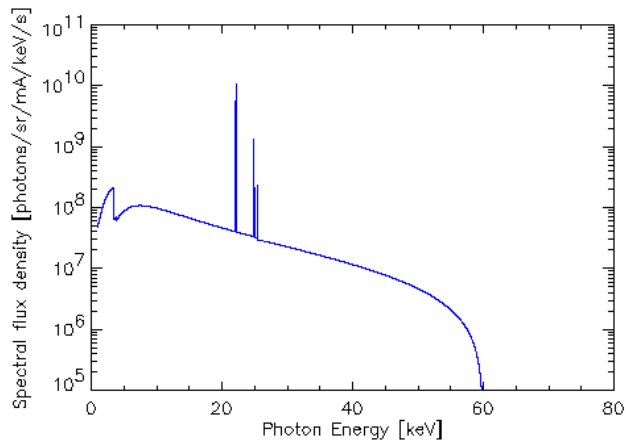

*Figure B1.3.6:* Ag target: Bremsstrahlung with characteristic lines of $K_{\alpha 1}$, $K_{\alpha 2}$, $K_{\beta 1}$, $K_{\beta 2}$ at 60kV. Logarithmic graphic.

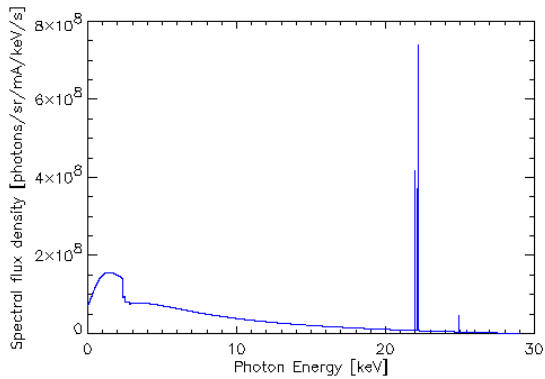
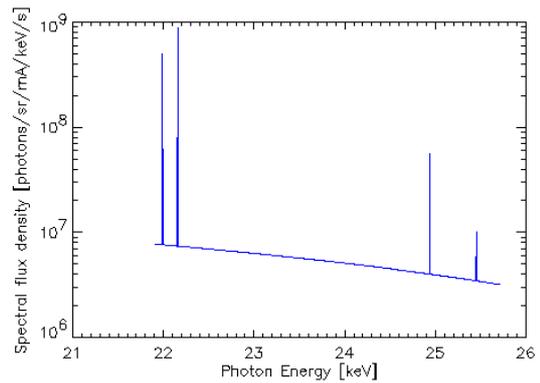

*Figure B1.3.7:* Ag target: Bremsstrahlung and Characteristic lines at 30 kV

*Figure B1.3.8:* Ag target: Characteristic lines at 30 kV

## B2. X-ray tube spectrum simulation program

### *B2.1. X-ray Tube Spectra: Bremsstrahlung*

pro x_ray_spectrum7F,el,energy,mu_energy=mu_energy

; NAME:
;      x_ray_spectrum7F.pro
; INPUTS:
;      **current** - x-ray tube current,[mA]
;    **solid_angle** - solid angle of photon flux from point source, [sr]
;      **voltage** - x-ray tube high voltage,[kV]
;      **epsilon** - take-off angle of x-rays with regard to the target surface
;      **phy** - an incidence angle with regard to the target surface
;    **el** – element name
;    **a** – photon energy vector size
; PROCEDURES:
;      other procedures to work with: murho.pro, el_info.pro, readabs.pro, mucalc.pro,
;                                     sigcalc.pro
; OUTPUTS:
;      dN - number of pseudomonochromatic photons,[1/s]
; HISTORY:
;      written by A.Khromova, October 2003

h=6.6260755D-39;*Planck's constant*
e=1.60217733D-19;*charge of the electron*
const=1.35D9;  *sr^(-1)mA^(-1)keV^(-1)s^(-1)*

;*To input initial parameters*
read, prompt='Enter x-ray tube current, [mA]:',current
read, prompt='Enter Solid Angle of photon flux from point source, [sr]:',solid_angle
read, prompt='Enter voltage, [kV]:',voltage
read, prompt='Enter take-off angle,[degree]:',epsilon
read, prompt='Enter incident angle, [degree]:',phy

el=''
read,prompt='Element name:',el
read,prompt='Photon energy vector size:',a
c=a/voltage
mu_energy=(FINDGEN(a)/c+1)*10D-1; *photon energy in [keV]*

mutotal=murho(el,mu_energy,density=density,atm_wght=atm_wght,name=name,el_z=el_z,p
hotomu=photomu); *mass absorption coefficient for photons of energy* **mu_energly** *in element* **el**

; *to make output for the element name, density and atomic weight*
print,format='("density",14x,"atm_wght",9x,"name",11x,"el_z")'
print, '======================================='
print, density,  atm_wght,  name,  el_z, format='(f0,4x,f,4x,a10,4x,i3,q)'

Z=el_z; *atomic number*
E0=voltage;*high energy limitation*

x=1.109-0.00435*Z+0.00175*E0; *exponent for a better fitting of experimental data to theory*
U0=E0/mu_energy
J=0.0135*Z
m=0.1382-(0.9211/sqrt(Z))
rho=density; in g cm-3

A=atm_wght ; *relative atomic weight*

nu=(E0^m)*(0.1904-0.2236*ALOG(Z)+0.1292*(ALOG(Z))^2-0.0149*(ALOG(Z))^3)

; z1-*linear dimension of depth in cm*
; rhoz1- *depth expressed in mass per unit area (g\*cm^(-2))*

rhoz1m=(A/Z)*(0.787D-5*sqrt(J)*E0^(3/2)+0.735D-6*E0^2)

rhoz=rhoz1m*((0.49269-1.0987*nu+0.78557*nu^2)/(0.70256.09865*nu+1.0046*nu^2+
+ALOG(U0)))*ALOG(U0)

dE=0.1D

;dN – *spectral distibution of the number of photons*
dN=(solid_angle*current*const*Z*(E0/mu_energy-1)^x)*((1-exp(-mutotal*2*rhoz*
*sin(!pi*phy/180)/sin(!pi*epsilon/180)))/(mutotal*2*rhoz*sin(!pi*phy/180)/sin(!pi*epsilon/1
80)))*dE

; *to plot the distribution of the number of photons versus the photon energy*
plot,mu_energy,dN

End

### B2.2. X-ray Tube Spectra: Bremsstrahlung + Characteristic lines

```
; NAME:
;       x_ray_spectrum10.pro
; INPUTS:
;       current - x-ray tube current,[mA]
;    solid_angle - solid angle of photon flux from point source, [sr]
;       voltage - x-ray tube high voltage,[kV]
;       epsilon - take-off angle of x-rays with regard to the target surface
;       phy - an incidence angle with regard to the target surface
;       el – element name
;       avector – photon energy vector size
; PROCEDURES:
;       other procedures to work with: murho.pro, el_info.pro, readabs.pro, mucalc.pro, sigcalc.pro
; OUTPUTS:
;       dN - number of pseudomonochromatic photons,[1/s]
; HISTORY:
;       written by A.Khromova, October 2003

pro x_ray_spectrum10,el, energy,mu_energy=mu_energy

   h=6.6260755D-39 ; Planck's constant
   e=1.60217733D-19 ; Charge of the electron
   const=1.35D9 ;  sr^(-1)mA^(-1)keV^(-1)s^(-1)

   ;To input initial parameters
   read, prompt='Enter x-ray tube current,[mA]:',current
   read, prompt='Enter Solid Angle of photon flux from point source, [sr]:',solid_angle
   read, prompt='Enter voltage, [kV]:',voltage
   read, prompt='Enter take-off angle, [degree]:',epsilon
   read, prompt='Enter incident angle,[degree]:',phy
   el=''
   read, prompt='Element name:',el
   read, prompt='Photon energy vector size:', avector

   c=avector/voltage

   mu_energy=(FINDGEN(avector)/c+1)*10D-1 ;  photon energy in [keV]

mutotal=murho(el,mu_energy,density=density,atm_wght=atm_wght,name=name,el_z=el_z,photomu=photomu) ; mass absorption coefficient for photons of energy mu_energly in element el

; to make output for the element name, density and atomic weight
print,format='("density",14x,"atm_wght",9x,"name",11x,"el_z")'
print,'======================================='
print,density, atm_wght,  name,  el_z, format='(f0,4x,f,4x,a10,4x,i3,q)'

Z=el_z ; atomic number
```

E0=voltage ; *high energy limitation*

x=1.109-0.00435*Z+0.00175*E0 ; *exponent for a better fitting of experimental data to theory*
U0=E0/mu_energy
J=0.0135*Z
m=0.1382-(0.9211/sqrt(Z))
rho=density

A=atm_wght ; *relative atomic weight*

nu=(E0^m)*(0.1904-0.2236*ALOG(Z)+0.1292*(ALOG(Z))^2-0.0149*(ALOG(Z))^3)

; *z1 - linear dimension of depth in cm*
; *rhoz1 - depth expressed in mass per unit area (g*cm^(-2))*

rhoz1m=(A/Z)*(0.787D-5*sqrt(J)*E0^(3/2)+0.735D-6*E0^2)

rhoz=rhoz1m*((0.49269-1.0987*nu+0.78557*nu^2)/(0.70256-1.09865*nu+1.0046*nu^2+
+ALOG(U0)))*ALOG(U0)

dE=0.1D

;dN – *spectral distibution of the number of photons*

dN=(solid_angle*current*const*Z*(E0/mu_energy-1)^x)*((1-exp(-mutotal*2*rhoz*
*sin(!pi*phy/180)/sin(!pi*epsilon/180)))/(mutotal*2*rhoz*sin(!pi*phy/180)/sin(!pi*epsilon/1
80)))*dE

; *to plot the distribution of the number of photons versus the photon energy*
;plot,mu_energy,dN

;**Characteristic radiation for silver**-----------------------------------------------------------------

mu_energy=[22.163D,21.991D, 24.942D] ; *absorption edge of the corresponding characteristic radiation*
Array=mu_energy[0:2]*10D2

w_jk=0.842 ; *fluorescence*
p_Ag_alfa=[2.961,1.571, 0.502] ; *transition probabilities*

U0_Ag_alfa=E0/mu_energy
mutotal=murho(el,mu_energy,density=density) ; *mass absorption coefficient for photons of energy*
**mu_energly**
                                        ; *in element* **el**
;*Stopping power factor*
S_Ag_alfa=2*0.35D/Z*(U0_Ag_alfa*alog(U0_Ag_alfa)+1-U0_Ag_alfa)*
*(1+16.05D*sqrt(J/mu_energy)*(sqrt(U0_Ag_alfa)*alog(U0_Ag_alfa)+
+2*(1-sqrt(U0_Ag_alfa)))/(U0_Ag_alfa*alog(U0_Ag_alfa)+1-U0_Ag_alfa))

;*Backscattering factor*
R_Ag_alfa=1-0.0081517*Z+3.613D-5*Z^2+0.009583*Z*exp(-U0_Ag_alfa)+0.001141*E0

; *Absorption term f_Ag_alfa*
f_Ag_alfa=(1-exp(-mutotal*2*rhoz*sin(!pi*phy/180)/sin(!pi*epsilon/180)))/
/(mutotal*2*rhoz*sin(!pi*phy/180)/sin(!pi*epsilon/180))

;*Characteristic count-rate*
Nch=solid_angle*current*5D13*S_Ag_alfa*R_Ag_alfa*w_jk*p_Ag_alfa*f_Ag_alfa

;a=600

d=(voltage/avector)*10D2 ;step

print,d
index_energy=FIX(Array/d)
print,index_energy

; *To put Bremsstrahlung and Characteristic lines together*

mu_energyfull=findgen(avector)
mu_energyfull[*]=0
mu_energyfull[index_energy]=Nch
;print,mu_energyfull
dNtotal=dN+mu_energyfull

c=avector/voltage

mu_energy=(FINDGEN(avector)/c+1)*10D-1
plot,mu_energy,dNtotal,/ylog

;live_plot,/ylog,dNtotal

End

**Erklärung**

  Hiermit erkläre ich, daß ich die vorligende Masterarbeit selbständig verfaßt und keine anderen als die angegebenen Quellen und Hilfsmittel benutzt, sowie Zitate und Ergebnisse Anderer kenntlich gemacht habe.

| | |
|---|---|
| ……………………………………………… | ………………………………………… |
| (Ort)　　　　(Datum) | (Unterschrift) |